\documentclass[twocolumn,letterpaper,aps,prc,longbibliography,superscriptaddress,showpacs,nofootinbib,floatfix]{revtex4-1}

\usepackage{graphicx}	
\usepackage{xspace}	
\usepackage{amsmath}

\newcommand{\pt}{\mbox{$p_T$}\xspace}

\newcommand{\Npart}{\mbox{$N_{\rm part}$}\xspace}
\newcommand{\Ncoll}{\mbox{$N_{\rm coll}$}\xspace}
\newcommand{\Nch}{\mbox{$N_{\rm ch}$}\xspace}
\newcommand{\Et}{\mbox{$E_T$}\xspace}

\def\mean#1{\left<#1\right>}

\newcommand{\sqsn}{\mbox{$\sqrt{s_{_{NN}}}$}\xspace}

\newcommand{\pp}{\mbox{$p$$+$$p$}\xspace}
\newcommand{\dau}{\mbox{$d$$+$Au}\xspace}
\newcommand{\ha}{\mbox{$h$$+$A}\xspace}

\newcommand{\auau}{\mbox{Au$+$Au}\xspace}
\newcommand{\pbpb}{\mbox{Pb$+$Pb}\xspace}
\newcommand{\cucu}{\mbox{Cu$+$Cu}\xspace}
\newcommand{\cuau}{\mbox{Cu$+$Au}\xspace}
\newcommand{\uu}{\mbox{U$+$U}\xspace}
\newcommand{\heau}{\mbox{$^{3}$He$+$Au}\xspace}

\newcommand{\Etemc}{\mbox{$E_{T {\rm EMC}}$}\xspace}
\newcommand{\Nqp}{\mbox{$N_{q{\rm p}}$}\xspace}

\newcommand{\dEt}{\mbox{$dE_T/d\eta$}\xspace}
\newcommand{\dNch}{\mbox{$dN_{\rm ch}/d\eta$}\xspace}
\newcommand{\dEtNorm}{\mbox{$(dE_T/d\eta) / (0.5 \Npart)$}\xspace}
\newcommand{\dNchNorm}{\mbox{$(dN_{\rm ch}/d\eta) / (0.5 \Npart)$}\xspace}
\newcommand{\dEtNormQ}{\mbox{$(dE_T/d\eta) / (0.5 \Nqp)$}\xspace}
\newcommand{\dNchNormQ}{\mbox{$(dN_{\rm ch}/d\eta) / (0.5 \Nqp)$}\xspace}
\newcommand{\ebj}{\mbox{$\varepsilon_{\rm BJ}$}\xspace}

\begin{document}

\title{Transverse energy production and charged-particle multiplicity at 
midrapidity in various systems from $\sqrt{s_{NN}}=$7.7 to 200~GeV}

\newcommand{\abilene}{Abilene Christian University, Abilene, Texas 79699, USA}
\newcommand{\augie}{Department of Physics, Augustana University, Sioux Falls, South Dakota 57197, USA}
\newcommand{\banaras}{Department of Physics, Banaras Hindu University, Varanasi 221005, India}
\newcommand{\barc}{Bhabha Atomic Research Centre, Bombay 400 085, India}
\newcommand{\baruch}{Baruch College, City University of New York, New York, New York, 10010 USA}
\newcommand{\bnlcoll}{Collider-Accelerator Department, Brookhaven National Laboratory, Upton, New York 11973-5000, USA}
\newcommand{\bnlphys}{Physics Department, Brookhaven National Laboratory, Upton, New York 11973-5000, USA}
\newcommand{\caucr}{University of California-Riverside, Riverside, California 92521, USA}
\newcommand{\charlesczech}{Charles University, Ovocn\'{y} trh 5, Praha 1, 116 36, Prague, Czech Republic}
\newcommand{\chonbuk}{Chonbuk National University, Jeonju, 561-756, Korea}
\newcommand{\ciae}{Science and Technology on Nuclear Data Laboratory, China Institute of Atomic Energy, Beijing 102413, P.~R.~China}
\newcommand{\cns}{Center for Nuclear Study, Graduate School of Science, University of Tokyo, 7-3-1 Hongo, Bunkyo, Tokyo 113-0033, Japan}
\newcommand{\colorado}{University of Colorado, Boulder, Colorado 80309, USA}
\newcommand{\columbia}{Columbia University, New York, New York 10027 and Nevis Laboratories, Irvington, New York 10533, USA}
\newcommand{\czechtech}{Czech Technical University, Zikova 4, 166 36 Prague 6, Czech Republic}
\newcommand{\dapnia}{Dapnia, CEA Saclay, F-91191, Gif-sur-Yvette, France}
\newcommand{\debrecen}{Debrecen University, H-4010 Debrecen, Egyetem t{\'e}r 1, Hungary}
\newcommand{\elte}{ELTE, E{\"o}tv{\"o}s Lor{\'a}nd University, H-1117 Budapest, P{\'a}zm{\'a}ny P.~s.~1/A, Hungary}
\newcommand{\ewha}{Ewha Womans University, Seoul 120-750, Korea}
\newcommand{\fit}{Florida Institute of Technology, Melbourne, Florida 32901, USA}
\newcommand{\fsu}{Florida State University, Tallahassee, Florida 32306, USA}
\newcommand{\gsu}{Georgia State University, Atlanta, Georgia 30303, USA}
\newcommand{\hanyang}{Hanyang University, Seoul 133-792, Korea}
\newcommand{\hiroshima}{Hiroshima University, Kagamiyama, Higashi-Hiroshima 739-8526, Japan}
\newcommand{\howard}{Department of Physics and Astronomy, Howard University, Washington, DC 20059, USA}
\newcommand{\ihepprot}{IHEP Protvino, State Research Center of Russian Federation, Institute for High Energy Physics, Protvino, 142281, Russia}
\newcommand{\illuiuc}{University of Illinois at Urbana-Champaign, Urbana, Illinois 61801, USA}
\newcommand{\inrras}{Institute for Nuclear Research of the Russian Academy of Sciences, prospekt 60-letiya Oktyabrya 7a, Moscow 117312, Russia}
\newcommand{\instpasczech}{Institute of Physics, Academy of Sciences of the Czech Republic, Na Slovance 2, 182 21 Prague 8, Czech Republic}
\newcommand{\isu}{Iowa State University, Ames, Iowa 50011, USA}
\newcommand{\jaea}{Advanced Science Research Center, Japan Atomic Energy Agency, 2-4 Shirakata Shirane, Tokai-mura, Naka-gun, Ibaraki-ken 319-1195, Japan}
\newcommand{\jinrdubna}{Joint Institute for Nuclear Research, 141980 Dubna, Moscow Region, Russia}
\newcommand{\jyvaskyla}{Helsinki Institute of Physics and University of Jyv{\"a}skyl{\"a}, P.O.Box 35, FI-40014 Jyv{\"a}skyl{\"a}, Finland}
\newcommand{\kaeri}{KAERI, Cyclotron Application Laboratory, Seoul, Korea}
\newcommand{\karoly}{K\'aroly R\'oberts University College, H-3200 Gy\"ngy\"os, M\'atrai \'ut 36, Hungary}
\newcommand{\kek}{KEK, High Energy Accelerator Research Organization, Tsukuba, Ibaraki 305-0801, Japan}
\newcommand{\korea}{Korea University, Seoul, 136-701, Korea}
\newcommand{\kurchatov}{National Research Center ``Kurchatov Institute", Moscow, 123098 Russia}
\newcommand{\kyoto}{Kyoto University, Kyoto 606-8502, Japan}
\newcommand{\labllr}{Laboratoire Leprince-Ringuet, Ecole Polytechnique, CNRS-IN2P3, Route de Saclay, F-91128, Palaiseau, France}
\newcommand{\lahorelums}{Physics Department, Lahore University of Management Sciences, Lahore 54792, Pakistan}
\newcommand{\lawllnl}{Lawrence Livermore National Laboratory, Livermore, California 94550, USA}
\newcommand{\losalamos}{Los Alamos National Laboratory, Los Alamos, New Mexico 87545, USA}
\newcommand{\lpc}{LPC, Universit{\'e} Blaise Pascal, CNRS-IN2P3, Clermont-Fd, 63177 Aubiere Cedex, France}
\newcommand{\lund}{Department of Physics, Lund University, Box 118, SE-221 00 Lund, Sweden}
\newcommand{\maryland}{University of Maryland, College Park, Maryland 20742, USA}
\newcommand{\mass}{Department of Physics, University of Massachusetts, Amherst, Massachusetts 01003-9337, USA}
\newcommand{\michigan}{Department of Physics, University of Michigan, Ann Arbor, Michigan 48109-1040, USA}
\newcommand{\muenster}{Institut f\"ur Kernphysik, University of Muenster, D-48149 Muenster, Germany}
\newcommand{\muhlenberg}{Muhlenberg College, Allentown, Pennsylvania 18104-5586, USA}
\newcommand{\myongji}{Myongji University, Yongin, Kyonggido 449-728, Korea}
\newcommand{\nagasaki}{Nagasaki Institute of Applied Science, Nagasaki-shi, Nagasaki 851-0193, Japan}
\newcommand{\nara}{Nara Women's University, Kita-uoya Nishi-machi Nara 630-8506, Japan}
\newcommand{\natmephi}{National Research Nuclear University, MEPhI, Moscow Engineering Physics Institute, Moscow, 115409, Russia}
\newcommand{\newmex}{University of New Mexico, Albuquerque, New Mexico 87131, USA}
\newcommand{\nmsu}{New Mexico State University, Las Cruces, New Mexico 88003, USA}
\newcommand{\ohio}{Department of Physics and Astronomy, Ohio University, Athens, Ohio 45701, USA}
\newcommand{\ornl}{Oak Ridge National Laboratory, Oak Ridge, Tennessee 37831, USA}
\newcommand{\orsay}{IPN-Orsay, Univ.~Paris-Sud, CNRS/IN2P3, Universit\'e Paris-Saclay, BP1, F-91406, Orsay, France}
\newcommand{\peking}{Peking University, Beijing 100871, P.~R.~China}
\newcommand{\pnpi}{PNPI, Petersburg Nuclear Physics Institute, Gatchina, Leningrad region, 188300, Russia}
\newcommand{\riken}{RIKEN Nishina Center for Accelerator-Based Science, Wako, Saitama 351-0198, Japan}
\newcommand{\rikjrbrc}{RIKEN BNL Research Center, Brookhaven National Laboratory, Upton, New York 11973-5000, USA}
\newcommand{\rikkyo}{Physics Department, Rikkyo University, 3-34-1 Nishi-Ikebukuro, Toshima, Tokyo 171-8501, Japan}
\newcommand{\saispbstu}{Saint Petersburg State Polytechnic University, St.~Petersburg, 195251 Russia}
\newcommand{\saopaulo}{Universidade de S{\~a}o Paulo, Instituto de F\'{\i}sica, Caixa Postal 66318, S{\~a}o Paulo CEP05315-970, Brazil}
\newcommand{\seoulnat}{Department of Physics and Astronomy, Seoul National University, Seoul 151-742, Korea}
\newcommand{\stonybrkc}{Chemistry Department, Stony Brook University, SUNY, Stony Brook, New York 11794-3400, USA}
\newcommand{\stonycrkp}{Department of Physics and Astronomy, Stony Brook University, SUNY, Stony Brook, New York 11794-3800, USA}
\newcommand{\subatech}{SUBATECH (Ecole des Mines de Nantes, CNRS-IN2P3, Universit{\'e} de Nantes) BP 20722-44307, Nantes, France}
\newcommand{\sungskku}{Sungkyunkwan University, Suwon, 440-746, Korea}
\newcommand{\tenn}{University of Tennessee, Knoxville, Tennessee 37996, USA}
\newcommand{\titech}{Department of Physics, Tokyo Institute of Technology, Oh-okayama, Meguro, Tokyo 152-8551, Japan}
\newcommand{\tsukuba}{Center for Integrated Research in Fundamental Science and Engineering, University of Tsukuba, Tsukuba, Ibaraki 305, Japan}
\newcommand{\vandy}{Vanderbilt University, Nashville, Tennessee 37235, USA}
\newcommand{\waseda}{Waseda University, Advanced Research Institute for Science and Engineering, 17  Kikui-cho, Shinjuku-ku, Tokyo 162-0044, Japan}
\newcommand{\weizmann}{Weizmann Institute, Rehovot 76100, Israel}
\newcommand{\wigner}{Institute for Particle and Nuclear Physics, Wigner Research Centre for Physics, Hungarian Academy of Sciences (Wigner RCP, RMKI) H-1525 Budapest 114, POBox 49, Budapest, Hungary}
\newcommand{\yonsei}{Yonsei University, IPAP, Seoul 120-749, Korea}
\newcommand{\zagreb}{University of Zagreb, Faculty of Science, Department of Physics, Bijeni\v{c}ka 32, HR-10002 Zagreb, Croatia}
\affiliation{\abilene}
\affiliation{\augie}
\affiliation{\banaras}
\affiliation{\barc}
\affiliation{\baruch}
\affiliation{\bnlcoll}
\affiliation{\bnlphys}
\affiliation{\caucr}
\affiliation{\charlesczech}
\affiliation{\chonbuk}
\affiliation{\ciae}
\affiliation{\cns}
\affiliation{\colorado}
\affiliation{\columbia}
\affiliation{\czechtech}
\affiliation{\dapnia}
\affiliation{\debrecen}
\affiliation{\elte}
\affiliation{\ewha}
\affiliation{\fit}
\affiliation{\fsu}
\affiliation{\gsu}
\affiliation{\hanyang}
\affiliation{\hiroshima}
\affiliation{\howard}
\affiliation{\ihepprot}
\affiliation{\illuiuc}
\affiliation{\inrras}
\affiliation{\instpasczech}
\affiliation{\isu}
\affiliation{\jaea}
\affiliation{\jinrdubna}
\affiliation{\jyvaskyla}
\affiliation{\kaeri}
\affiliation{\karoly}
\affiliation{\kek}
\affiliation{\korea}
\affiliation{\kurchatov}
\affiliation{\kyoto}
\affiliation{\labllr}
\affiliation{\lahorelums}
\affiliation{\lawllnl}
\affiliation{\losalamos}
\affiliation{\lpc}
\affiliation{\lund}
\affiliation{\maryland}
\affiliation{\mass}
\affiliation{\michigan}
\affiliation{\muenster}
\affiliation{\muhlenberg}
\affiliation{\myongji}
\affiliation{\nagasaki}
\affiliation{\nara}
\affiliation{\natmephi}
\affiliation{\newmex}
\affiliation{\nmsu}
\affiliation{\ohio}
\affiliation{\ornl}
\affiliation{\orsay}
\affiliation{\peking}
\affiliation{\pnpi}
\affiliation{\riken}
\affiliation{\rikjrbrc}
\affiliation{\rikkyo}
\affiliation{\saispbstu}
\affiliation{\saopaulo}
\affiliation{\seoulnat}
\affiliation{\stonybrkc}
\affiliation{\stonycrkp}
\affiliation{\subatech}
\affiliation{\sungskku}
\affiliation{\tenn}
\affiliation{\titech}
\affiliation{\tsukuba}
\affiliation{\vandy}
\affiliation{\waseda}
\affiliation{\weizmann}
\affiliation{\wigner}
\affiliation{\yonsei}
\affiliation{\zagreb}
\author{A.~Adare} \affiliation{\colorado} 
\author{S.~Afanasiev} \affiliation{\jinrdubna} 
\author{C.~Aidala} \affiliation{\columbia} \affiliation{\losalamos} \affiliation{\mass} \affiliation{\michigan} 
\author{N.N.~Ajitanand} \affiliation{\stonybrkc} 
\author{Y.~Akiba} \affiliation{\riken} \affiliation{\rikjrbrc} 
\author{R.~Akimoto} \affiliation{\cns} 
\author{H.~Al-Bataineh} \affiliation{\nmsu} 
\author{J.~Alexander} \affiliation{\stonybrkc} 
\author{M.~Alfred} \affiliation{\howard} 
\author{A.~Al-Jamel} \affiliation{\nmsu} 
% ...
\collaboration{PHENIX Collaboration} \noaffiliation

\date{\today}


\begin{abstract}


Measurements of midrapidity charged particle multiplicity distributions, 
$dN_{\rm ch}/d\eta$, and midrapidity transverse-energy distributions, 
$dE_T/d\eta$, are presented for a variety of collision systems and 
energies. Included are 
distributions for Au$+$Au collisions 
at $\sqrt{s_{_{NN}}}=200$, 130, 62.4, 39, 27, 19.6, 14.5, and 7.7~GeV, 
Cu$+$Cu collisions at $\sqrt{s_{_{NN}}}=200$ and 62.4~GeV, 
Cu$+$Au collisions at $\sqrt{s_{_{NN}}}=200$~GeV, 
U$+$U collisions at $\sqrt{s_{_{NN}}}=193$~GeV, 
$d$$+$Au collisions at $\sqrt{s_{_{NN}}}=200$~GeV, 
$^{3}$He$+$Au collisions at $\sqrt{s_{_{NN}}}=200$~GeV, and 
$p$$+$$p$ collisions at $\sqrt{s_{_{NN}}}=200$~GeV.  
Centrality-dependent distributions at midrapidity are presented in terms 
of the number of nucleon participants, $N_{\rm part}$, and the number of 
constituent quark participants, $N_{q{\rm p}}$.  For all $A$$+$$A$ 
collisions down to $\sqrt{s_{_{NN}}}=7.7$~GeV, it is observed that the midrapidity 
data are better described by scaling with $N_{q{\rm p}}$ than scaling with 
$N_{\rm part}$.  Also presented are estimates of the Bjorken energy 
density, $\varepsilon_{\rm BJ}$, and the ratio of $dE_T/d\eta$ to 
$dN_{\rm ch}/d\eta$, the latter of which is seen to be constant as a 
function of centrality for all systems.

\end{abstract}

	
\maketitle

\section{Introduction}
\label{sec:intro}

Systematic measurements of the centrality dependence of transverse energy 
production and charged particle multiplicity at midrapidity provide 
excellent characterization of the nuclear geometry of the reaction and are 
sensitive to the dynamics of the colliding system.  For example, 
measurements of \dNch and \dEt in \auau collisions at \sqsn = 200~GeV and 
130~GeV as a function of centrality expressed as the number of participant 
nucleons, \Npart, exhibit a nonlinear increase with increasing \Npart. 
This has been explained by a two-component model proportional to a linear 
combination of the number of collisions, \Ncoll, and 
\Npart~\cite{Wang:2000bf,Kharzeev:2000ph}. In a previous study by the 
PHENIX collaboration, measurements of \dEt and \dNch for \auau collisions 
at 200, 130, and 62.4~GeV are presented along with comparisons to the 
results of several models~\cite{Adler:2004zn}. The models that were 
examined included HIJING~\cite{Wang:1991hta}, a final state parton 
saturation model called EKRT~\cite{Eskola:1999fc}, an initial state parton 
saturation model called KLN~\cite{Kharzeev:2000ph}, and a multiphase 
transport model called AMPT~\cite{Lin:2000cx}. The comparisons showed that 
most models could reproduce some of the features of the data, but most 
failed in describing all of the data with the HIJING and AMPT models best 
describing the overall trends, including the nonlinear increase of \dEt 
and \dNch as a function of \Npart.

It was also proposed that \dNch is linearly proportional to the number of 
constituent-quark participants without a significant contribution from a 
hard scattering component~\cite{Eremin:2003qn}.  Recently, the PHENIX 
collaboration at Brookhaven National Laboratory's Relativistic Heavy Ion 
Collider (RHIC), presented \dEt distributions at midrapidity for \auau 
collisions at \sqsn = 200, 130, and 62.4~GeV, \dau collisions at \sqsn = 
200~GeV, and \pp collisions at \sqsn = 200~GeV~\cite{Adler:2013aqf}.  The 
data are better described by a model based upon the number of constituent 
quark participants than by the wounded nucleon model~\cite{Bialas:1976ed}.  
Here, this study will be extended to include both \dEt and \dNch 
measurements at midrapidity in \auau collisions down to \sqsn = 7.7~GeV. 
This study will also examine the centrality dependence of \dEt and \dNch 
for smaller systems including \cuau, \cucu, \dau, and \heau.

Recent lattice quantum chromodynamics (QCD) calculations indicate that 
the transition from quark to hadronic matter is a crossover transition 
at high temperature and small baryochemical potential, 
$\mu_{B}$~\cite{Aoki:2006we}.  At high values of $\mu_{B}$ and low 
temperatures, model calculations indicate the presence of a first-order 
phase transition and the possibility of a critical end point in the QCD 
phase diagram~\cite{Ejiri:2008xt}. Relativistic heavy ion collisions 
serve as excellent probes of the QCD phase 
diagram~\cite{Stephanov:1998dy}. The region of the QCD phase diagram 
sampled by the collisions can be controlled by changing the beam energy. 
Lowering the beam energy corresponds to raising the value of $\mu_{B}$.  
From 2010 to 2014, RHIC executed a beam energy scan program to explore 
the QCD phase diagram, look for evidence of the phase boundaries and 
search for evidence of the critical end point.  Presented here are \dEt 
and \dNch measurements from the beam energy scan as a function of 
centrality expressed as the number of nucleon participants, \Npart, from 
\auau collisions at \sqsn = 200, 130, 62.4, 39, 27, 19.6, 14.5, and 
7.7~GeV.

Over the past 15 years, PHENIX has collected a comprehensive dataset 
covering a wide variety of colliding nuclei and collision energies, 
including the \auau collision beam energy scan mentioned above. Presented 
here will be charged particle multiplicity and transverse energy 
measurements from the following systems: \auau collisions at \sqsn = 200, 
130, 62.4, 39, 27, 19.6, 14.5, and 7.7~GeV; \cucu collisions at \sqsn = 
200 and 62.4~GeV; \cuau collisions at \sqsn = 200~GeV; \uu collisions at 
\sqsn = 193~GeV; \heau collisions at \sqsn = 200~GeV; \dau collisions at 
\sqsn = 200~GeV; and \pp collisions at \sqsn = 200~GeV. The results will 
be discussed in the context of scaling with the number of participant 
nucleons (\Npart) and the number of participant quarks (\Nqp).

PHENIX has previously published charged particle multiplicity 
distributions from \auau collisions at \sqsn = 
200~GeV~\cite{Adler:2004zn}, \auau collisions at \sqsn = 
130~GeV~\cite{Adcox:2000sp, Adler:2004zn}, and \auau collisions at \sqsn = 
19.6~GeV~\cite{Adler:2004zn}. PHENIX has also previously published 
transverse energy distributions from \auau collisions at \sqsn = 
200~GeV~\cite{Adler:2004zn}, \auau collisions at \sqsn = 130 
GeV~\cite{Adcox:2001ry}, \auau collisions at \sqsn = 62.4 
GeV~\cite{Adler:2013aqf}, \auau collisions at \sqsn = 19.6 
GeV~\cite{Adler:2004zn}, and minimum-bias distributions for \dau and \pp 
collisions at \sqsn = 200~GeV~\cite{Adler:2013aqf}.  Here the previously 
published PHENIX results will be presented along with data from the many 
new collision systems in a consistent format to facilitate comparisons.

Similar measurements have been published by the other RHIC experiments.  
Charged particle multiplicity distributions have been published by BRAHMS 
for \auau collisions at \sqsn = 200 and 130~GeV~\cite{Arsene:2004fa}, STAR 
for \auau collisions at \sqsn = 130~GeV~\cite{Adler:2001yq}, and PHOBOS 
for \auau collisions at \sqsn = 200, 130, 62.4, 56, and 19.6~GeV along 
with \cucu collisions at \sqsn = 200 and 62.4~GeV, \dau collisions at 
\sqsn = 200~GeV, and \pp collisions at \sqsn = 410 and 200 
GeV~\cite{Alver:2010ck}. Transverse energy distributions have been 
published by STAR for \auau collisions at \sqsn = 200 
GeV~\cite{Adams:2004cb}. Presented here are many collision systems and 
energies that have not been previously published by PHENIX or the other 
RHIC experiments, especially for the transverse energy measurements. The 
first complete results on charged particle multiplicity and transverse 
energy from the RHIC beam energy scan program conducted from 2010 to 2014 
are also included.

This paper is organized as follows. The PHENIX detector and the methods 
used for centrality determination in each dataset will be described in 
Section~\ref{sec:detector}.  The analysis of the data to measure \dEt and 
\dNch including a description of estimates of the systematic uncertainties 
is described in Section~\ref{sec:analysis}. The centrality dependent 
results at midrapidity from the \auau beam energy scan in terms of 
\Npart 
are presented in Section~\ref{sec:besResults}. A description of the 
centrality dependent results at midrapidity for \cucu and \cuau collisions 
in terms of \Npart are found in Section~\ref{sec:cuResults}. A description 
of the centrality dependent results at midrapidity for \uu collisions in 
terms of \Npart are found in Section~\ref{sec:uResults}. 
Section~\ref{sec:smallResults} contains a description of the centrality 
dependent results at midrapidity of \heau and \dau collisions in terms of 
\Npart.  A review all of the centrality dependent results in terms of \Nqp 
is presented in Section~\ref{sec:quarks}. Section~\ref{sec:summary} 
contains a summary of the results. Data tables for all data sets are 
tabulated in the Appendix.

		\section{The PHENIX Detector}
		\label{sec:detector}

The PHENIX detector comprises two central spectrometer arms, two muon 
spectrometer arms, and a set of forward detectors. All of the detector 
components and their performance is described 
elsewhere~\cite{Adcox:2003zm}. The analysis of charged particle 
multiplicity utilizes detectors in the central arm 
spectrometer~\cite{Adcox:2003zp} including the drift chamber (DC) and pad 
chamber 1 (PC1) detectors. The drift chambers are cylindrically shaped and 
located radially from 2.0 to 2.4 meters.  The DC covers the pseudorapidity 
region $|\eta|<$0.35 and $90^{o}$ in azimuth for each arm. The DC has a 
resolution better than 150 $\mu$m in r-$\phi$, better than 2 mm in the z 
direction, and a two track separation better than 1.5 mm.  The PC1 
detector is a multiwire proportional chamber mounted on the outer radius 
of the drift chamber at 2.5 m from the beam axis.  PC1 covers the full 
central arm acceptance. PC1 measures minimum ionizing particles with an 
efficiency greater than 99.5\% with a position resolution of 1.7 mm by 3 
mm and a two track separation of 4 cm.  Reconstructed tracks from the 
drift chamber with an associated hit from PC1 are counted as charged 
particle tracks in the multiplicity measurement.

The analysis of transverse energy utilizes five of the lead-scintillator
(PbSc) electromagnetic-calorimeter (EMCal) sectors in the central arm 
spectrometers~\cite{Aphecetche:2003zr}.  Each calorimeter sector covers a 
pseudorapidity range of $|\eta|<$0.38 and subtends $22.5^{o}$ in azimuth 
for a total azimuthal coverage of $112.5^{o}$. The front face of each 
sector is located 5.1 m from the beam axis. Each sector contains 2592 PbSc 
towers arranged in a 36 $\times$ 72 array. Each tower has a 5.535 $\times$ 
5.535 cm surface area and a thickness of 0.85 nuclear interaction lengths 
or 18 radiation lengths.  The PbSc EMCal energy resolution has been 
measured using test beam electrons to be $\frac{\Delta E}{E} = 
\frac{8.1\%}{\sqrt{E (GeV)}} \oplus$ 2.1\%, with a measured response 
proportional to the incident electron energy to within $ \pm 2\%$ over the 
range $0.3 \leq E_{e} \leq 40$~GeV.

For all data sets, a minimum-bias trigger is provided by a pair of 
beam-beam counters (BBC)~\cite{Allen:2003zt}. Each BBC comprises 64 
individual \v{C}erenkov counters. Each BBC covers 2$\pi$ azimuthally and a 
pseudorapidity range of 3.0 $< |\eta| <$ 3.9. For \pp, \dau, and \heau 
collisions, an event is required to have at least one counter fire in each 
BBC. For all other collisions, at least two counters must fire in each 
BBC. The event vertex is reconstructed with a resolution of 2.0~cm in \pp 
collisions and 0.5~mm in central \auau collisions using the timing 
information from the BBCs.  All events are required to have an event 
vertex within 20 cm of the center of the detector.

Centrality determination in the original \sqsn = 200~GeV and \sqsn = 
130~GeV \auau PHENIX analysis is based upon the total charge deposited in 
the BBCs and the total energy deposited in the zero-degree calorimeters 
(ZDC)~\cite{Allen:2003zt}.  The ZDCs are a pair of hadronic calorimeters 
that cover the pseudorapidity range $|\eta|>$6. For subsequent data sets 
taken after 2002, only the BBC information is used for the centrality 
determination, including the following data sets: \cuau at \sqsn = 
200~GeV, \cucu at \sqsn = 200~GeV, \uu at \sqsn = 193~GeV, \heau at \sqsn 
= 200~GeV, and \dau at \sqsn = 200~GeV. As the collision energy decreases, 
the width of the pseudorapidity distribution of produced particles becomes 
more narrow~\cite{Back:2004dy}. As a result, for energies below \sqsn = 
130~GeV, the acceptance of the ZDC is reduced, therefore only the BBC 
information is used for \auau collisions at \sqsn = 62.4 and 39~GeV, and 
for \cucu collisions at \sqsn = 62.4~GeV.  Below \sqsn = 39~GeV, the BBC 
acceptance becomes sensitive to the presence of beam fragments, which 
affects the linear response of the BBC to the centrality. To avoid this 
nonlinear response, the reaction-plane detector 
(RXNP)~\cite{Richardson:2010hm} is used for the centrality determination 
for \auau collisions at \sqsn = 7.7~GeV, which was taken during the 2010 
running period. The RXNP comprises two sets of plastic scintillators 
positioned on either side of the collision vertex. Each RXNP detector is 
arranged in 12 azimuthal segments separated into an inner and outer ring. 
The RXNP has an azimuthal coverage of 2$\pi$. The pseudorapidity coverage 
is 1.5$<|\eta|<$2.8 and 1.0$<|\eta|<$1.5 for the inner and outer ring, 
respectively. A 2 cm thick lead converter is located directly in front of 
the RXNP scintillators with respect to the collision region, which allows 
the RXNP to also measure contributions from neutral particles through 
conversion electrons. The RXNP is designed to measure the reaction plane 
angle, but it can also function well as a centrality detector, because the 
magnitude of the total charge measured by the RXNP is dependent on the 
centrality of the collision. In order to minimize contamination from beam 
fragments, only the outer ring of the RXNP is used for centrality 
determination for \auau collisions at \sqsn = 7.7~GeV. For the 2011 data 
taking period and later when the \auau data sets at \sqsn = 27, 19.6, and 
14.5~GeV were collected, the RXNP was removed in order to install a 
silicon vertex detector, which was being commissioned during this time. 
So, for these two data sets, the multiplicity of hits in the PC1 detector 
were used to determine the centrality.  A summary of the centrality 
detectors used for each dataset is included in 
Table~\ref{tab:datasetSummary}.

		\section{Data Analysis}
		\label{sec:analysis}

Table~\ref{tab:datasetSummary} provides a summary of the data sets used in 
this analysis. For \auau collisions at \sqsn = 62.4~GeV, the \dEt analysis 
uses data taken in 2004~\cite{Adler:2013aqf} and the \dNch analysis uses 
data taken in 2010. The number of events are those events that pass the 
minimum-bias trigger condition for the dataset and have an event vertex 
within 20 cm of the center of the detector.

\begin{table}[htb]
\caption{Summary of the data sets used in this analysis.}
\label{tab:datasetSummary}
\begin{ruledtabular}
\begin{tabular}{cccccc}
\sqsn (GeV) & System & Year & $N_{events}$ & Centrality & Trigger eff.\\[0.2pc]\hline
200  & \auau      & 2002 & 270 k  & BBC+ZDC & $93 \pm 3$\%\\
200  & \auau      & 2004 & 133 M  & BBC+ZDC & $93 \pm 3$\%\\
130  & \auau      & 2000 & 160 k  & BBC+ZDC & $93 \pm 3$\%\\
62.4 & \auau      & 2004 & 20 M   & BBC     & $86 \pm 3$\%\\
62.4 & \auau      & 2010 & 12 M   & BBC     & $86 \pm 3$\%\\
39   & \auau      & 2010 & 132 M  & BBC     & $86 \pm 3$\%\\
27   & \auau      & 2011 & 24.5 M & PC1     & $86 \pm 3$\%\\
19.6 & \auau      & 2011 & 6.3 M  & PC1     & $86 \pm 3$\%\\
14.5 & \auau      & 2014 & 6.8 M  & PC1     & $85 \pm 3$\%\\
7.7  & \auau      & 2010 & 803 k  & RXNP    & $75 \pm 3$\%\\
200  & \cucu      & 2005 & 558 M  & BBC     & $93 \pm 3$\%\\
62.4 & \cucu      & 2005 & 175 M  & BBC     & $88 \pm 3$\%\\
200  & \cuau      & 2012 & 2.6 B  & BBC     & $93 \pm 3$\%\\
193  & \uu        & 2012 & 317 M  & BBC     & $93 \pm 3$\%\\
200  & \heau      & 2014 & 1.6 B  & BBC     & $88 \pm 4$\%\\
200  & \dau       & 2008 & 1.4 B  & BBC     & $88 \pm 4$\%\\
200  & \pp        & 2003 & 14.6 M & ---     & $54.8 \pm 5.3$\%\\
\end{tabular}
\end{ruledtabular}
\end{table}

\subsection{Transverse Energy Analysis}

\begin{figure}[!htb] 
  \includegraphics[width=0.98\linewidth]{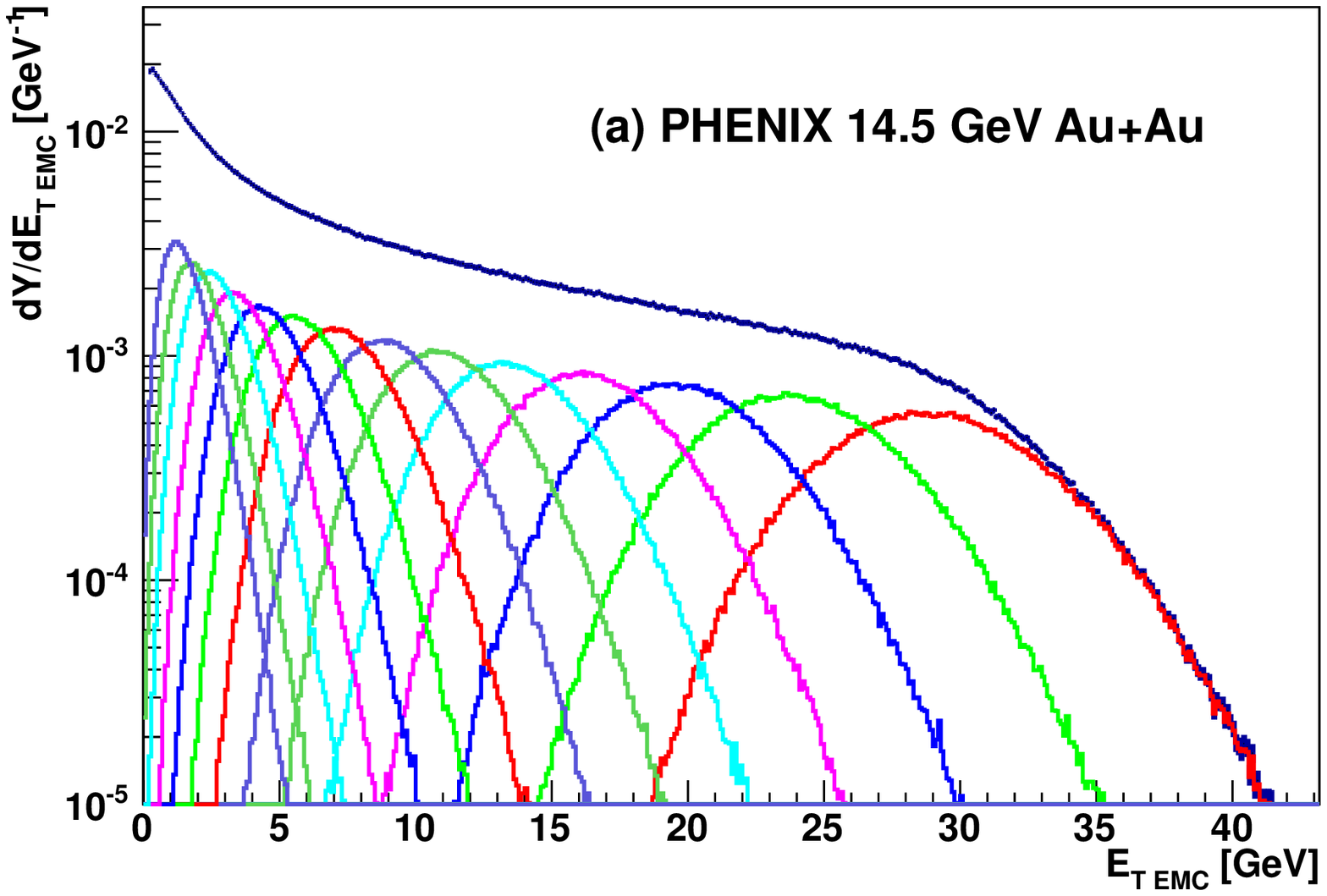}
  \includegraphics[width=0.98\linewidth]{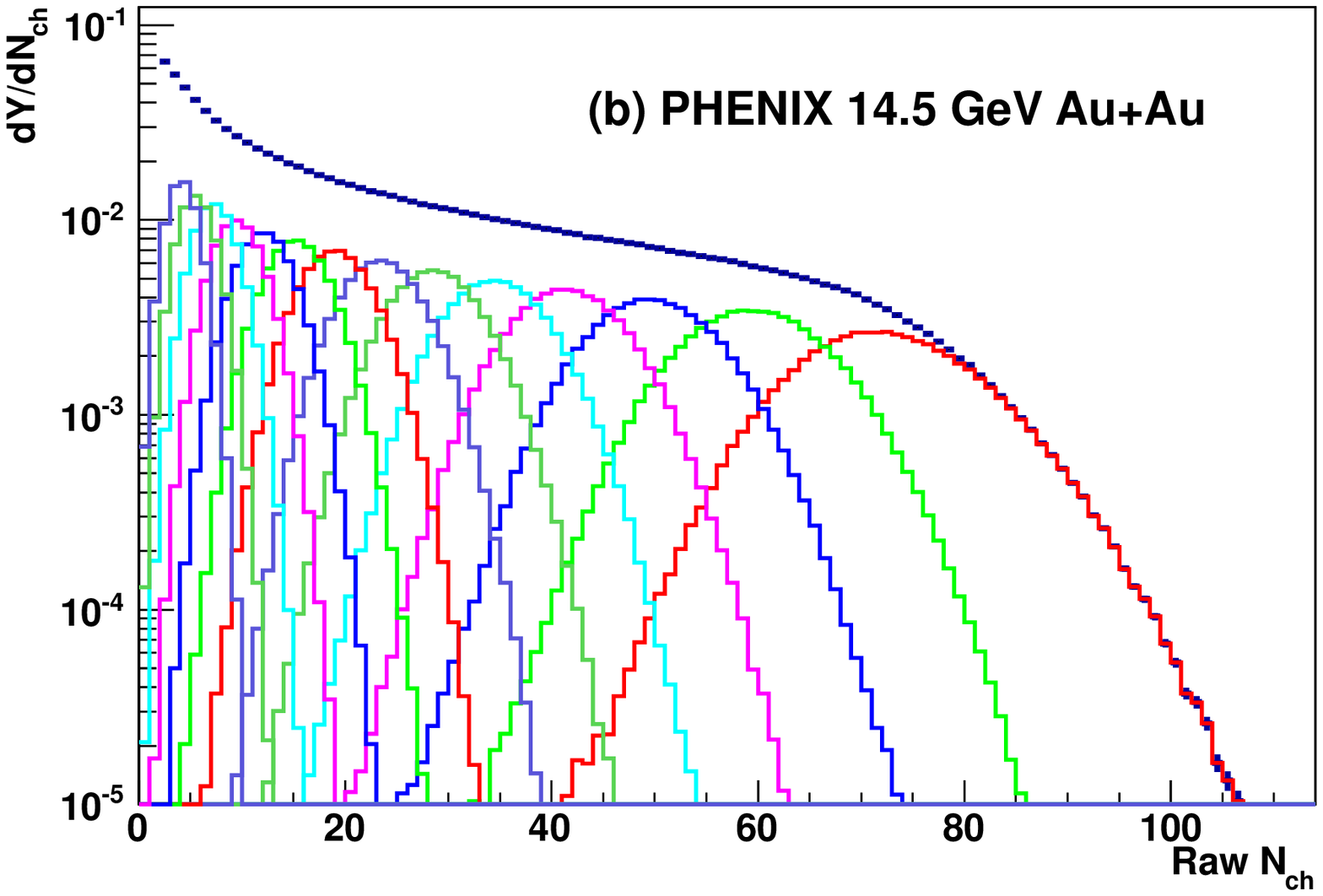}
  \caption{(Color online) 
Raw \Etemc (a) and \Nch (b) distributions for $\sqrt{s_{\rm NN}}=14.5$~GeV 
\auau collisions. Shown are the minimum-bias distribution along with the 
distributions in 5\% wide centrality bins.  All the plots are normalized 
so that the integral of the minimum-bias distribution is unity.}
      \label{fig:rawEtNch}
\end{figure}

The analysis procedure for \dEt is described in detail in 
Ref.~\cite{Adler:2013aqf} and summarized here.  The absolute energy scale 
for each EMCal sector is calibrated using the $\pi^{0}$ mass peak from 
pairs of reconstructed EMCal clusters for each dataset. The transverse 
energy for each event was computed using clusters in the EMCal with an 
energy greater than 30 MeV composed of adjacent towers each with a 
deposited energy of more than 10 MeV. Faulty towers and all towers in a $3 
\times 3$ tower area around any faulty tower are excluded from the 
analysis.  The transverse energy \Et is a multiparticle variable defined 
as the sum
   \begin{eqnarray}
   \Et&=&\sum_i E_i\,\sin\theta_i\cr
   d\Et(\eta)/d\eta&=&\sin\theta(\eta)\, dE(\eta)/d\eta \quad, 
   \label{eq:ETdef}
   \end{eqnarray}

where $\theta_{i}$ is the polar angle, $\eta=-\ln \tan(\theta/2)$ is the 
pseudorapidity, $E_i$ is by convention taken as the kinetic energy for 
baryons, the kinetic energy + 2 $m_{N}$ for antibaryons, and the total 
energy for all other particles, where $m_{N}$ is the nucleon mass. The sum 
is taken over all particles emitted into a fixed solid angle for each 
event. An example of the raw \Etemc distributions as a function of 
centrality for \auau collisions at \sqsn = 14.5~GeV are shown in 
Fig.~\ref{fig:rawEtNch}(a).

In order to obtain the total hadronic \Et within a reference acceptance of 
$\Delta\eta=1.0, \Delta\phi=2\pi$ from the measured raw transverse energy, 
\Etemc, the total correction can be decomposed into three main components. 
First is a correction by a factor of 4.188 to account for the fiducial 
acceptance in azimuth and pseudorapidity. Second, a correction factor is 
applied to account for disabled calorimeter towers not used in the 
analysis. Third is a factor, $k$, which is the ratio of the total hadronic 
\Et in the fiducial aperture to the measured \Etemc. Details on the 
estimate of the values of the $k$ factor are given below.

The $k$ factor comprises three components. The first component, denoted 
$k_{response}$, is due to the fact that the EMCal was designed for the 
detection of electromagnetic particles~\cite{Adcox:2001ry}. Hadronic 
particles passing through the EMCal only deposit a fraction of their total 
energy. The average EMCal response is estimated for the various particle 
species using the HIJING~\cite{Wang:1991hta} event generator for \sqsn 
above 7.7~GeV and the URQMD~\cite{Bleicher:1999xi} event generator for 
Au$+$Au collisions at \sqsn = 7.7~GeV. The event generator output is 
processed through a {\sc geant}-based Monte Carlo simulation of the PHENIX 
detector. For all of the data sets, 75\% of the total energy incident on 
the EMCal is measured, thus $k_{response}$ = 1/0.75 = 1.33. The second 
component of the $k$ factor, denoted $k_{inflow}$, is a correction for 
energy inflow from outside the fiducial aperture of the EMCal. This energy 
inflow has two sources: from parent particles with an original trajectory 
outside of the fiducial aperture whose decay products are incident within 
the fiducial aperture, and from particles that reflect off of the PHENIX 
magnet poles into the EMCal fiducial aperture. The energy inflow 
contribution is 24\% of the measured energy, thus $k_{inflow}$ = 1-0.24 = 
0.76. The third component of the $k$ factor, denoted $k_{losses}$, is due 
to energy losses. There are three components to the energy loss: from 
particles with an original trajectory inside the fiducial aperture of the 
EMCal whose decay products are outside of the fiducial aperture (10\%), 
from energy losses at the edges of the EMCal (6\%), and from energy losses 
due to the energy thresholds (6\%). The total contribution from energy 
losses is 22\%, thus $k_{losses}$ = 1/(1-0.22) = 1.282. The total $k$ 
factor correction is $k = k_{response} \times k_{inflow} 
\times k_{losses}$ = 1.30. This value varies by less than 1\% for 
all data sets.

There are several contributions to the systematic uncertainties for the 
\dEt measurement which are added in quadrature to obtain the total 
uncertainty.  These contributions include the following: uncertainties due 
to the energy response of the EMCal, uncertainties due to the estimate of 
the EMCal acceptance, uncertainties due to the estimate of losses and 
inflow, uncertainties due to sector-by-sector variations, uncertainties 
due to the noise background estimate, uncertainties due to the trigger 
background estimate, and uncertainties due to the trigger efficiency 
estimate. A summary of the systematic uncertainties for the \dEt analysis 
of each dataset is listed in Table~\ref{tab:etErrors} for each dataset and 
further explained below.

There is an uncertainty due to the energy response of the EMCal. This 
includes uncertainties in the absolute energy scale, uncertainties in the 
estimate of the hadronic response, uncertainties from energy losses on the 
EMCal edges and uncertainties from energy thresholds. The uncertainties in 
the hadronic response include a 3\% uncertainty estimated using a 
comparison of the simulated energy deposited by hadrons with different 
momenta with test beam data~\cite{Aphecetche:2003zr} along with an 
additional 1\% uncertainty in the particle composition and momentum 
distribution. There is an estimated uncertainty of 2\% for the calculation 
of the EMCal acceptance. There is an estimated uncertainty of 3\% for the 
calculation of the fraction of the total energy incident on the EMCal 
fiducial area (losses and inflow). There is an uncertainty due to 
sector-by-sector variations in the energy measurement. There is an 
uncertainty due to the noise, or background, contribution which is 
estimated to be consistent with zero with uncertainties determined by 
measuring the average energy deposited per sector in events where all the 
particles are screened by the central magnet pole tips by requiring an 
interaction z-vertex of $+50 < z < +60$ cm and $-60 < z < -50$ cm. There 
is a centrality-dependent uncertainty for background due to multiple 
interactions and trigger effects.

There is also an uncertainty in the trigger efficiency determination.  
The method by which the trigger efficiency is calculated is described 
in~\cite{Adler:2004zn}. The BBC trigger efficiency for \auau collisions 
ranges from 93\% at \sqsn = 200~GeV to 75\% at \sqsn = 7.7~GeV. The 
trigger efficiencies for each dataset are summarized in 
Table~\ref{tab:datasetSummary}. Note that the trigger inefficiency leads 
to a partial loss of the more peripheral collisions while the trigger is 
fully efficient for midcentral and central collisions.  Because the 
centrality is defined for a given event as a percentage of the total 
geometrical cross section, an uncertainty in the trigger efficiency 
translates into an uncertainty in the centrality definition. This 
uncertainty is estimated by measuring the variation in \dEt by redefining 
the centrality using trigger efficiencies that vary by $\pm 1$ standard 
deviation.

The trigger efficiency uncertainty allows for bending or inclination of 
the points. So, when plotting \dEtNorm and \dNchNorm, the trigger 
efficiency will be represented by error bands about the points within 
which the points can be tilted. The other systematic and statistical 
uncertainties are represented by error bars.

\begin{table*}[htb]
\caption{Summary of the systematic uncertainties for the \dEt measurement 
for each dataset, given in percent (\%). If a range is specified, the 
value for central collisions is listed first and the value for the most 
peripheral collisions presented for the dataset is listed second.  If no 
value is specified, then there is no contribution to the systematic 
uncertainty for that dataset.}
\label{tab:etErrors}
\begin{ruledtabular}
\begin{tabular}{cccccccc}
Dataset               & energy resp. & acceptance & losses \& inflow & sector-by-sector & noise & trigger bkg. & trigger eff.\\[0.2pc]\hline
200~GeV \auau       & 3.9 & 2.0 & 3.0 & --- & 0.2-6.0  & ---        & 0.3-16.0 \\
130~GeV \auau       & 3.8 & 2.0 & 3.0 & --- & 0.4-10.0 & ---        & 0.3-16.0 \\
62.4~GeV \auau      & 4.3 & 2.0 & 3.0 & 2.2 & 0.4-4.1  & 0.01-0.06  & 0.3-16.1 \\
39~GeV \auau        & 4.5 & 2.0 & 3.0 & 1.6 & 0.5-3.6  & 0.002-0.02 & 0.2-16.3 \\
27~GeV \auau        & 4.5 & 2.0 & 3.0 & 2.2 & 0.5-3.5  & 0.006-0.04 & 0.3-13.1 \\
19.6~GeV \auau      & 4.7 & 2.0 & 3.0 & 2.8 & 0.5-3.5  & 0.008-0.07 & 0.3-13.4 \\
14.5~GeV \auau      & 4.7 & 2.0 & 3.0 & 2.9 & 0.5-3.4  & 0.007-0.04 & 0.3-9.8  \\
7.7~GeV \auau       & 4.7 & 2.0 & 3.0 & 3.7 & 0.5-3.4  & 0.002-0.05 & 0.4-10.6 \\
200~GeV \cucu       & 3.9 & 2.0 & 3.0 & 5.9 & 0.2-6.0  & 0.002-0.04 & 0.3-6.5  \\
62.4~GeV \cucu      & 4.5 & 2.0 & 3.0 & 2.2 & 0.4-4.1  & 0.006-0.02 & 0.3-8.1  \\
200~GeV \cuau       & 3.9 & 2.0 & 3.0 & 2.8 & 0.5-3.5  & 0.02-0.20  & 0.2-8.8  \\
193~GeV \uu         & 3.9 & 2.0 & 3.0 & 2.5 & 0.2-6.0  & 0.001-0.03 & 0.4-9.3  \\
200~GeV \dau        & 3.9 & 2.0 & 3.0 & 6.5 & 0.2-0.2  & 0.13-0.21  & 0.3-5.1  \\
200~GeV \heau       & 3.9 & 2.0 & 3.0 & 3.9 & 0.2-0.2  & 0.08-0.16  & 0.2-5.2  \\
200~GeV \pp      & 3.9 & 2.0 & 3.0 & 3.9 & 0.2         & 0.60       & ---      \\
\end{tabular}
\end{ruledtabular}
\end{table*}

	\subsection{Charged Particle Multiplicity Analysis}

In previous PHENIX publications~\cite{Adcox:2000sp, Adler:2004zn} for 
\auau collisions at \sqsn = 200 and 130~GeV, charged particle multiplicity 
was measured using cluster pairs reconstructed from the PC1 and PC3 
detectors in the absence of a magnetic field.  The \dNch values quoted 
here for \auau collisions at \sqsn = 200 and 130~GeV are from the previous 
analyses. For all other collision species and collision energies, charged 
particle multiplicity is measured using reconstructed tracks from the 
drift chamber that have an unambiguous match to a reconstructed cluster in 
the PC1 detector with the magnetic field turned on.  In order to remove 
multiple counting of incorrectly reconstructed tracks in the drift 
chamber, commonly referred to as ghost tracks, a charge-dependent track 
proximity cut is applied.  The two methods give consistent results for 
200~GeV \auau collisions.  An example of the raw \Nch distributions as a 
function of centrality for the \auau collisions at \sqsn = 14.5~GeV are 
shown in Fig.~\ref{fig:rawEtNch} (b).

In order to obtain the total charged particle \Nch within a reference 
acceptance of $\Delta\eta=1.0, \Delta\phi=2\pi$ from the measured raw 
multiplicity, five corrections are applied.  First is a correction of 3.74 
to account for the fiducial acceptance in azimuth and pseudorapidity.  
The second correction is applied to account for drift chamber and PC1 
inefficiencies within the fiducial acceptance.  The third correction is 
applied to account for particles with a transverse momentum below the 200 
MeV/c minimum \pt cut applied to reconstructed tracks. This correction is 
determined using the average of results from the HIJING event 
generator~\cite{Wang:1991hta} and the URQMD event 
generator~\cite{Bleicher:1999xi} to estimate the fraction of the total 
charged particle multiplicity lying below \pt = 200 MeV/c. The collision 
energy cutoff for the HIJING event generator lies above \sqsn = 7.7~GeV, 
so only URQMD is used for \auau collisions at \sqsn = 7.7~GeV. This 
correction is 22\% for \auau collisions at \sqsn = 62.4~GeV and 23\% for 
\auau collisions at \sqsn = 7.7~GeV.  There is an estimated 2\% 
uncertainty for this correction. The fourth correction is a 
centrality-dependent correction for the track reconstruction efficiency.

The last correction is an in-flight decay correction that accounts for 
particle decays after the collision interaction that can add or remove 
charged particles from the measured multiplicity. This includes primary 
charged particles that decay and miss the detector. It also includes 
feed-down from neutral primary particle decays that go into the detector. 
This correction is determined by processing simulated events from the 
HIJING~\cite{Wang:1991hta} event generator for \sqsn above 7.7~GeV, and 
the URQMD~\cite{Bleicher:1999xi} event generator at \sqsn = 7.7~GeV. Below 
\sqsn = 62.4~GeV, results from the two event generators are consistent 
with each other within the uncertainties. The event generator output is 
processed through a {\sc geant}-based simulation of the PHENIX detector 
response. For \auau collisions, this correction varies from 0.99 at \sqsn 
= 200~GeV to 1.061 at \sqsn = 7.7~GeV. The energy dependence is primarily 
due to the decrease of the particle momenta and the narrowing of the width 
of the $\eta$ distribution at lower energies that affects the number of 
tracks from the decay of particles coming from comparable rapidities.

\begin{table*}[htb]
\caption{Summary of the systematic uncertainties for charged particle 
multiplicity for each dataset given in percent (\%). If a range is 
specified, the value for central collisions is listed first and the value 
for the most peripheral collisions presented for the dataset is listed 
second. If no value is specified, then there is no contribution to the 
systematic uncertainty for that dataset.}
\label{tab:nchErrors}
\begin{ruledtabular}
\begin{tabular}{cccccccc}
Dataset         & acceptance & decays & low \pt & occupancy & tracking eff. & trigger bkg. & trigger eff.\\[0.2pc]\hline
200~GeV \auau   & 2.3 & 2.9 & 2.0 & 3.5-0.10 & --- & 1.0         & 0.3-16.0 \\
130~GeV \auau   & 2.5 & 2.5 & 2.0 & 3.1-0.10 & --- & 1.0         & 0.3-16.0 \\
62.4~GeV \auau  & 4.0 & 5.0 & 2.0 & 3.5-0.10 & 5.0 & 0.001-0.03  & 0.2-16.1 \\
39~GeV \auau    & 4.0 & 5.4 & 2.0 & 3.0-0.03 & 5.0 & 0.001-0.009 & 0.2-13.0 \\
27~GeV \auau    & 4.0 & 5.6 & 2.0 & 2.0-0.01 & 5.0 & 0.01-0.03   & 0.2-13.3 \\
19.6~GeV \auau  & 4.0 & 5.7 & 2.0 & 1.9-0.01 & 5.0 & 0.002-0.003 & 0.2-9.3  \\
14.5~GeV \auau  & 4.0 & 5.8 & 2.0 & 1.9-0.01 & 5.0 & 0.001-0.007 & 0.3-9.8  \\
7.7~GeV \auau   & 4.0 & 5.9 & 2.0 & 1.2-0.01 & 5.0 & 0.001-0.03  & 0.4-12.3 \\
200~GeV \cucu   & 4.0 & 2.9 & 2.0 & 1.5-0.01 & 5.0 & 0.03-0.08   & 0.3-8.0  \\
62.4~GeV \cucu  & 4.0 & 5.0 & 2.0 & 1.0-0.01 & 5.0 & 0.02-0.01   & 0.3-9.2  \\
200~GeV \cuau   & 4.0 & 2.9 & 2.0 & 2.6-0.05 & 5.0 & 0.001-0.07  & 0.9-10.1 \\
193~GeV \uu     & 4.0 & 2.9 & 2.0 & 3.5-0.10 & 5.0 & 0.001-0.01  & 0.4-9.3  \\
200~GeV \dau    & 4.0 & 2.9 & 2.0 & 0.1-0.01 & 5.0 & 0.001-0.001 & 0.3-7.2  \\
200~GeV \heau   & 4.0 & 2.9 & 2.0 & 0.1-0.01 & 5.0 & 0.001-0.001 & 0.2-6.5  \\
200~GeV \pp     & 4.0 & 2.9 & 2.0 & 0.01     & 5.0 & 0.0015      & ---      \\
\end{tabular}
\end{ruledtabular}
\end{table*}

There are several contributions to the systematic uncertainties for the 
\dNch measurement which are added in quadrature to obtain the total 
uncertainty.  A summary of the systematic uncertainties for the \dNch 
analysis for all data sets is listed in Table~\ref{tab:nchErrors}. There 
is an estimated uncertainty of 4\% for the acceptance correction. There is 
an uncertainty for the estimate of the correction for in-flight decays 
that varies from 2.9\% at \sqsn = 200~GeV to 5.9\% at \sqsn = 7.7~GeV.  
There is a 2\% uncertainty for the estimate of charged particle 
multiplicity for low \pt below 200 MeV/c. There is a centrality dependent 
uncertainty due to the occupancy of the PC1 detector that varies from 
3.5\% to 1.2\% for \auau central collisions from \sqsn = 200 to 7.7~GeV.  
There is an estimated 5\% uncertainty for the tracking efficiency 
estimate. There is a centrality-dependent uncertainty for background due 
to trigger effects and multiple interactions.  Finally, there is an 
uncertainty for the determination of the trigger efficiency, which is 
estimated in the same manner as for the \dEt analysis.

\begin{figure*}[!htb] 
	\begin{minipage}{1.0\linewidth}
  \includegraphics[width=0.4\linewidth]{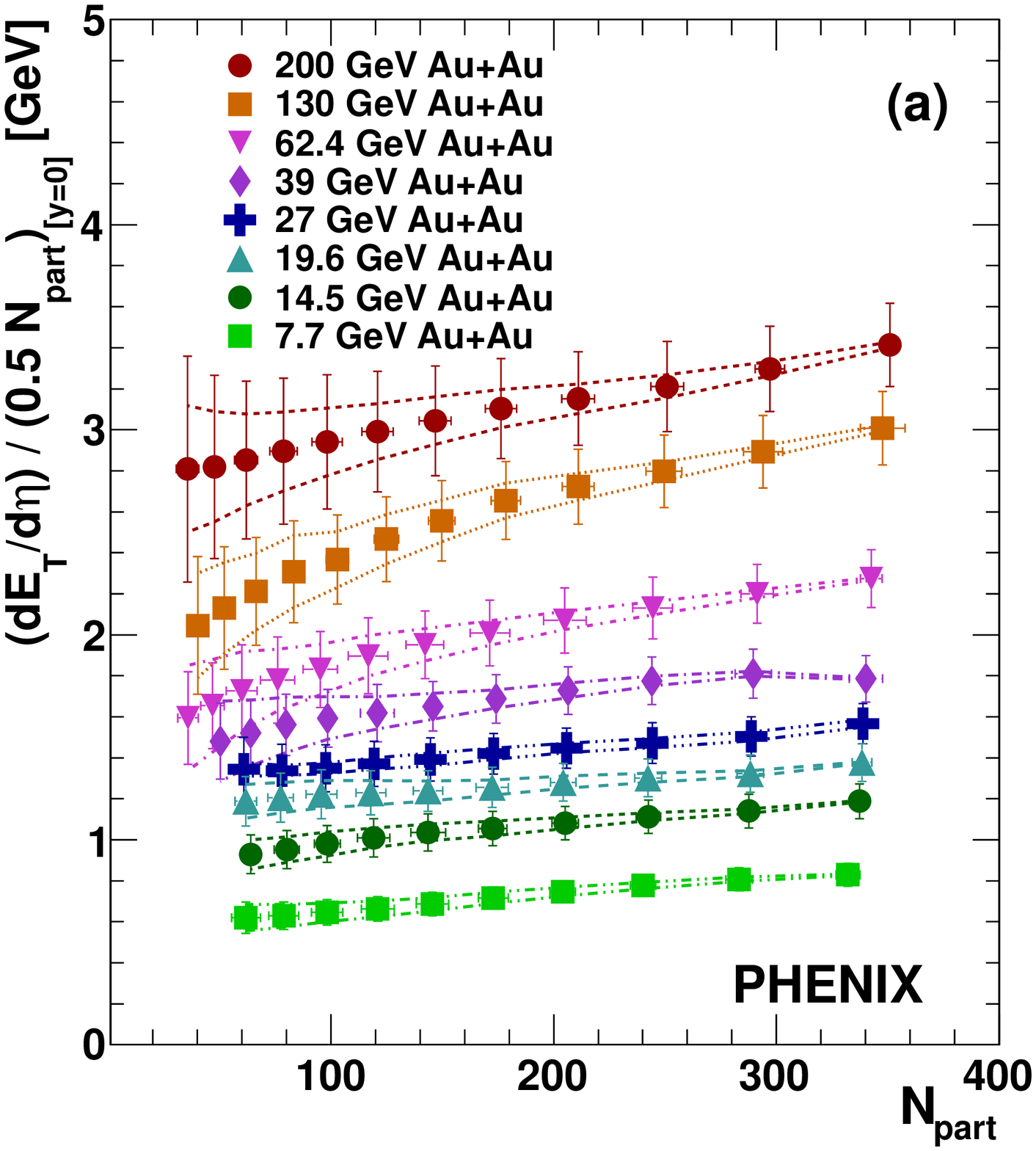}
  \includegraphics[width=0.4\linewidth]{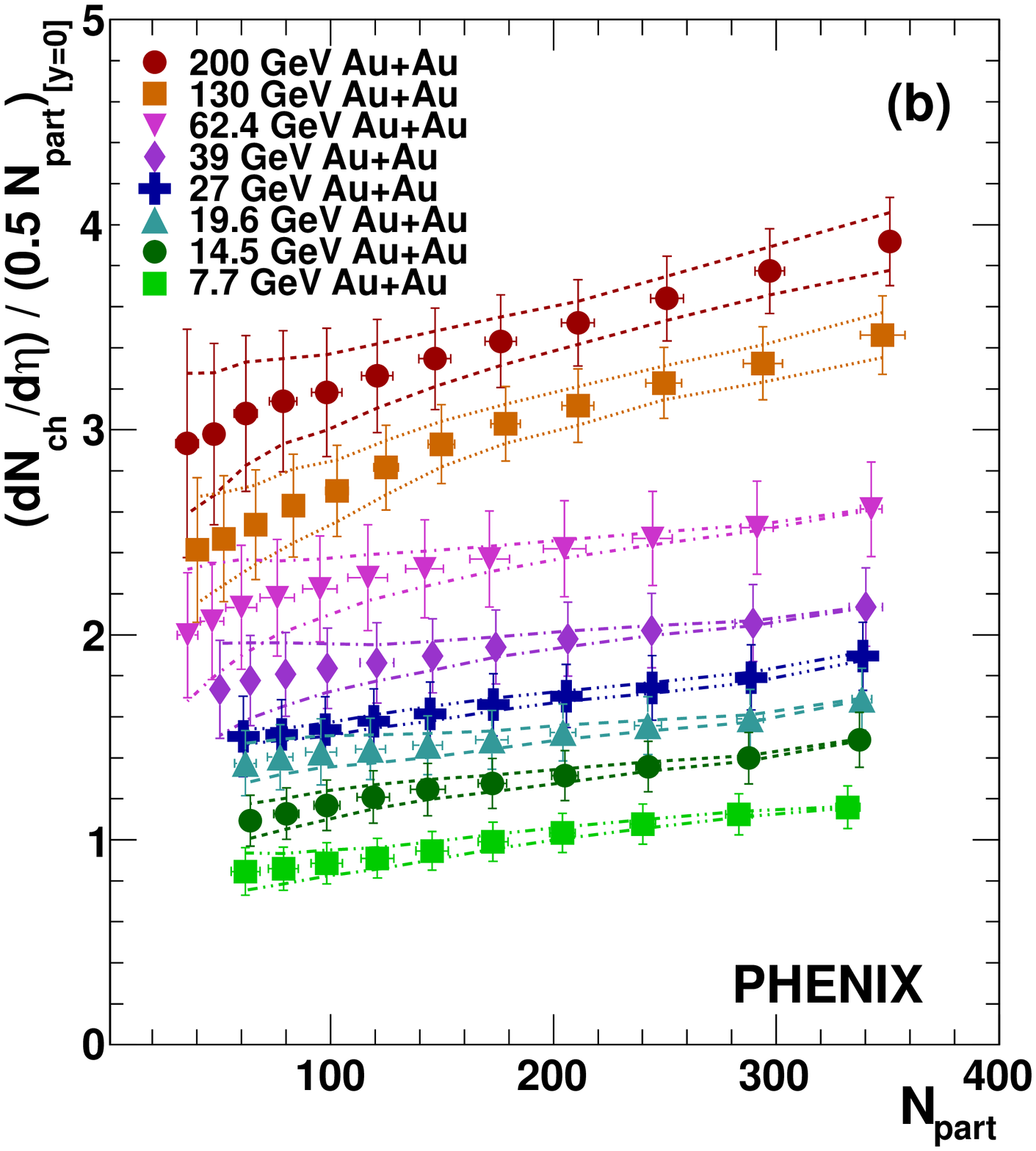}
  \caption{(Color online) 
\dEtNorm (a) and \dNchNorm (b) at midrapidity as a function of \Npart for 
\auau collisions 200, 130, 62.4, 39, 27, 19.6, 14.5, and 7.7~GeV. The 
lines bounding the points represent the trigger efficiency uncertainty 
within which the points can be tilted. The error bars represent the 
remaining total statistical and systematic uncertainty.}
  \label{fig:detdnNormBES}
	\end{minipage}
	\begin{minipage}{1.0\linewidth}
  \includegraphics[width=0.48\linewidth]{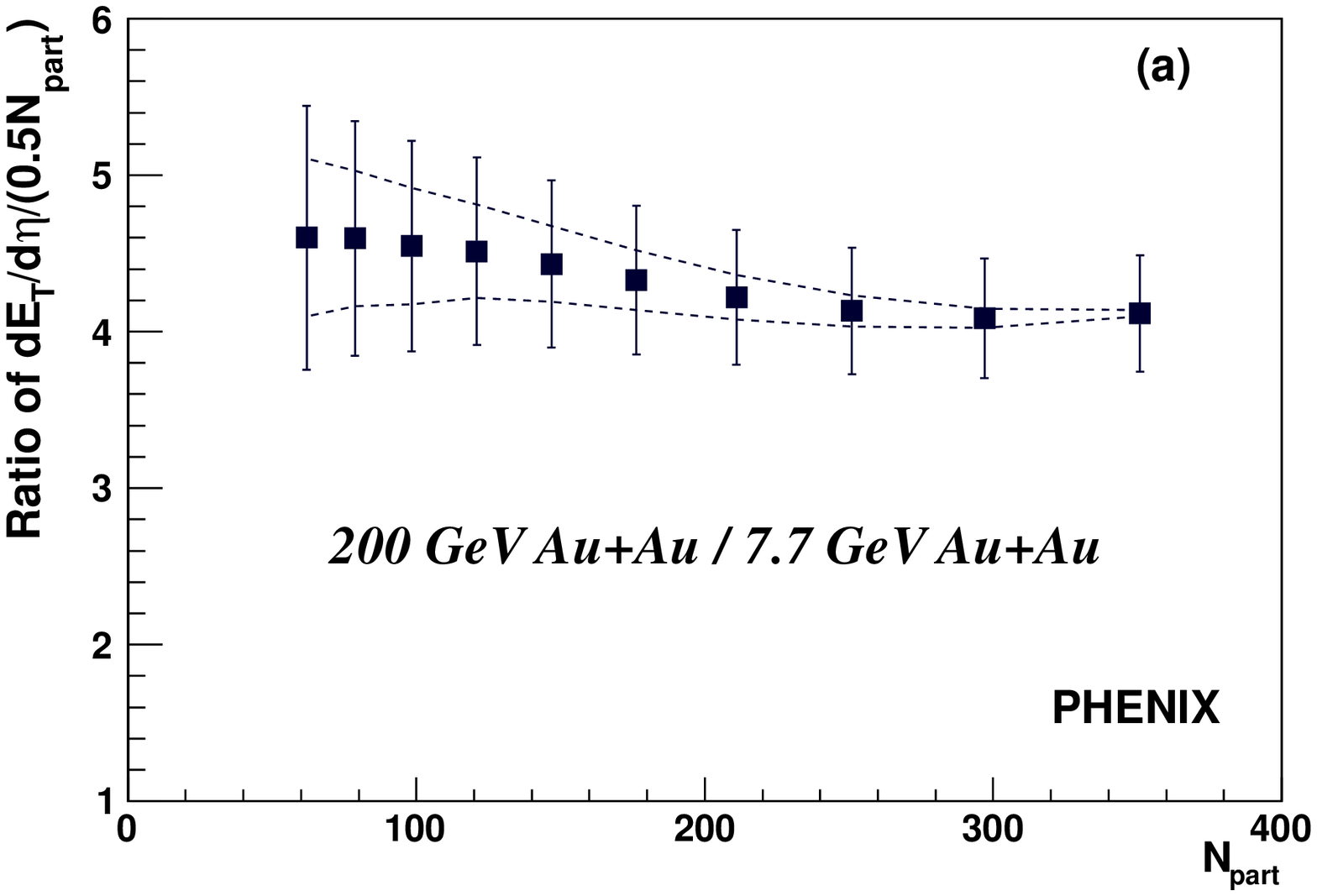}
  \includegraphics[width=0.48\linewidth]{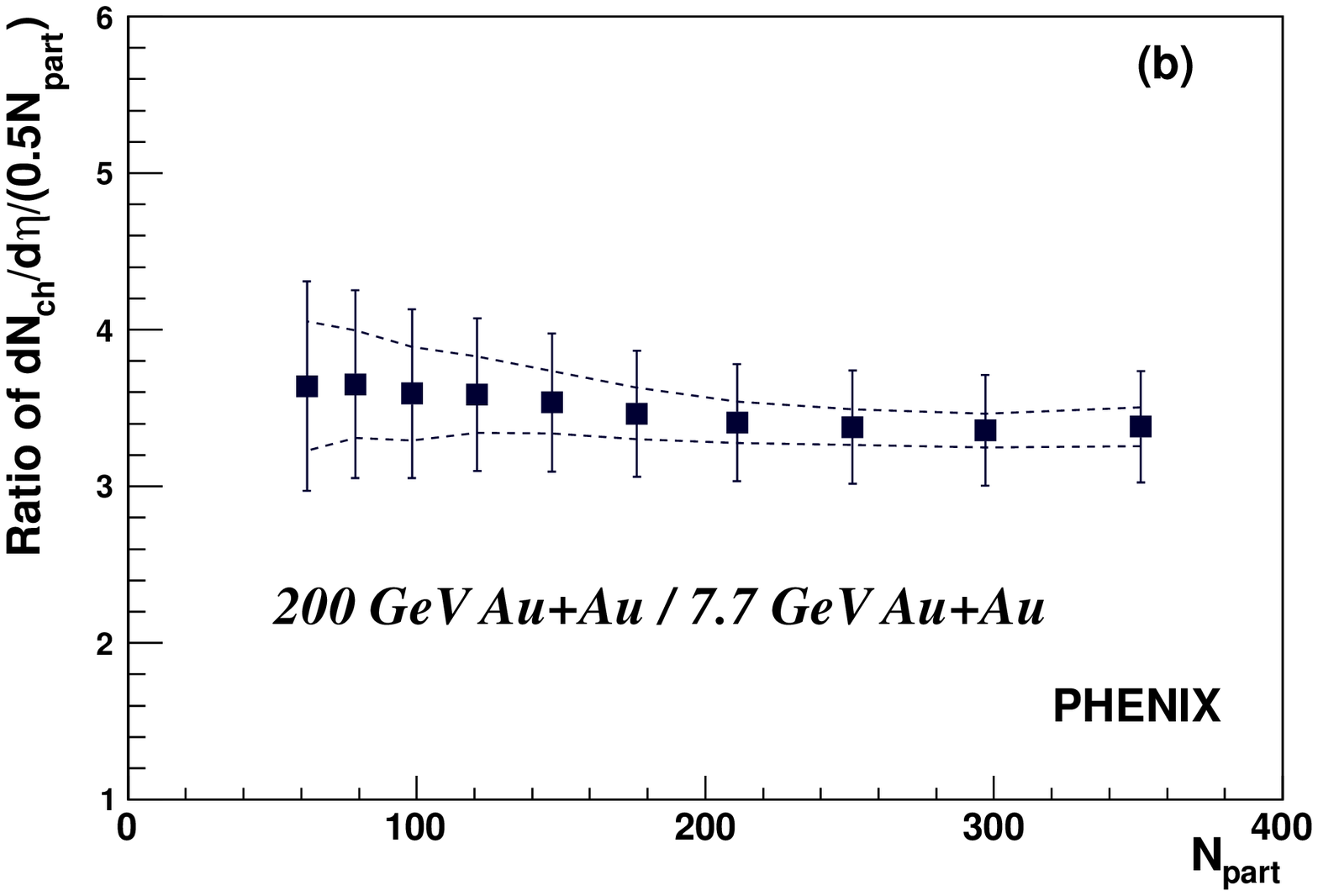}
  \caption{The ratio of \sqsn = 200~GeV \auau collisions to \sqsn = 
7.7~GeV \auau collisions for \dEtNorm (a) and for \dNchNorm (b).  The 
lines bounding the points represent the trigger efficiency uncertainty 
within which the points can be tilted. The error bars represent the total 
statistical and systematic uncertainties.}
  \label{fig:detdnRatio200}
	\end{minipage}
\end{figure*}

		\section{Au$+$Au Beam Energy Scan Results}
		\label{sec:besResults}

This section will present \dEt and \dNch measurements as a function of 
centrality expressed as the number of nucleon participants, \Npart, from 
the RHIC beam energy scan that includes \auau collisions at \sqsn = 200, 
130, 62.4, 39, 27, 19.6, 14.5, and 7.7~GeV. A Monte Carlo Glauber model 
calculation is used to obtain estimates of \Npart as a function of 
centrality using the procedure outlined in Ref.~\cite{Miller:2007ri}.  At 
each collision energy, the Glauber model is run using the inelastic 
nucleon-nucleon cross sections, $\sigma_{nn}^{\rm inel}$, listed in 
Table~\ref{tab:crossSections}.

\begin{table}[htb]
\caption{Summary of the cross sections as a function of \sqsn.}
\label{tab:crossSections}
\begin{ruledtabular}
\begin{tabular}{cccc}
Energy & $\sigma_{nn}^{\rm tot}$ [mb] & $\sigma_{nn}^{\rm inel}$ [mb] & $\sigma^{\rm inel}_{qq}$ [mb]\\[0.2pc]\hline
200  & 52.5 & 42.3 & 8.17\\
130  & 48.7 & 39.6 & 7.54\\
62.4 & 43.6 & 36.0 & 6.56\\
39   & 41.2 & 34.3 & 6.15\\
27   & 39.8 & 33.2 & 5.86\\
19.6 & 39.0 & 32.5 & 5.70\\
15.0 & 38.5 & 32.0 & 5.58\\
7.7  & 38.6 & 31.2 & 5.35\\
\end{tabular}
\end{ruledtabular}
\end{table}

When plotting \dEt and \dNch, systematic uncertainties are decomposed into 
two types. Type A uncertainties include point-to-point uncertainties that 
are uncorrelated between bins and include only statistical uncertainties 
in this analysis. The remaining uncertainties are classified as Type B 
uncertainties that are correlated bin-by-bin such that all points move in 
the same direction, but not necessarily by the same factor.  Because the 
magnitudes of the Type A statistical uncertainties are small compared to 
the magnitudes of the Type B uncertainties, the error bars in the plots 
presented below will represent the total statistical and systematic 
uncertainties added in quadrature. The trigger efficiency uncertainty is 
represented separately by error bands bounding the points within which the 
points can be tilted, as described in Section~\ref{sec:analysis}.

\begin{figure*}[!htb]
  \includegraphics[width=0.48\linewidth]{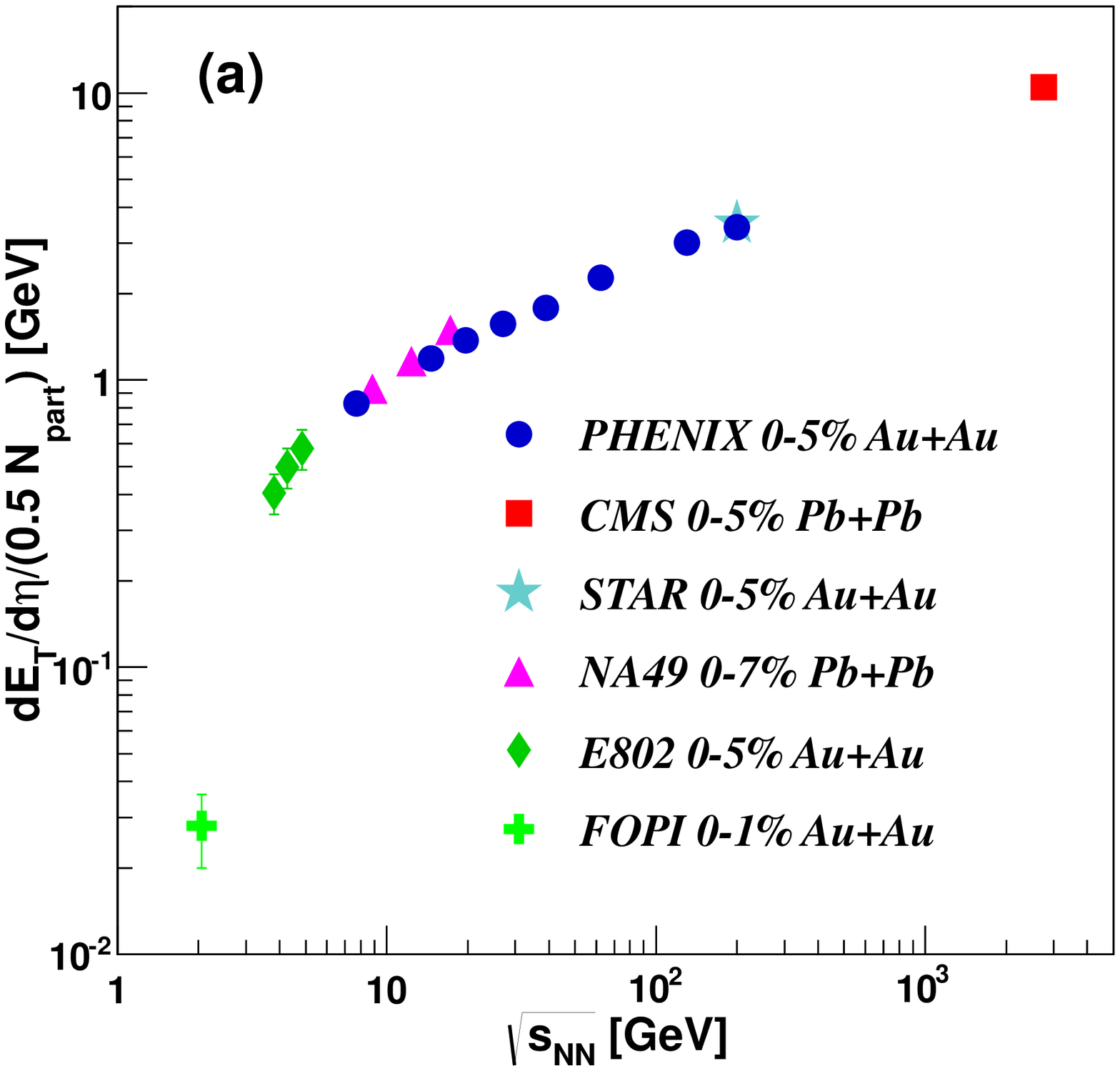}
  \includegraphics[width=0.48\linewidth]{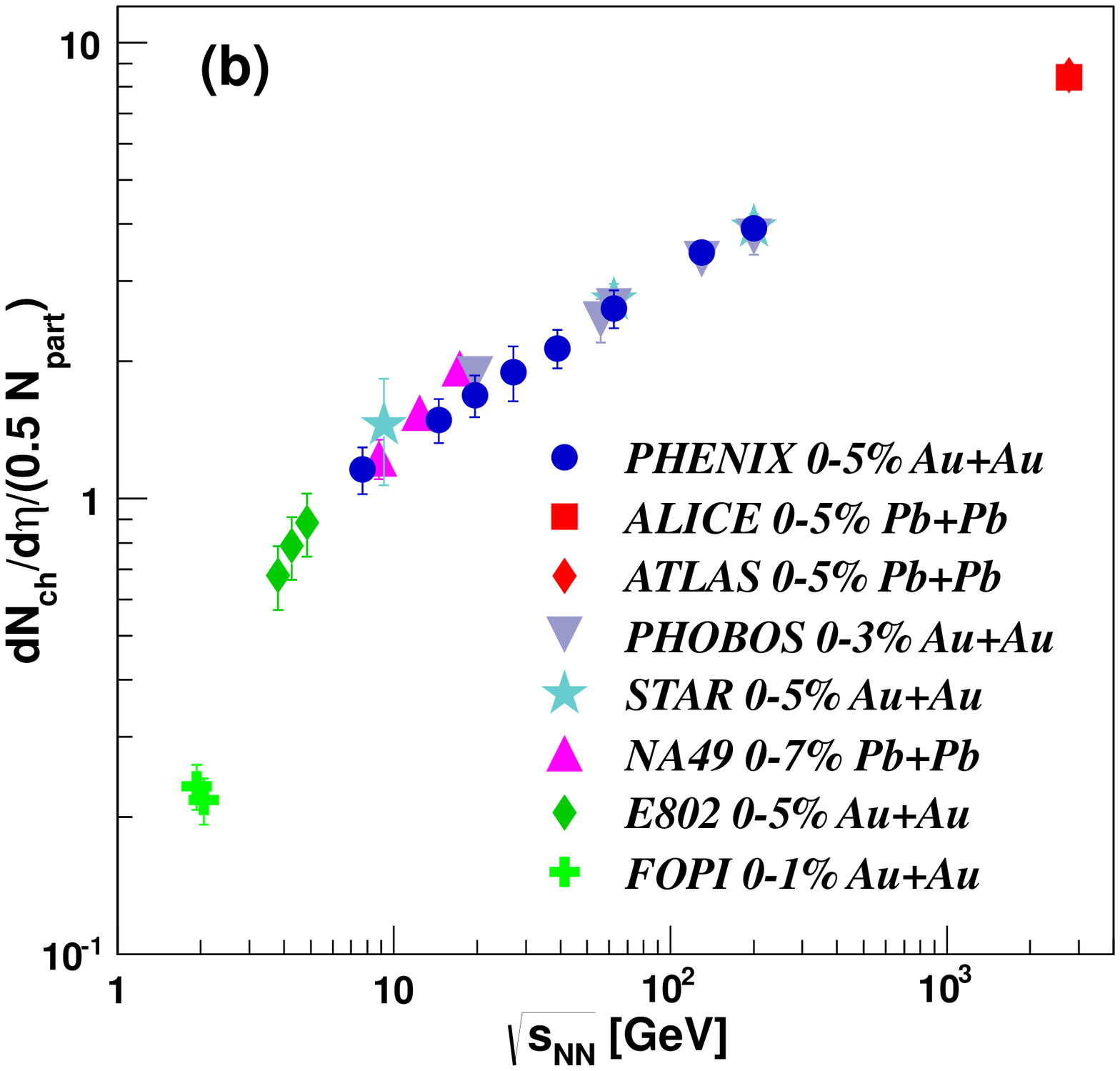}
  \caption{(Color online) 
The excitation function of \dEtNorm (a) and \dNchNorm (b) for central 
collisions at midrapidity as a function of \sqsn. The error bars represent 
the total statistical and systematic uncertainties.  For \dEtNorm (a), 
data are shown from FOPI~\cite{Reisdorf:1996qj}, E802~\cite{Ahle:1999jm}, 
NA49~\cite{Afanasiev:2002fk, Chatrchyan:2012mb}, STAR~\cite{Adams:2004cb}, 
and CMS~\cite{Chatrchyan:2012mb}. For \dNchNorm (b), data are shown from 
FOPI~\cite{Reisdorf:1996qj}, E802~\cite{Ahle:1999jm, Ahle:1998gv, 
Ahle:2000wq}, NA49~\cite{Afanasiev:2002fk}, STAR~\cite{Adams:2004cb, 
Abelev:2009bw}, PHOBOS~\cite{Alver:2010ck}, ALICE~\cite{Aamodt:2010cz}, 
and ATLAS~\cite{ATLAS:2011ag}.}
    \label{fig:detdnNormExcite}
\end{figure*}
\begin{figure*}[!htb] 
  \includegraphics[width=0.48\linewidth]{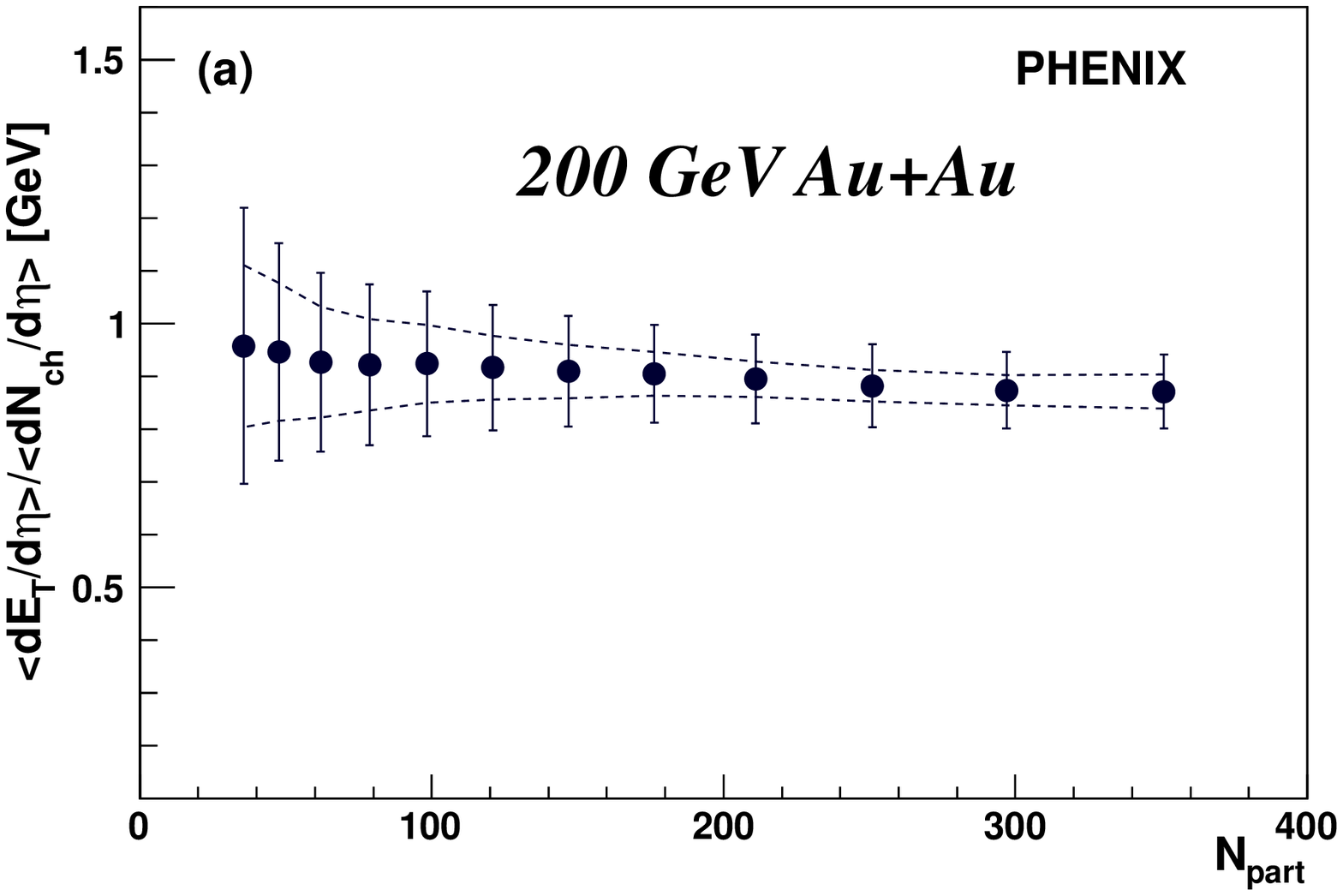}
  \includegraphics[width=0.48\linewidth]{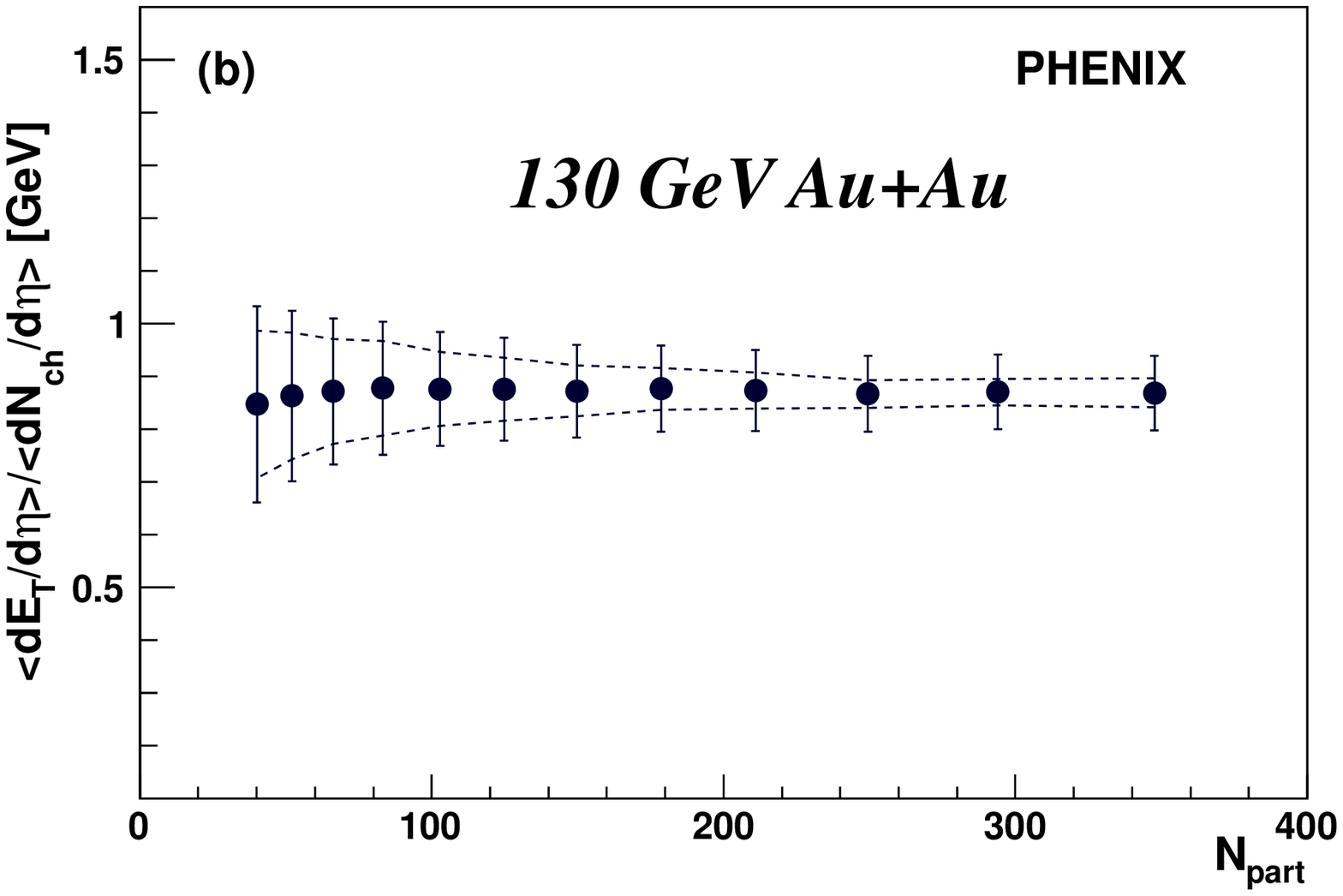}
  \includegraphics[width=0.48\linewidth]{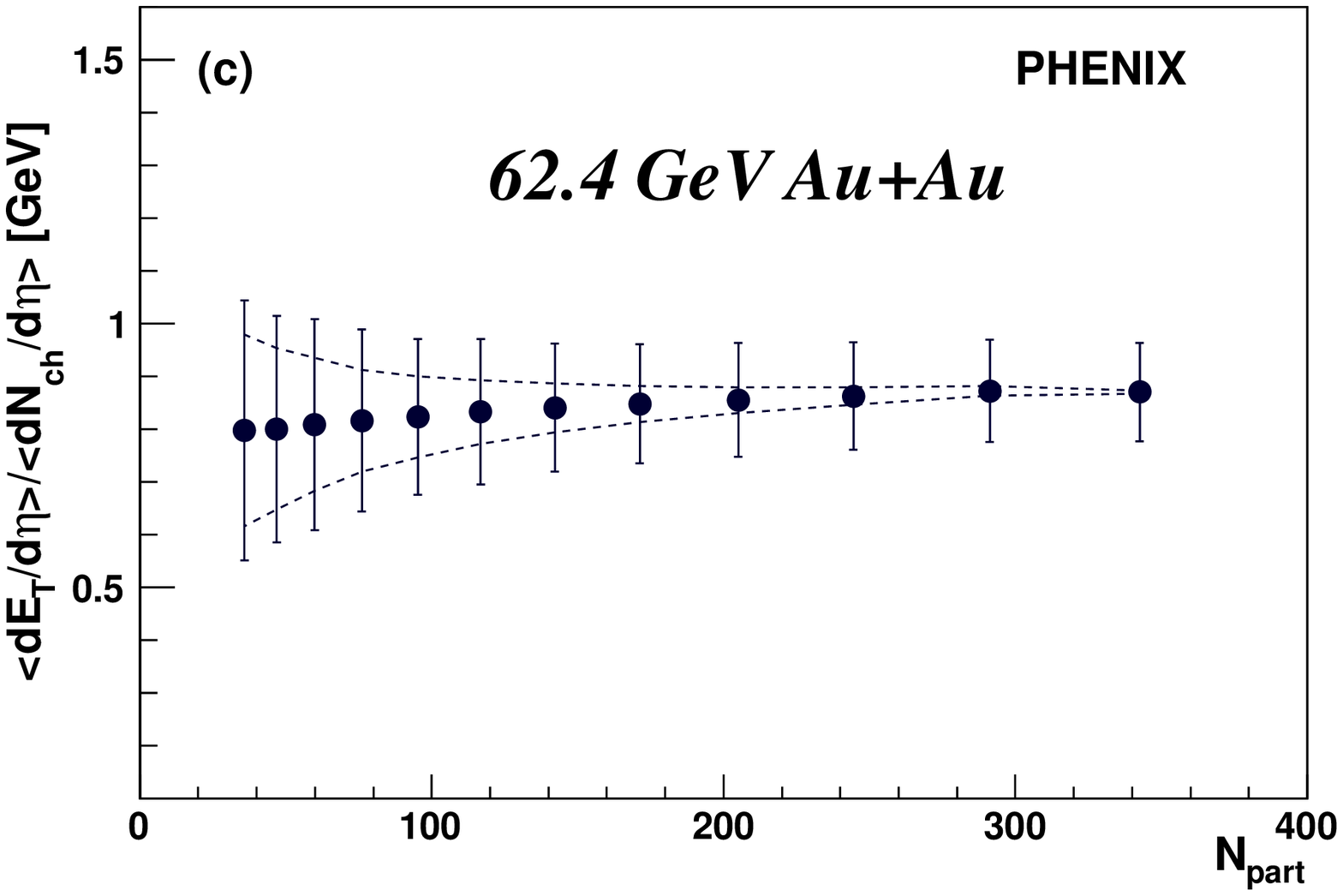}
  \includegraphics[width=0.48\linewidth]{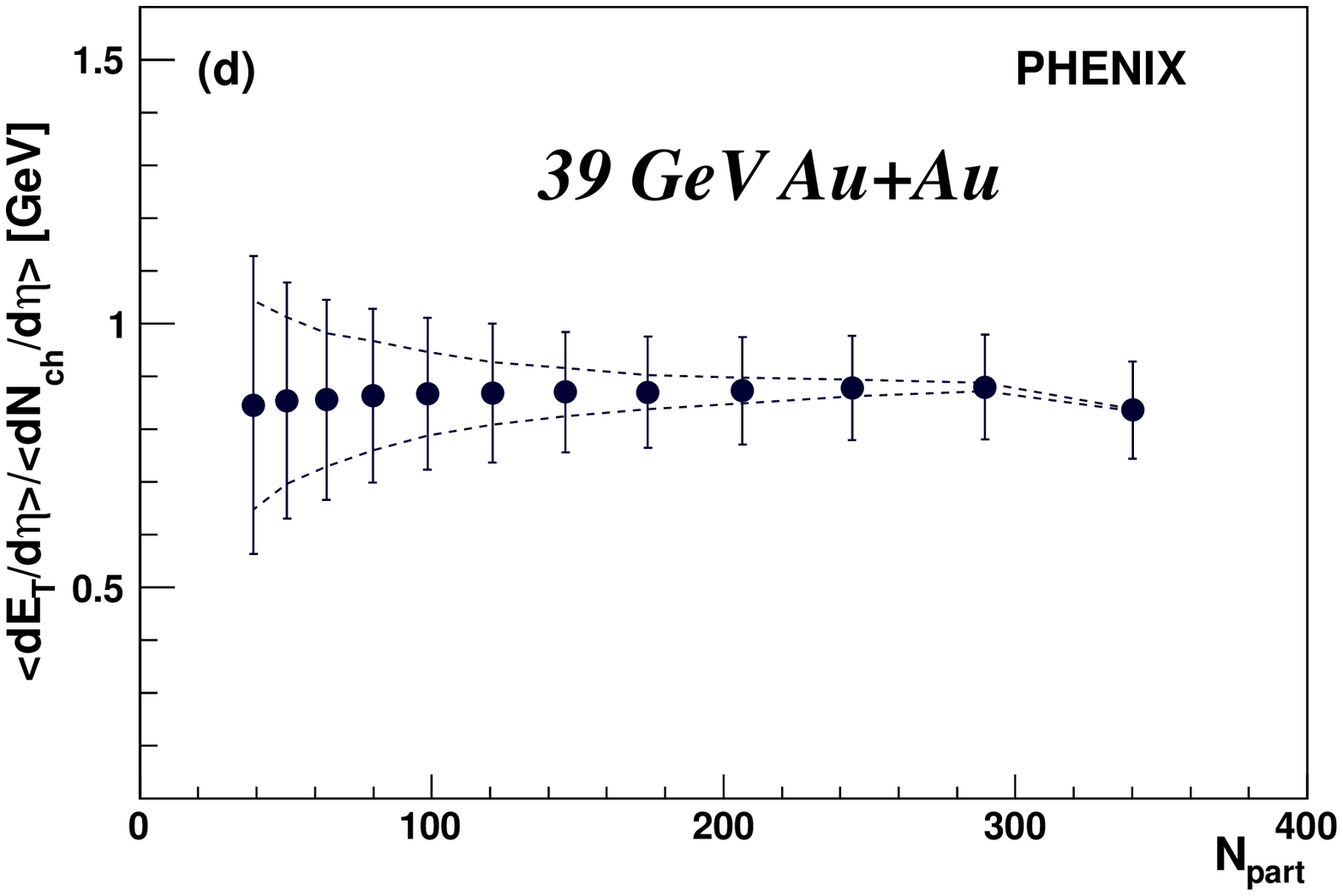}
  \includegraphics[width=0.48\linewidth]{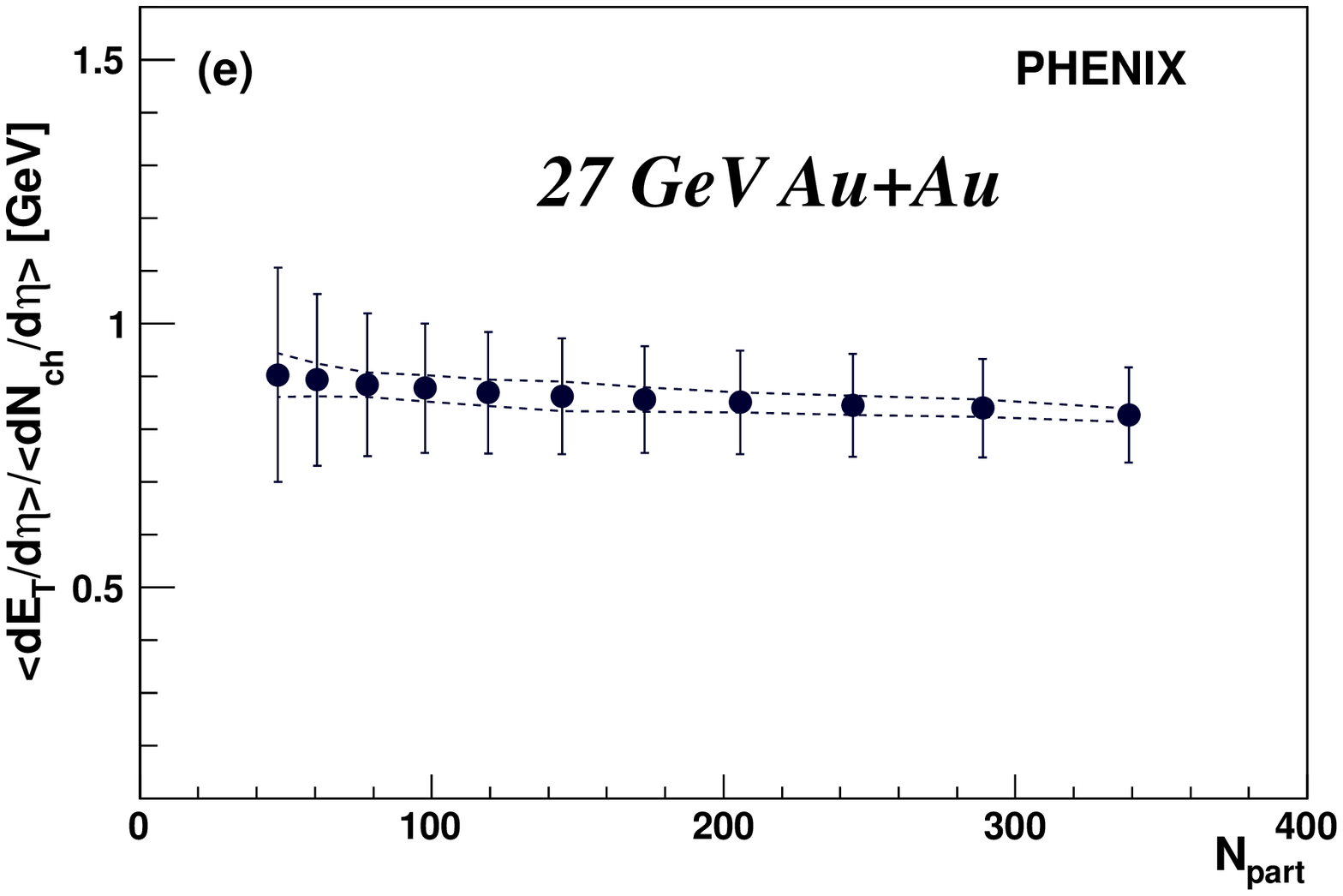}
  \includegraphics[width=0.48\linewidth]{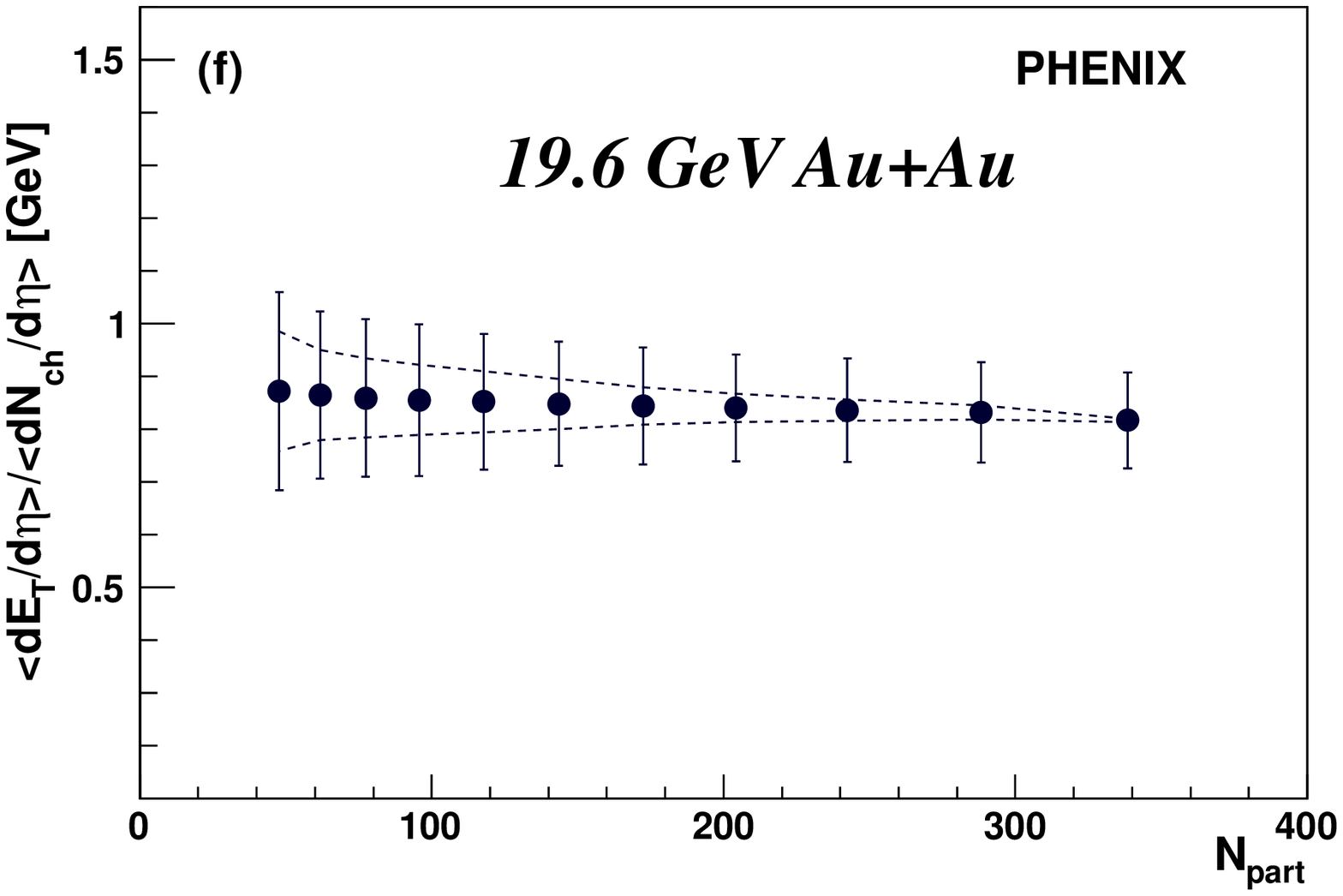}
  \includegraphics[width=0.48\linewidth]{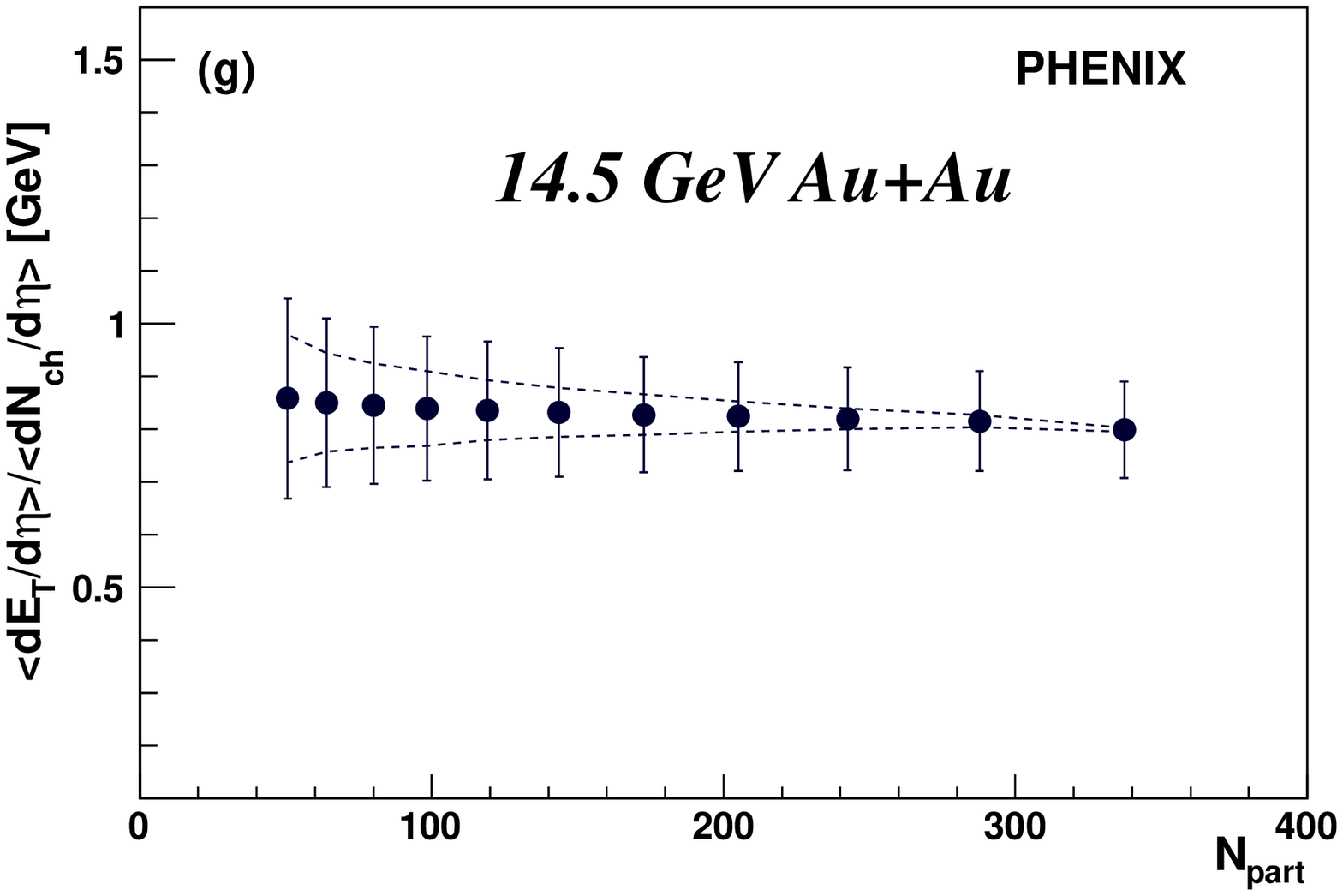}
  \includegraphics[width=0.48\linewidth]{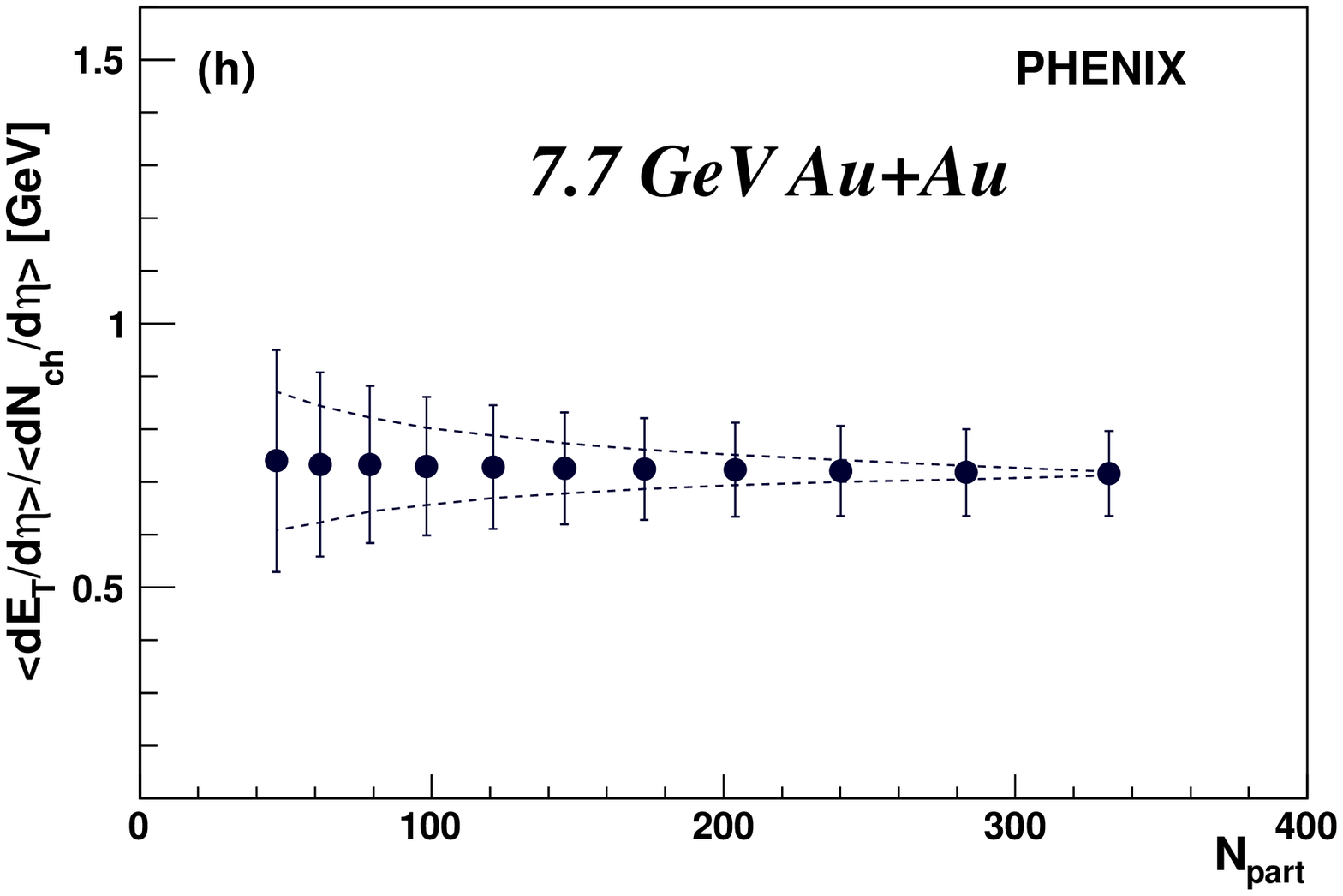}
  \caption{
The \Et/\Nch ratio as a function of \Npart for \auau collisions at varying 
values of \sqsn. The lines bounding the points represent the trigger 
efficiency uncertainty within which the points can be tilted. The error 
bars represent the total statistical and systematic uncertainties.}
  \label{fig:etOverNchBES}
\end{figure*}

\begin{figure}[!htb] 
  \includegraphics[width=0.98\linewidth]{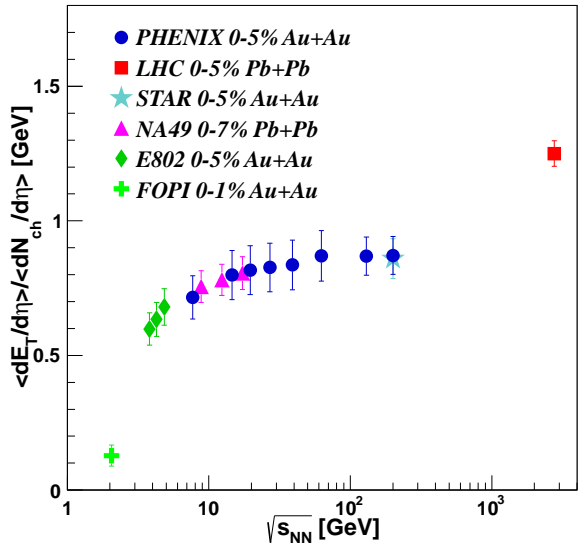}
  \caption{(Color online) The \Et/\Nch ratio as a function of \sqsn for central \auau 
collisions and \pbpb collisions at midrapidity. The error bars represent 
the total statistical and systematic uncertainties. The Large-Hadron-Collider, 
LHC, data point has been obtained by taking the ratio of the CMS \dEt 
data~\cite{Chatrchyan:2012mb} with the average of the 
ALICE~\cite{Aamodt:2010cz} and ATLAS~\cite{ATLAS:2011ag} data. For 
\dEtNorm, data are taken from FOPI~\cite{Reisdorf:1996qj}, 
E802~\cite{Ahle:1999jm}, NA49~\cite{Afanasiev:2002fk, Chatrchyan:2012mb}, 
STAR~\cite{Adams:2004cb}, and CMS~\cite{Chatrchyan:2012mb}. For \dNchNorm, 
data are taken from FOPI~\cite{Reisdorf:1996qj}, E802~\cite{Ahle:1999jm, 
Ahle:1998gv, Ahle:2000wq}, NA49~\cite{Afanasiev:2002fk}, 
STAR~\cite{Adams:2004cb, Abelev:2009bw}, PHOBOS~\cite{Alver:2010ck}, 
ALICE~\cite{Aamodt:2010cz}, and ATLAS~\cite{ATLAS:2011ag}.}
  \label{fig:etOverNchExcite}
\end{figure}

\begin{figure*}[!htb] 
\begin{minipage}{0.48\linewidth}
  \includegraphics[width=1.0\linewidth]{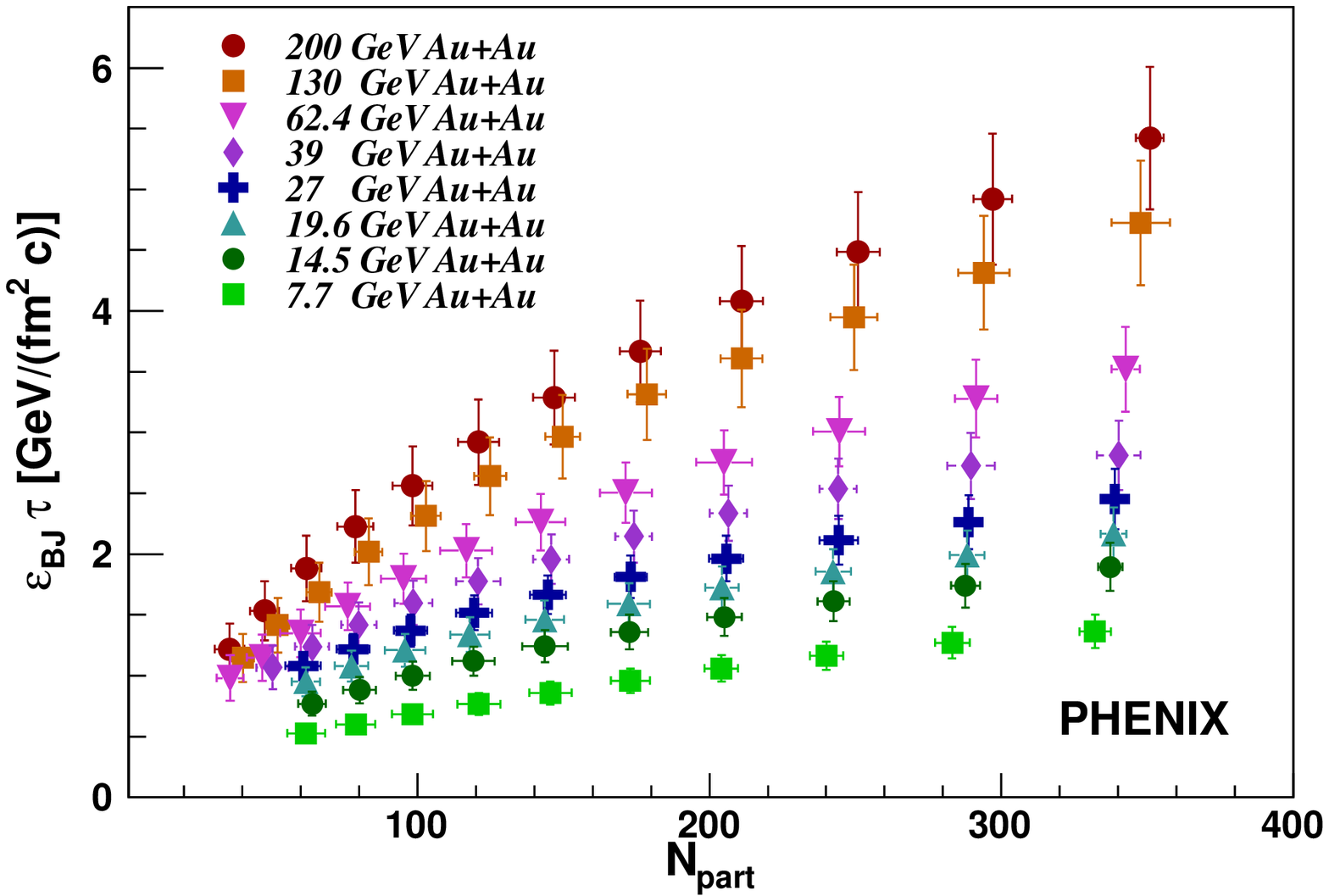}
  \caption{(Color online) 
The Bjorken energy density, \ebj, multiplied by $\tau$ as a 
function of \Npart for \auau collisions at varying values of \sqsn. The 
error bars represent the total statistical and systematic uncertainties.}
  \label{fig:ebjBES}
\end{minipage}
\begin{minipage}{0.48\linewidth}
  \includegraphics[width=1.0\linewidth]{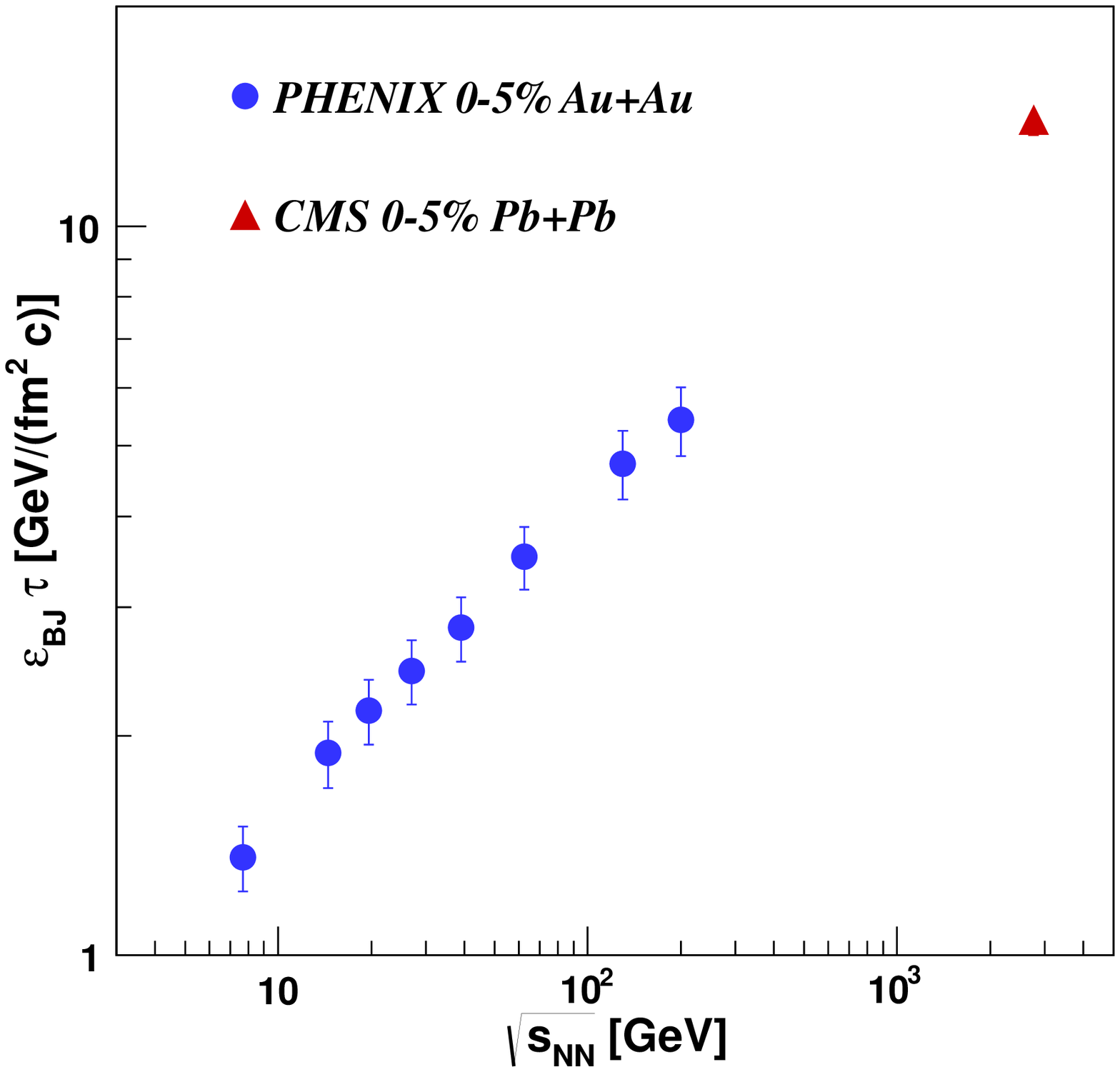}
  \caption{(Color online) 
The Bjorken energy density, \ebj, multiplied by $\tau$ as a 
function of \sqsn for central \auau (PHENIX) and \pbpb 
(CMS)~\cite{Chatrchyan:2012mb} collisions at midrapidity. The error bars 
represent the total statistical and systematic uncertainties.}
  \label{fig:ebjExcite}
\end{minipage}
\begin{minipage}{0.9\linewidth}
  \includegraphics[width=0.4\linewidth]{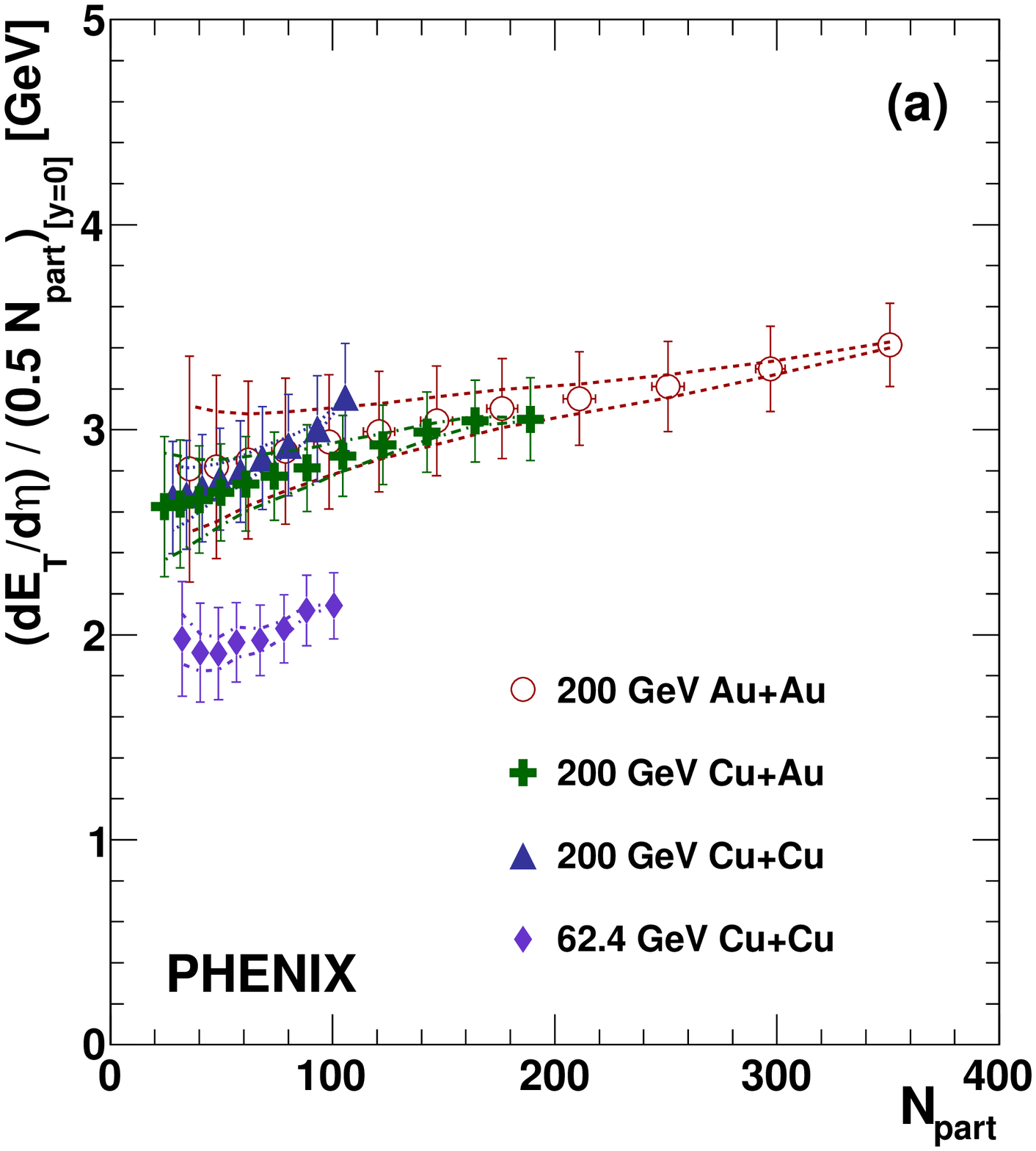}
  \includegraphics[width=0.4\linewidth]{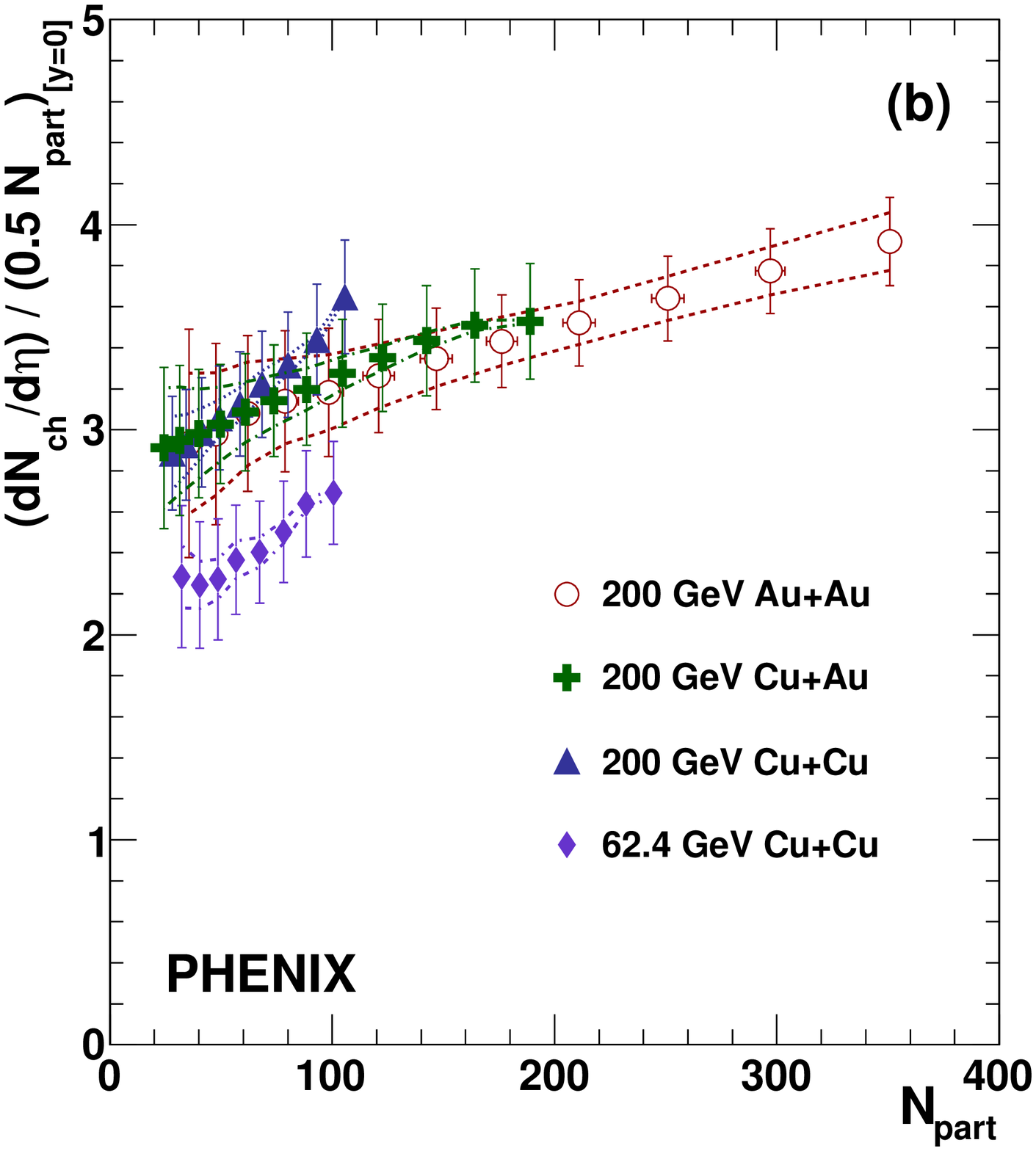}
  \caption{(Color online) 
\dEtNorm (a) and \dNchNorm (b) at midrapidity as a function of \Npart for 
\cucu and \cuau collisions. Also shown are results from \auau collisions 
at \sqsn = 200~GeV for comparison. The lines bounding the points represent 
the trigger efficiency uncertainty within which the points can be tilted. 
The error bars represent the remaining total statistical and systematic 
uncertainty.}
  \label{fig:detdnNormCu}
\end{minipage}
\end{figure*}

Examining the \Npart dependence of \dEt and \dNch normalized by the number 
of nucleon participant pairs at midrapidity is useful to determine if the 
data scales by \Npart and if the scaling changes as a function of \sqsn.  
The results for \auau collisions for all beam energies at midrapidity are 
shown in Fig.~\ref{fig:detdnNormBES} as a function of \Npart. 
For all energies, \dEt and \dNch do not scale with \Npart; the magnitudes 
of \dEt and \dNch increase as \Npart increases.
It has been previously observed that the shape of the distributions as a 
function of \Npart are preserved in \auau collisions from \sqsn = 200~GeV 
to \sqsn = 19.6~GeV~\cite{Back:2004dy, Adler:2004zn}.  
Figure~\ref{fig:detdnRatio200}a shows the ratio of \dEtNorm from \auau 
collisions at \sqsn = 200~GeV to \sqsn = 7.7~GeV, illustrating that the 
shapes of the distributions are preserved down to \sqsn = 7.7~GeV. 
Figure~\ref{fig:detdnRatio200}b shows the same for \dNchNorm. Previous 
measurements in fixed target \ha collisions showed that the total charged 
particle multiplicity does scale well as a function of \Npart in the
range of $10\leq$\sqsn$\leq20$~GeV~\cite{Busza:1975df}. However, this 
measurement was made over the full rapidity range rather than at 
midrapidity. For the midrapidity measurements presented here, the \Npart 
scaling behavior does not change significantly from \sqsn = 200~GeV down 
to \sqsn = 7.7~GeV.

Excitation functions of \dEtNorm and \dNchNorm are shown in 
Fig.~\ref{fig:detdnNormExcite}.  Shown are the PHENIX data along with 
results from other experiments.  The data points for the lower energies 
are from estimates described in Ref.~\cite{Adler:2004zn}.  For \dEtNorm, 
data are shown from FOPI 0-1\% centrality \auau 
collisions~\cite{Reisdorf:1996qj}, E802 0\%--5\% centrality \auau 
collisions~\cite{Ahle:1999jm}, NA49 0-7\% centrality \pbpb 
collisions~\cite{Afanasiev:2002fk, Chatrchyan:2012mb}, STAR 0\%--5\% 
centrality \auau collisions~\cite{Adams:2004cb}, and CMS 0\%--5\% centrality 
\pbpb collisions~\cite{Chatrchyan:2012mb}. For \dNchNorm, data are shown 
from FOPI~\cite{Reisdorf:1996qj}, E802~\cite{Ahle:1999jm, Ahle:1998gv, 
Ahle:2000wq}, NA49~\cite{Afanasiev:2002fk}, STAR~\cite{Adams:2004cb, 
Abelev:2009bw}, PHOBOS 0-3\% centrality \auau 
collisions~\cite{Alver:2010ck}, ALICE 0\%--5\% centrality \pbpb 
collisions~\cite{Aamodt:2010cz}, and ATLAS~\cite{ATLAS:2011ag} \pbpb 
collisions interpolated to 0\%--5\% centrality. The data are plotted on a 
log-log scale to illustrate the power law behavior of both \dEtNorm and 
\dNchNorm as a function of log(\sqsn) for \sqsn at or above 7.7~GeV. For 
\dEtNorm, the data between \sqsn = 7.7 and 200~GeV are described by 
$\dEtNorm(\sqsn) \propto e^{b \times log(\sqrt{s_{NN}})}$, where $b = 
0.428 \pm 0.021$. For \dNchNorm, the data between \sqsn = 7.7 and 200~GeV 
are described by $\dNchNorm(\sqsn) \propto e^{b \times 
log(\sqrt{s_{NN}})}$, where $b = 0.374 \pm 0.028$.  The data deviate from 
the power law behavior below the lowest PHENIX measurement at \sqsn = 
7.7~GeV.

The ratio of \dEt to \dNch, referred to here simply as \Et/\Nch, is a 
variable that is related to the average transverse mass of the produced 
particles~\cite{Adler:2004zn}. In previous measurements, this ratio has been 
observed to be independent of centrality and independent of \sqsn in \auau 
collisions from \sqsn = 200 to 19.6~GeV~\cite{Adler:2004zn}. 
Figure~\ref{fig:etOverNchBES} plots the \Et/\Nch ratio as a function of 
\Npart for \auau collisions at various values of \sqsn. For all cases, the 
ratio is constant with \Npart within the systematic uncertainties. The 
excitation function of \Et/\Nch is shown in Fig.~\ref{fig:etOverNchExcite}. 
Here, the Large-Hadron-Collider point has been obtained by taking the ratio 
of the CMS \dEt data~\cite{Chatrchyan:2012mb} with the average of the 
ALICE~\cite{Aamodt:2010cz} and ATLAS~\cite{ATLAS:2011ag} data.  The ratio 
increases below \sqsn $\approx$ 10~GeV, levels off, and then increases at 
\sqsn = 200~GeV.

The energy density per unit volume in nuclear collisions can be estimated 
from the energy density per unit rapidity~\cite{Bjorken:1982qr}. The 
Bjorken energy density can be calculated as follows:
\begin{equation}
  \varepsilon_{BJ} = \frac{1}{A_\perp \tau} J(y,\eta)  \frac{d\Et}{d\eta}
\end{equation}
where $A_\perp$ is the transverse overlap area of the nuclei determined 
from the Glauber model, $\tau$ is the formation time, and J($y,\eta$) is 
the Jacobian factor for converting pseudorapidity to rapidity.

The Jacobian factor depends on the momentum distributions of the produced 
particles, which are dependent on the beam energy. 
The Jacobian factor for each beam 
energy in the PHENIX acceptance has been estimated using the URQMD event 
generator, which well reproduces measured particle spectra over the RHIC 
beam energy range and, unlike HIJING, is valid at \sqsn = 7.7~GeV.  
Calculations of the Jacobian factor using URQMD are consistent with 
previous calculations using the HIJING event 
generator~\cite{Adler:2004zn}. There is an estimated uncertainty of 3\% 
for this calculation for all beam energies. The values of the Jacobian 
factors are summarized in Table~\ref{tab:jacobians}.

\begin{table}[htb]
\caption{
Summary of the Jacobian scale factor estimated for each beam energy.}
\label{tab:jacobians}
\begin{ruledtabular}
\begin{tabular}{cc}
Dataset               & J(y,$\eta$) \\[0.2pc]\hline
200~GeV \auau & 1.25 \\
130~GeV \auau & 1.25 \\
62.4~GeV \auau & 1.25 \\
39~GeV \auau   & 1.27 \\
27~GeV \auau   & 1.27 \\
19.6~GeV \auau & 1.28 \\
14.5~GeV \auau & 1.30 \\
7.7~GeV \auau  & 1.35 \\
\end{tabular}
\end{ruledtabular}
\end{table}

The transverse overlap area is estimated using the a Monte Carlo Glauber 
model as $A_\perp \sim \sigma_{x} \sigma_{y}$, where $\sigma_{x}$ and 
$\sigma_{y}$ are the widths of the x and y position distributions of the 
participating nucleons in the transverse plane. A normalization to $\pi 
R^{2}$, where $R$ is the sum of the $r_{n}$ radius and $a$ surface 
diffuseness parameters of the Woods-Saxon parametrization
\begin{equation}
  \rho(r) = 1/(1+e^{(r-r_{n})/a}),
\end{equation}
of the nuclear density profile, $\rho(r)$, was applied for the most 
central collisions at impact parameter $b=0$.

A compilation of the Bjorken energy density multiplied by $\tau$ for \auau 
collisions at various collision energies is shown in 
Fig.~\ref{fig:ebjBES}. The value of \ebj increases with increasing 
\sqsn and also with increasing \Npart. The value of \ebj for 
the most central \auau collisions at \sqsn = 7.7~GeV is $1.36 \pm 0.14$, 
which is still above the value of 1.0 for a formation time of 1 fm/c that 
had been the proposed value above which the Quark-Gluon Plasma can be 
formed in Bjorken's original paper~\cite{Bjorken:1982qr}. 
It is also above the result of $0.7 \pm 0.3$ GeV/fm$^{3}$ for the critical 
energy density obtained from lattice QCD 
calculations~\cite{Karsch:2001cy,Martinez:2013xka}. 
The excitation 
function of \ebj multiplied by $\tau$ is shown in 
Fig.~\ref{fig:ebjExcite}. The results are shown on a log-log scale to 
illustrate that \ebj follows a power law behavior from \sqsn = 7.7~GeV up 
to \sqsn = 2760~GeV, $\varepsilon_{BJ} \tau \propto e^{b \times 
log(\sqrt{s_{NN}})}$, where $b = 0.422 \pm 0.035$.

\begin{figure*}[!htb] 
\begin{minipage}{0.48\linewidth}
  \includegraphics[width=1.0\linewidth]{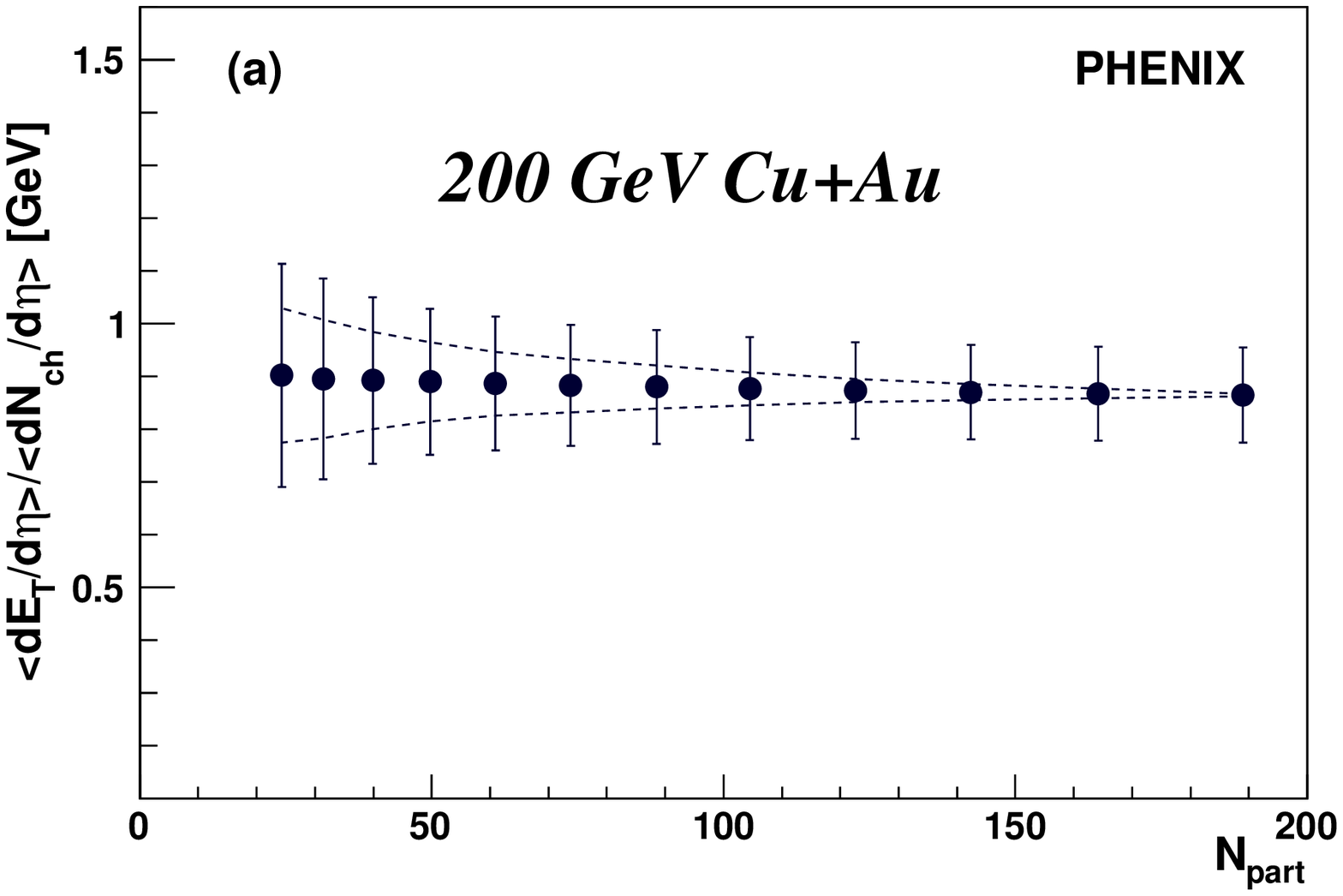}
  \includegraphics[width=1.0\linewidth]{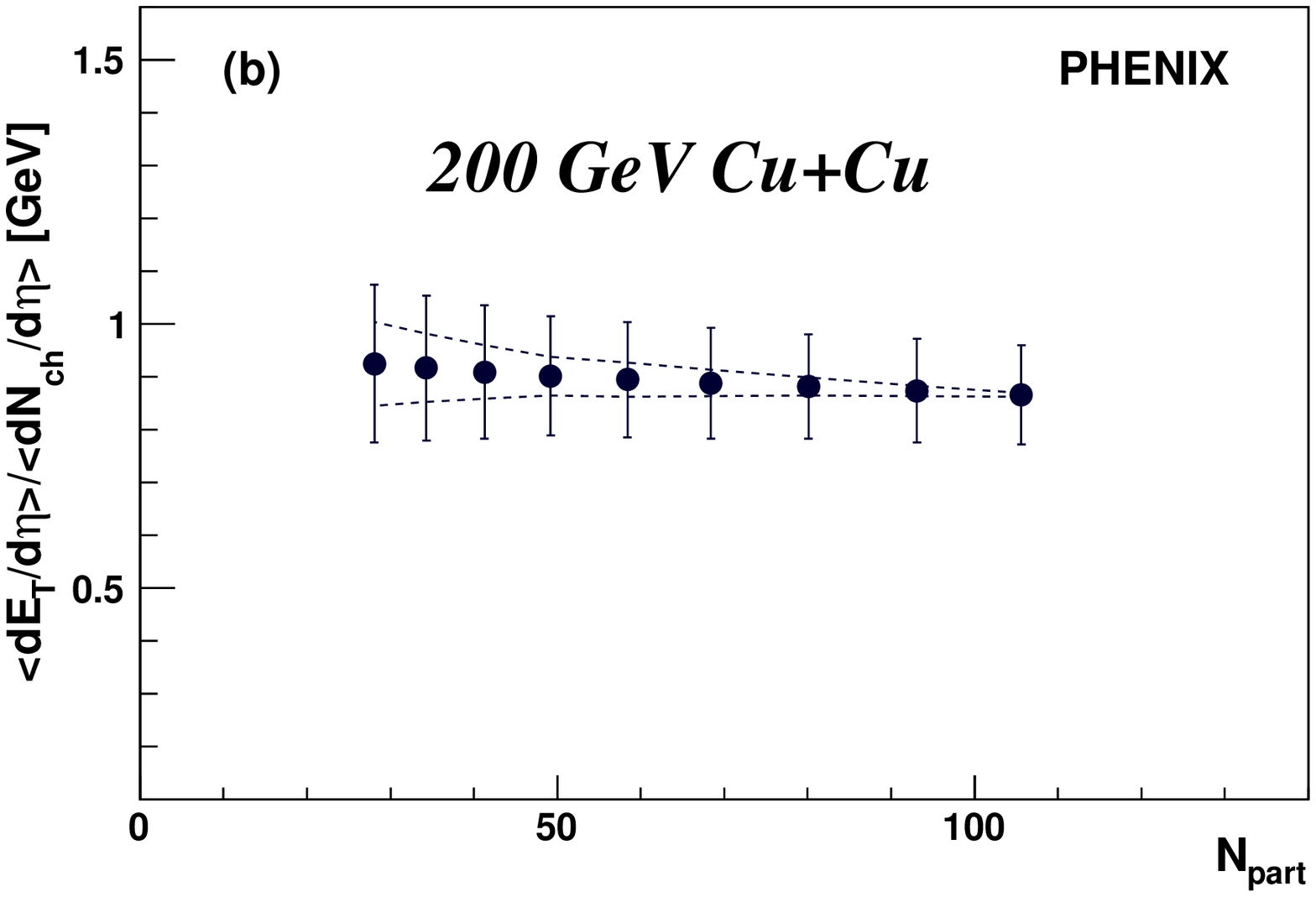}
  \includegraphics[width=1.0\linewidth]{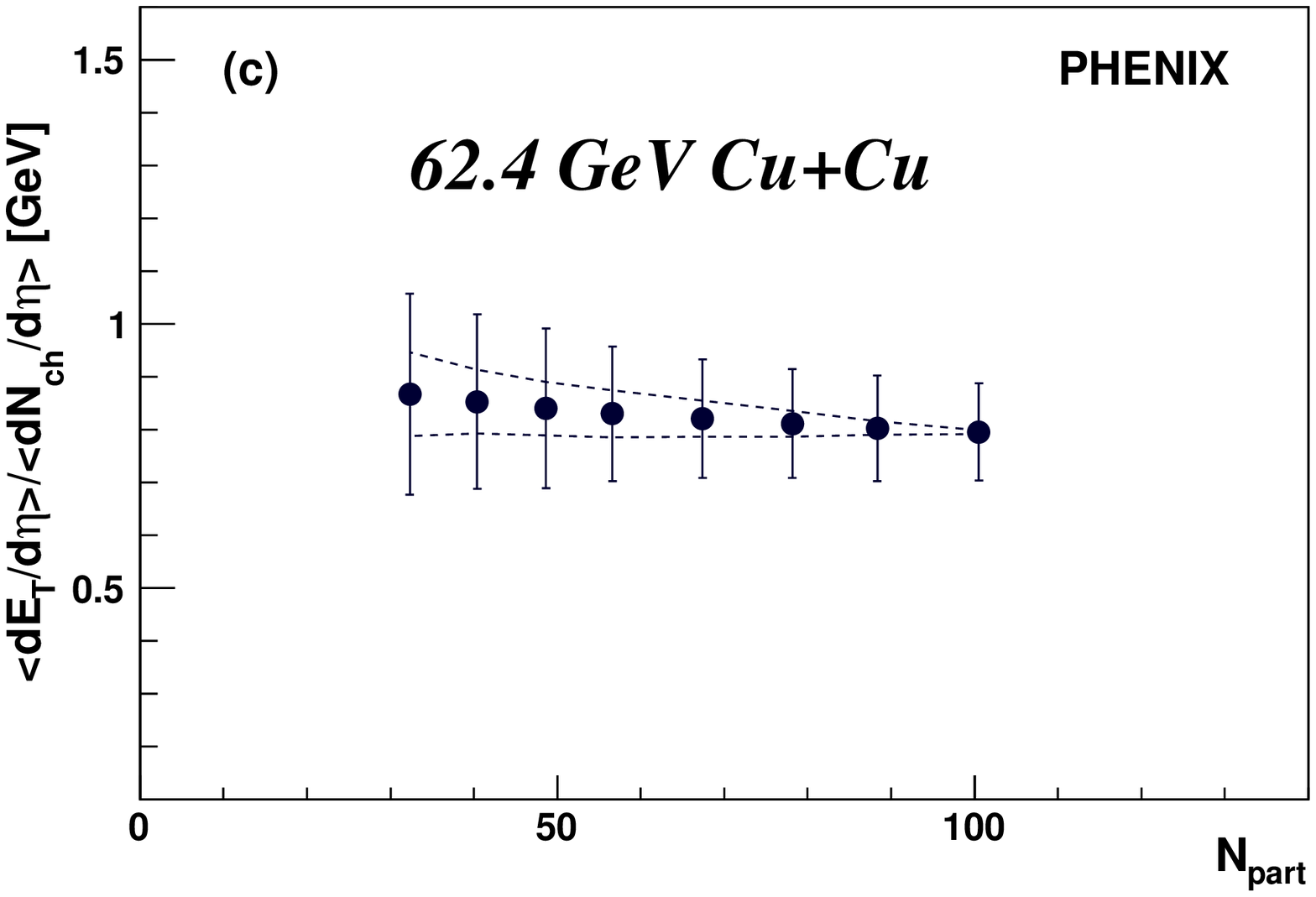}
  \caption{
The \Et/\Nch ratio as a function of \Npart for \cuau collisions at \sqsn = 
200~GeV (a), \cucu collisions at \sqsn = 200~GeV (b), and \cucu collisions 
at \sqsn = 62.4~GeV (c). The lines bounding the points represent the 
trigger efficiency uncertainty within which the points can be tilted. The 
error bars represent the total statistical and systematic uncertainties.}
  \label{fig:etOverNchCu}
\end{minipage}
\begin{minipage}{0.48\linewidth}
  \includegraphics[width=1.0\linewidth]{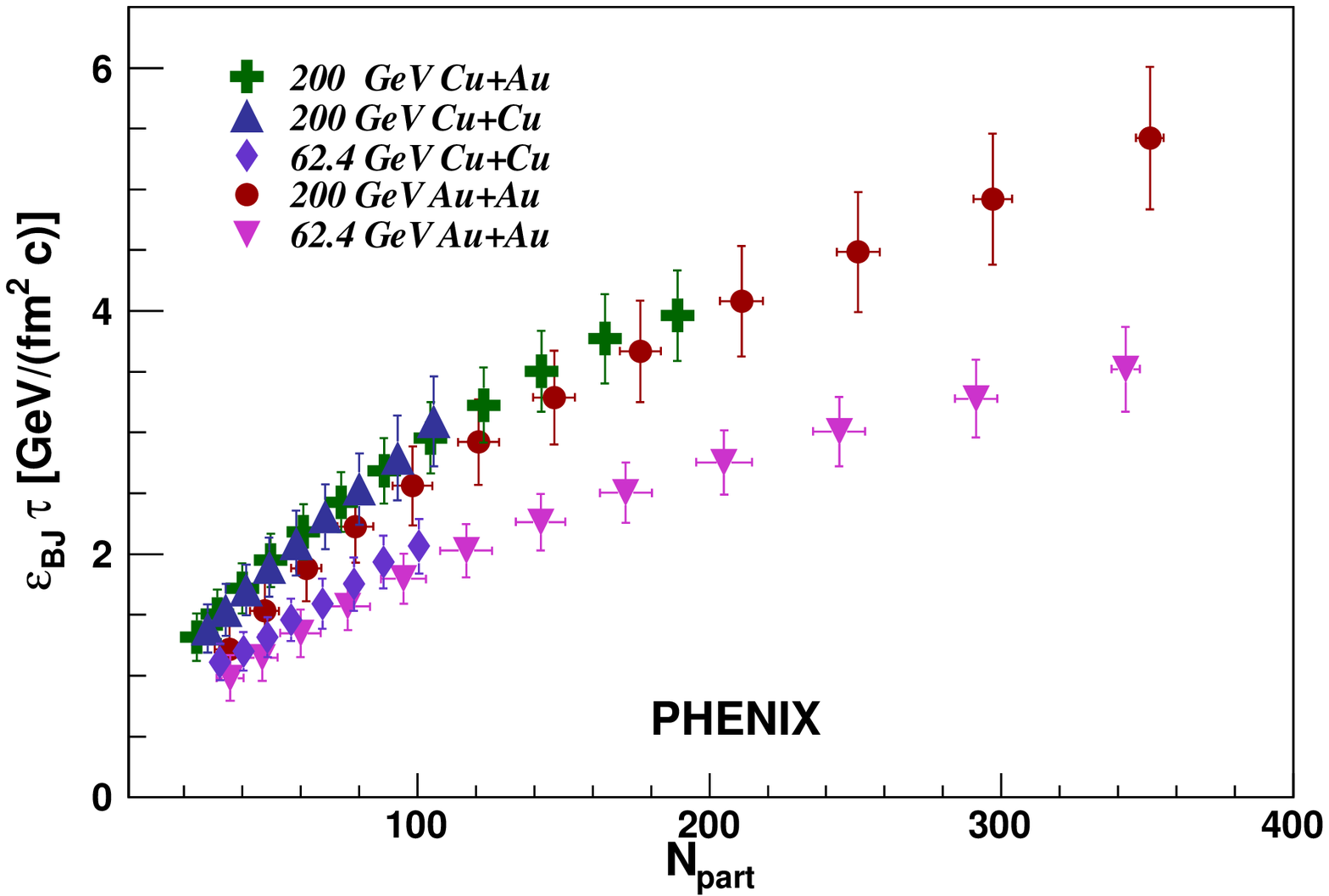}
  \caption{(Color online)
The Bjorken energy density, \ebj, multiplied by $\tau$ as a function of 
\Npart for \cucu , \cuau, and \auau collisions. The error bars represent 
the total statistical and systematic uncertainties.}
  \label{fig:ebjCu}
\end{minipage}
\end{figure*}


		\section{Results for Cu+Au and Cu+Cu Collisions}
		\label{sec:cuResults}

Measurements of \dNch in systems lighter than Au have been published by 
PHOBOS for 200~GeV and 62.4~GeV \cucu collisions~\cite{Alver:2010ck}, 
showing that the \cucu \dNch distribution as a function of \Npart exhibits 
similar features when compared to \auau collisions. Here, those 
measurements are extended to include measurements of \dEt and the addition 
of measurements from the asymmetric \cuau system at \sqsn = 200~GeV.

Figure~\ref{fig:detdnNormCu} shows \dEtNorm and \dNchNorm at midrapidity 
as a function of \Npart for \cucu and \cuau collisions.  Also shown for 
comparison are the data for \auau collisions at \sqsn = 200~GeV. Both 
plots exhibit the trend established in \auau collisions of increasing 
\dEtNorm and \dNchNorm with increasing \Npart and increasing \sqsn. The 
\cucu and \cuau distributions at \sqsn = 200~GeV are consistent with each 
other within the uncertainties of the measurement. All of the species 
(\auau, \cuau, and \cucu) at \sqsn = 200~GeV are consistent with each 
other for all overlapping values of \Npart.  This behavior had been 
previously noted when comparing \auau and \cucu data from 
PHOBOS~\cite{Nouicer:2006pr} and is now extended to include \cuau 
collisions. Figure~\ref{fig:etOverNchCu} shows that, as in the \auau 
collisions, the \Et/\Nch ratio in the lighter colliding system is 
consistent with being independent of \Npart.

Figure~\ref{fig:ebjCu} shows the \Npart dependence of \ebj multiplied by 
$\tau$ for \cucu and \cuau collisions.  Both the \cucu data at \sqsn = 
200~GeV and the \cucu data at \sqsn = 62.4~GeV increase with increasing 
\Npart. For all values of \Npart, \ebj for \cucu collisions at \sqsn = 
200~GeV and \cuau at \sqsn = 200~GeV are consistent with each other within 
the uncertainties of the measurement.  With the different collision 
geometries taken into account, there is a more consistent agreement 
between the most central \cucu and \cuau data points at \sqsn = 200~GeV 
than with \dEtNorm alone. Also shown for comparison are the \ebj values 
for Au$+$Au collisions at \sqsn = 200 and 62.4~GeV, illustrating that \ebj 
is independent of the size of the system.


		\section{Results for U+U Collisions}
		\label{sec:uResults}

During the 2012 data-taking period, RHIC delivered \uu collisions at 
\sqsn=193~GeV. \uu collisions can provide additional information about the 
dynamics of the 
system~\cite{Heinz:2004ir,Hirano:2010jg,Haque:2011aa,Schenke:2014tga} by 
varying the collision geometry of the nonspherical prolate uranium 
nuclei~\cite{Kuhlman:2005ts}. However, for this study, there is no 
collision geometry selection applied to the data. The results presented 
here are integrated over all orientations of the colliding nuclei.

The estimate of \Npart as a function of centrality for \uu collisions is 
made using the method described previously. However, the \uu collisions 
are now modeled in the Glauber Monte-Carlo calculation using a deformed 
Woods Saxon distribution for the uranium nucleus to describe its prolate 
shape,
\begin{equation}
  \rho(r) = \rho_{0}/(1+e^{(r-R')/a}),
\end{equation}
where $\rho_{0}$ is the normal nuclear density, $a$ is the surface 
diffuseness parameter, and $R'$ is a $\theta$-dependent description of the 
nuclear radius,
\begin{equation}
  R' = R[1+\beta_{2}Y_{2}^{0}(\theta) + \beta_{4}Y_{4}^{0}(\theta)],
\end{equation}
where $Y^{0}$ is a Legendre Polynomial. The Woods Saxon parameters used 
are taken from a previous study with $R$ = 6.81 fm, $a$ = 0.6 fm, 
$\beta_{2}$ = 0.28, and $\beta_{4}$ = 0.093~\cite{Masui:2009qk}.  There is 
an additional study that presents a different set of parameters ($R$ = 
6.86 fm, $a$ = 0.42 fm, $\beta_{2}$ = 0.265, and $\beta_{4}$ = 
0)~\cite{Shou:2014eya}. The two parametrizations result in \Npart 
estimates that are consistent within the uncertainties, so the \Npart 
values quoted here are from the former 
parametrization~\cite{Masui:2009qk}.


\begin{figure*}[!htb] 
\begin{minipage}{0.9\linewidth}
  \includegraphics[width=0.4\linewidth]{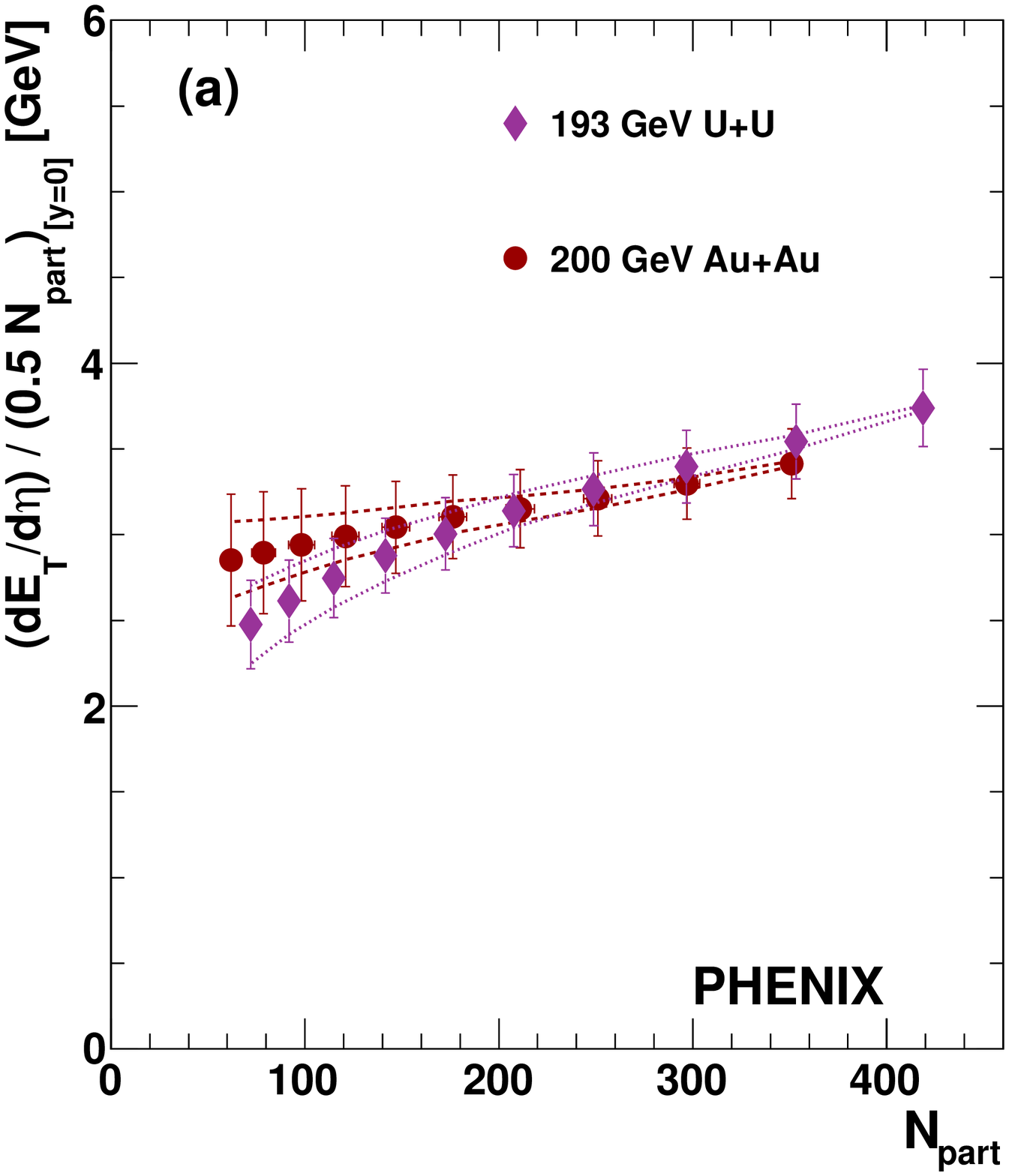}
  \includegraphics[width=0.4\linewidth]{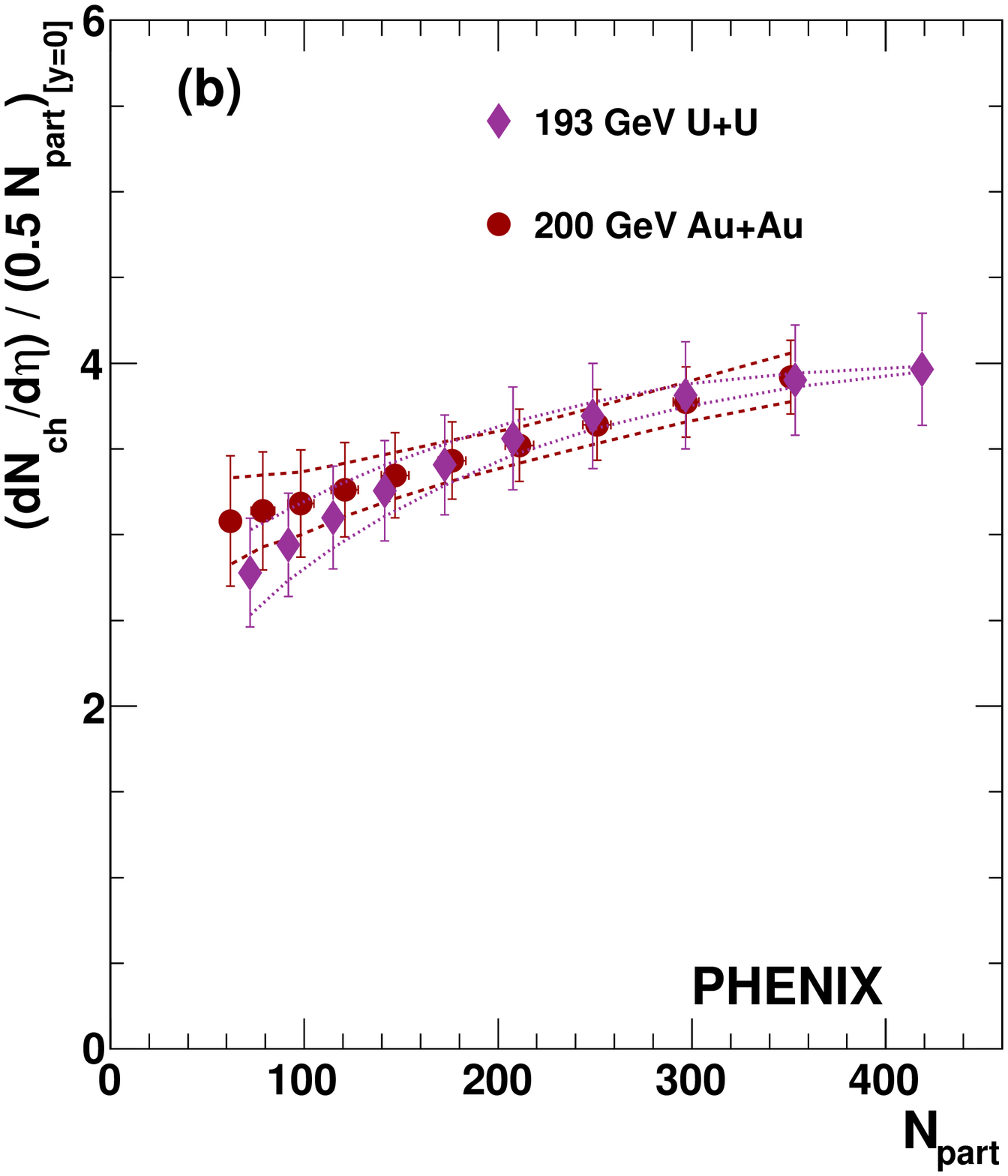}
  \caption{(Color online) 
\dEtNorm (a) and \dNchNorm (b) at midrapidity as a function of \Npart for 
\uu collisions. Also shown are results from \auau collisions at \sqsn = 
200~GeV for comparison. The lines bounding the points represent the 
trigger efficiency uncertainty within which the points can be tilted. The 
error bars represent the remaining total statistical and systematic 
uncertainty.}
  \label{fig:detdnNormU}
\end{minipage}
\begin{minipage}{0.9\linewidth}
  \includegraphics[width=0.4\linewidth]{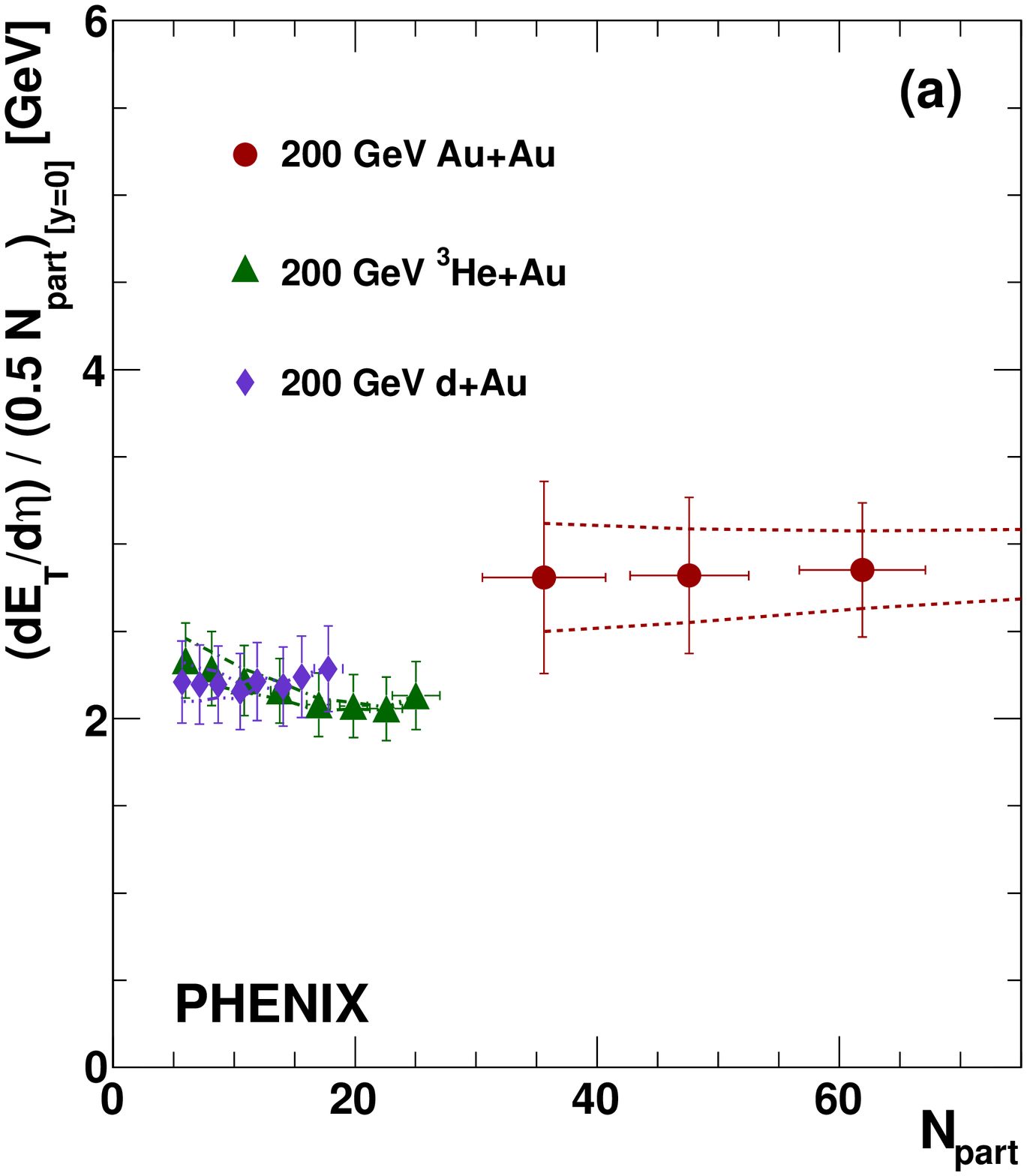}
  \includegraphics[width=0.4\linewidth]{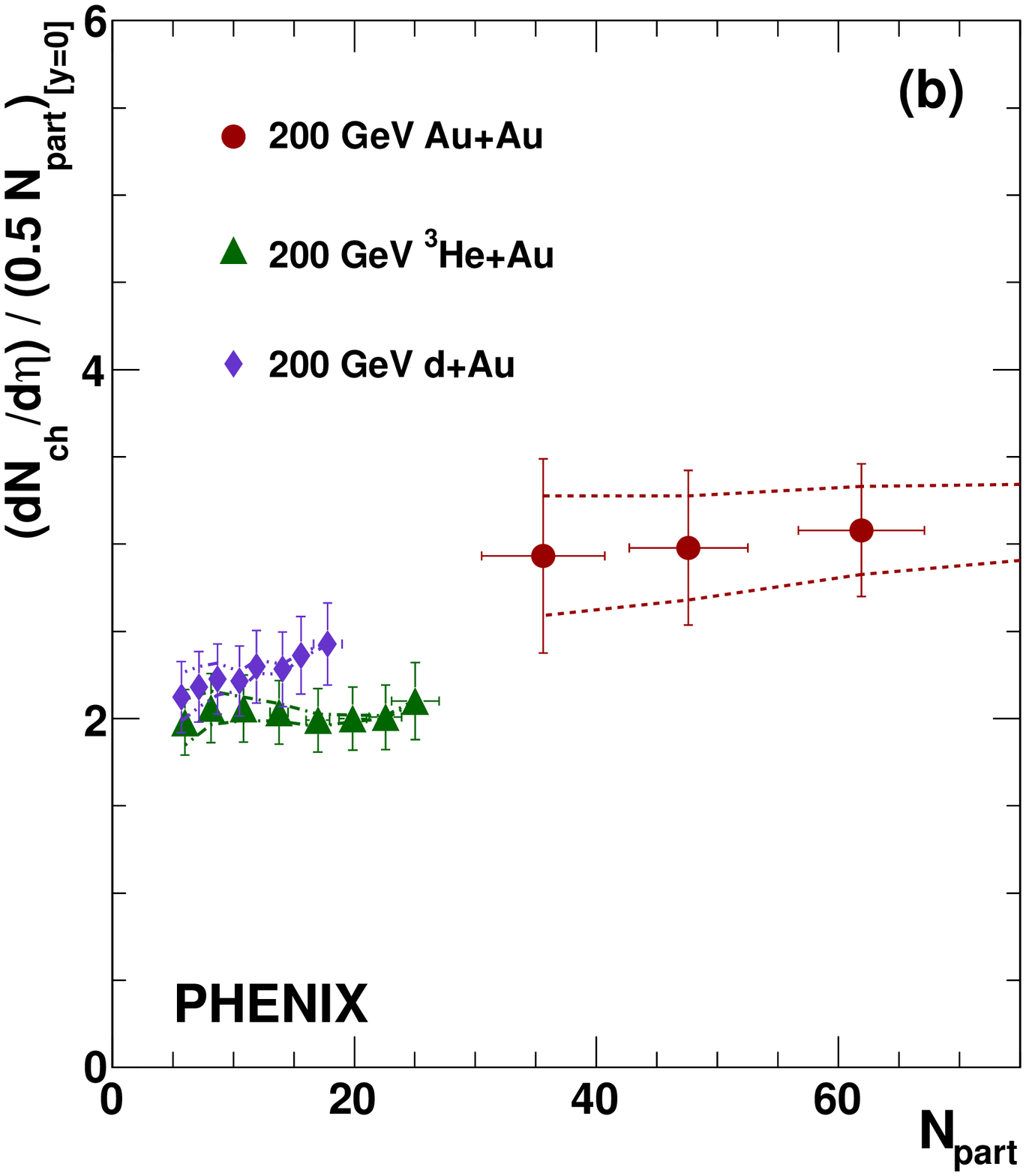}
  \caption{(Color online) 
\dEtNorm (a) and \dNchNorm (b) at midrapidity as a function of \Npart for 
$d$+Cu and \heau collisions. Also shown are results from the most 
peripheral \auau collisions at \sqsn = 200~GeV for comparison. The lines 
bounding the points represent the trigger efficiency uncertainty within 
which the points can be tilted. The error bars represent the remaining 
total statistical and systematic uncertainty.}
  \label{fig:detdnNormHe}
\end{minipage}
\end{figure*}

\begin{figure*}[!htb] 
  \includegraphics[width=0.48\linewidth]{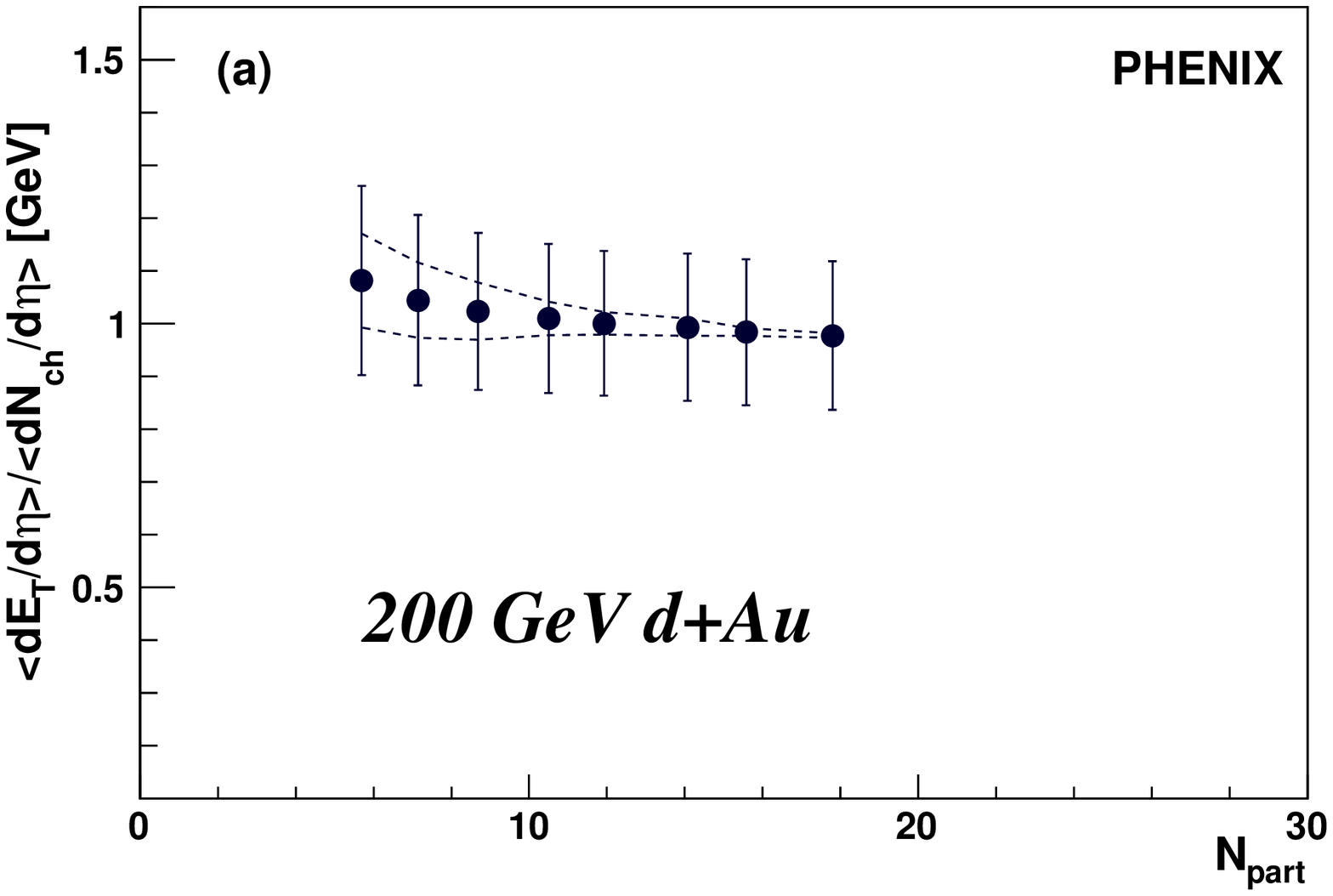}
  \includegraphics[width=0.48\linewidth]{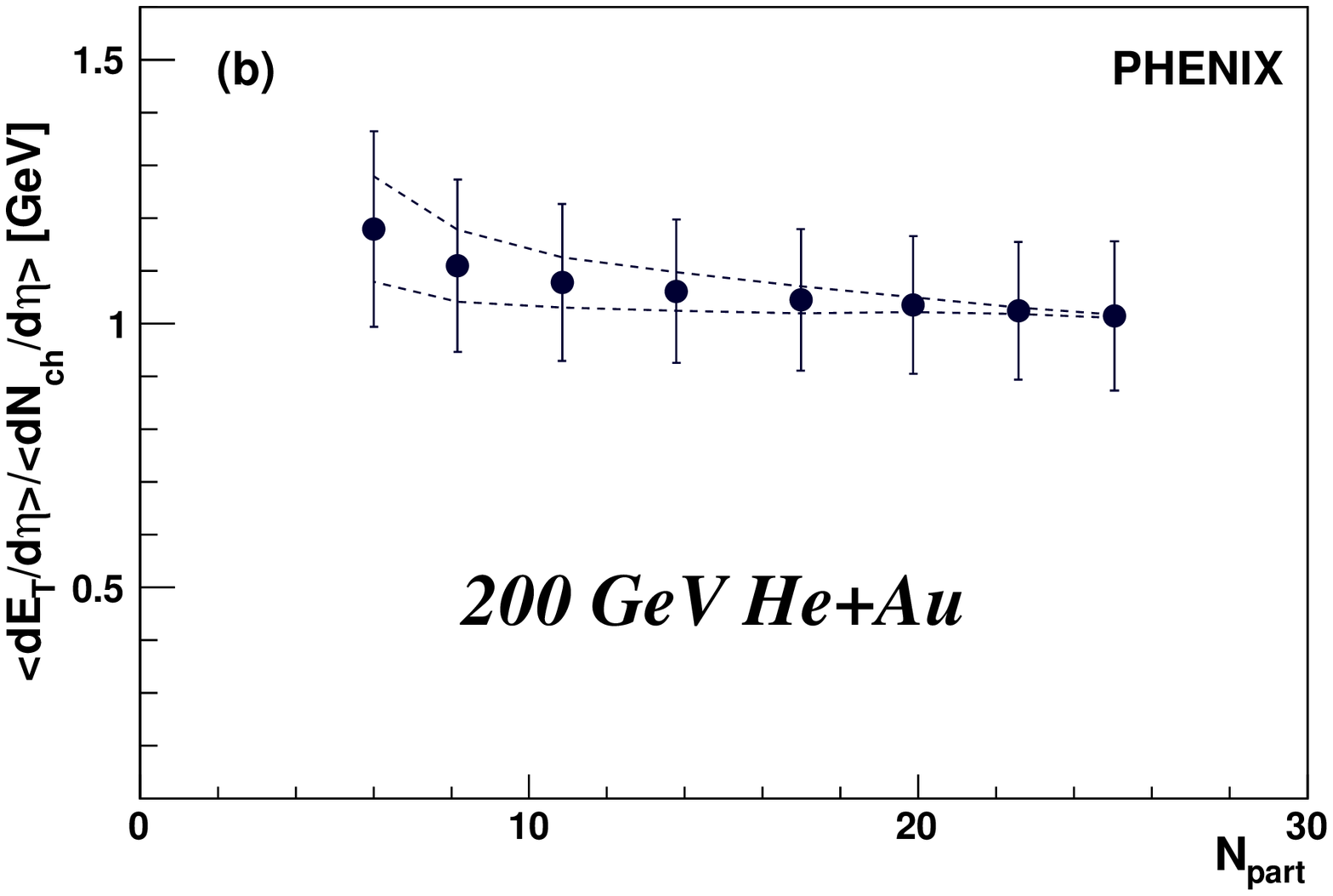}
  \caption{
The \Et/\Nch ratio as a function of \Npart for 200~GeV \dau (a) and 
200~GeV \heau (b) collisions. The error bars represent the total 
statistical and systematic uncertainties.}
  \label{fig:etOverNchHe}
\end{figure*}

\begin{table*}[htb]
\caption{
Summary of the results of the linear fits to the functions 
\dEt$=a_{E}$\Nqp+$b_{E}$ and \dNch$=a_{N}$\Nqp+$b_{N}$.}
\label{tab:linearFits}
\begin{ruledtabular}
\begin{tabular}{ccccccccc}
\sqsn    & System & $a_{E}$ [GeV] & $b_{E}$ [GeV] & $a_{N}$ & $b_{N}$ \\[0.2pc]\hline
200~GeV  & \auau & $0.629 \pm 0.021$ & $-6.1 \pm 5.4$  & $0.716 \pm 0.020$ & $-6.0  \pm 6.2$ \\
200~GeV  & \cuau & $0.612 \pm 0.021$ & $ 3.4 \pm 2.7$  & $0.706 \pm 0.029$ & $ 2.1  \pm 3.7$ \\
200~GeV  & \cucu & $0.632 \pm 0.039$ & $ 1.9 \pm 3.9$  & $0.735 \pm 0.040$ & $-1.1  \pm 3.9$ \\
130~GeV  & \auau & $0.555 \pm 0.017$ & $-1.9 \pm 4.3$  & $0.635 \pm 0.016$ & $-1.6  \pm 4.2$ \\
62.4~GeV & \auau & $0.435 \pm 0.015$ & $-1.9 \pm 3.7$  & $0.499 \pm 0.023$ & $ 2.2  \pm 5.2$ \\
62.4~GeV & \cucu & $0.449 \pm 0.026$ & $ 2.7 \pm 2.8$  & $0.578 \pm 0.043$ & $-0.9  \pm 4.5$ \\
39~GeV   & \auau & $0.356 \pm 0.013$ & $ 0.8 \pm 3.6$  & $0.409 \pm 0.020$ & $ 1.5  \pm 4.8$ \\
27~GeV   & \auau & $0.298 \pm 0.010$ & $ 2.9 \pm 2.2$  & $0.357 \pm 0.017$ & $ 0.3  \pm 3.4$ \\
19.6~GeV & \auau & $0.264 \pm 0.011$ & $ 3.0 \pm 2.8$  & $0.320 \pm 0.016$ & $ 1.5  \pm 3.9$ \\
14.5~GeV & \auau & $0.232 \pm 0.010$ & $-1.2 \pm 2.5$  & $0.287 \pm 0.015$ & $-3.2  \pm 3.5$ \\
7.7~GeV  & \auau & $0.163 \pm 0.007$ & $-1.8 \pm 1.8$  & $0.226 \pm 0.017$ & $-2.9  \pm 2.9$ \\
\end{tabular}
\end{ruledtabular}
\end{table*}

\begin{figure*}[!htb] 
\begin{minipage}{0.9\linewidth}
  \includegraphics[width=0.4\linewidth]{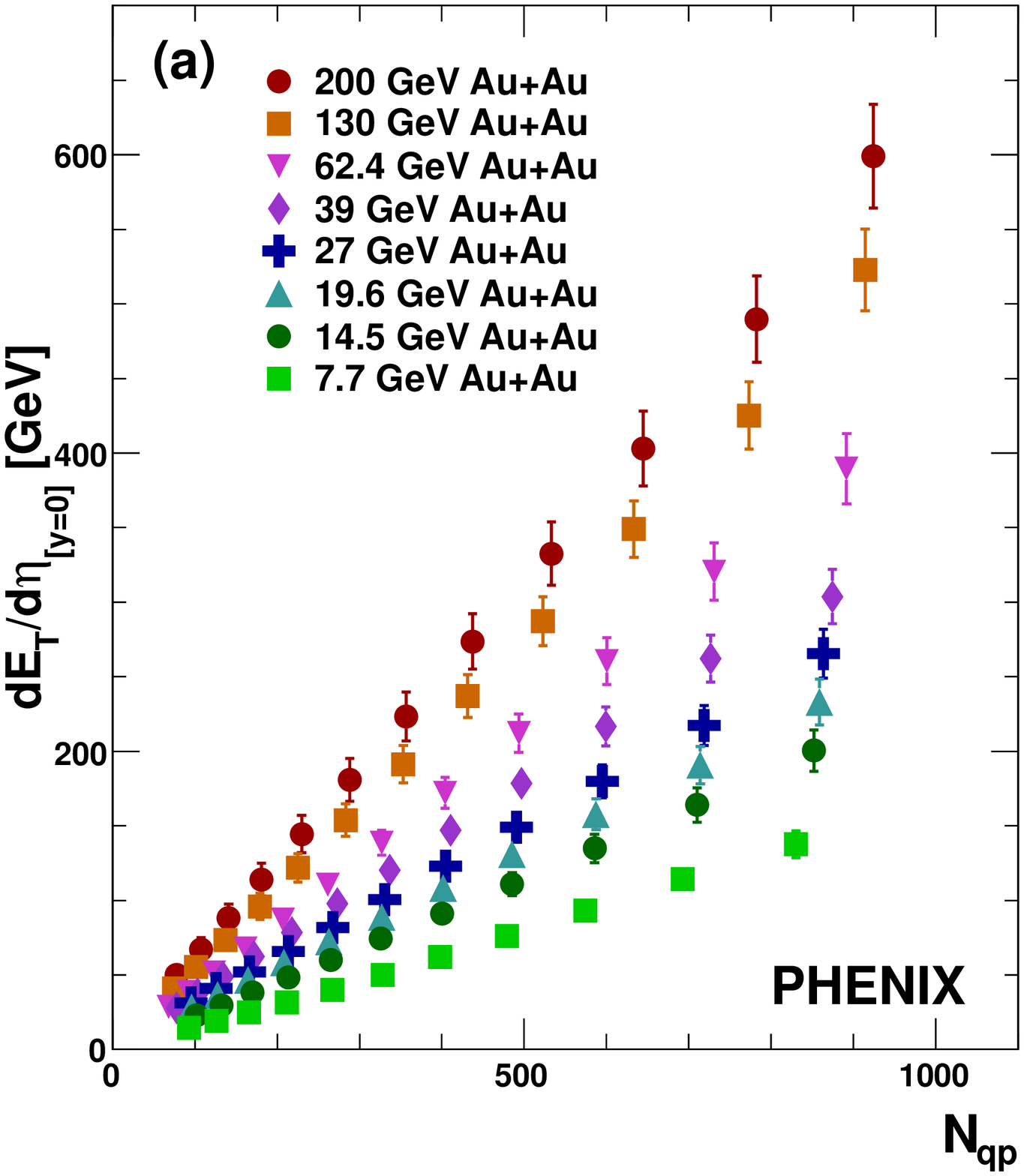}
  \includegraphics[width=0.4\linewidth]{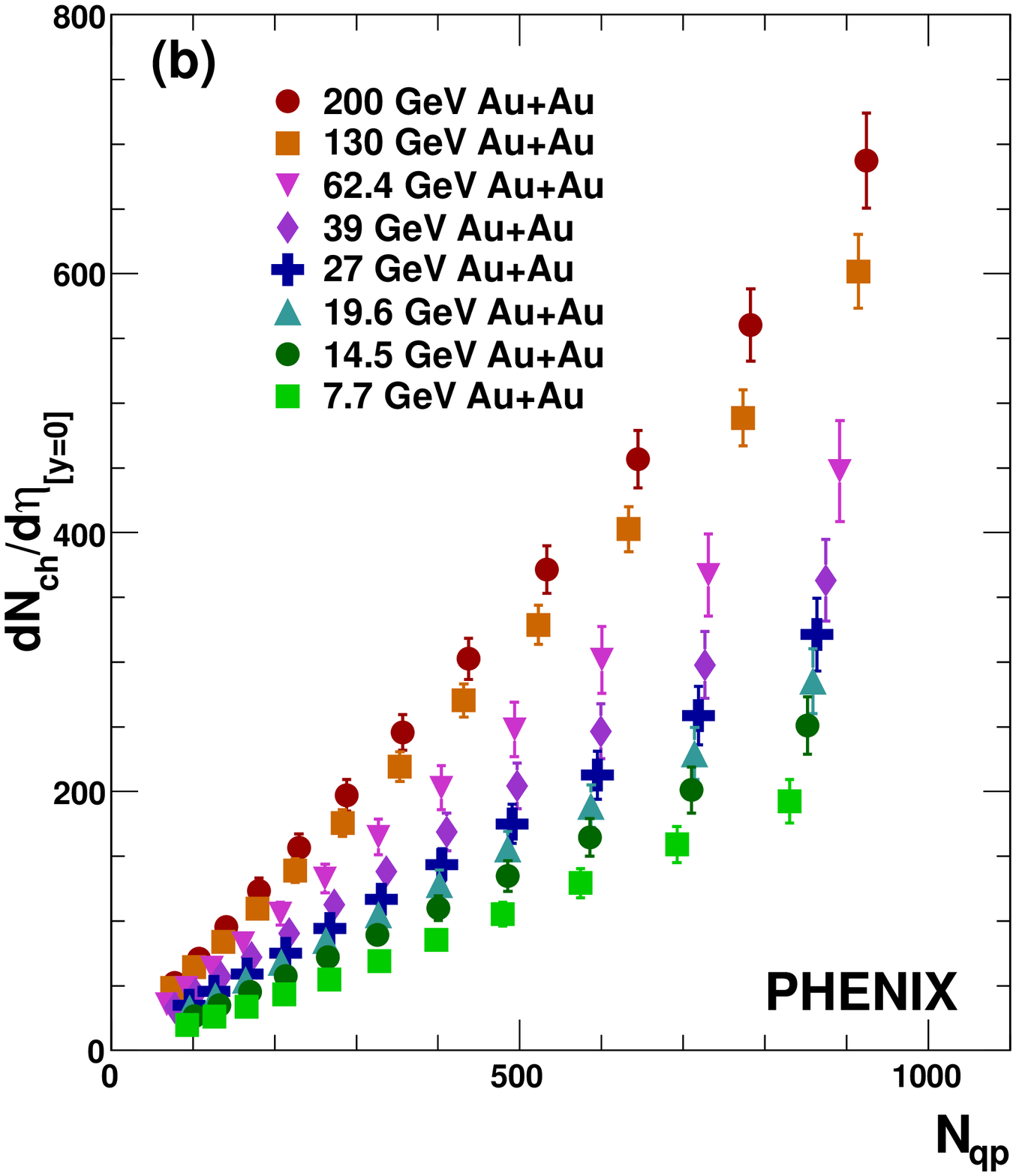}
  \caption{(Color online) 
\dEt (a) and \dNch (b) at midrapidity as a function of \Nqp for \auau 
collisions at \sqsn = 200, 130, 62.4, 39, 27, 19.6, 14.5, and 7.7~GeV. The 
error bars represent the total statistical and systematic uncertainties.}
  \label{fig:detdnNquarkBES}
\end{minipage}
\begin{minipage}{0.9\linewidth}
  \includegraphics[width=0.4\linewidth]{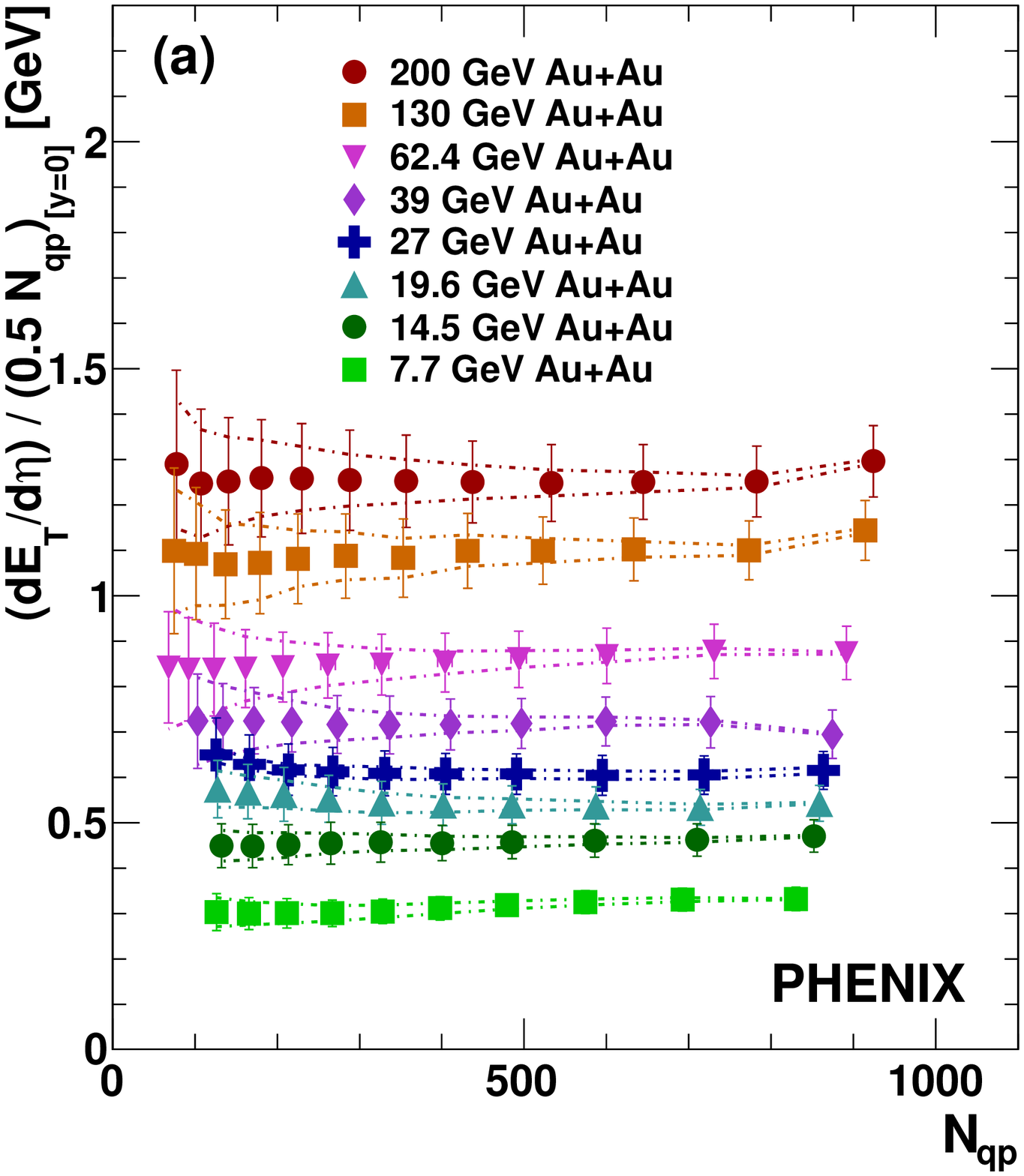}
  \includegraphics[width=0.4\linewidth]{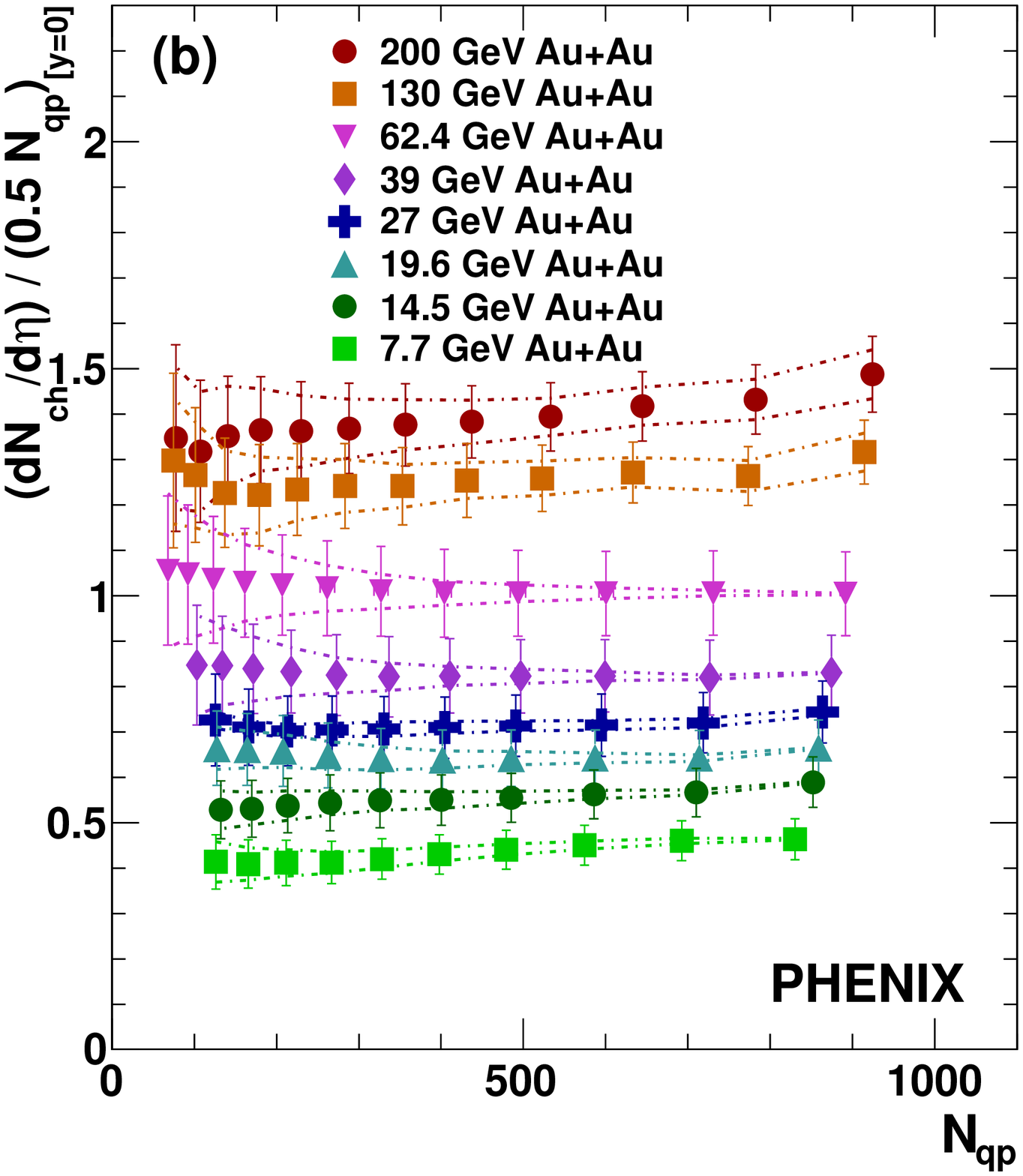}
  \caption{(Color online) 
\dEtNormQ (a) and \dNchNormQ (b) at midrapidity as a function of \Nqp for 
\auau collisions at \sqsn = 200, 130, 62.4, 39, 27, 19.6, 14.5, and 
7.7~GeV. The lines bounding the points represent the trigger efficiency 
uncertainty within which the points can be tilted. The error bars 
represent the remaining total statistical and systematic uncertainty.}
  \label{fig:detdnNormNquarkBES}
\end{minipage}
\end{figure*}

\begin{figure*}[!htb] 
  \includegraphics[width=0.48\linewidth]{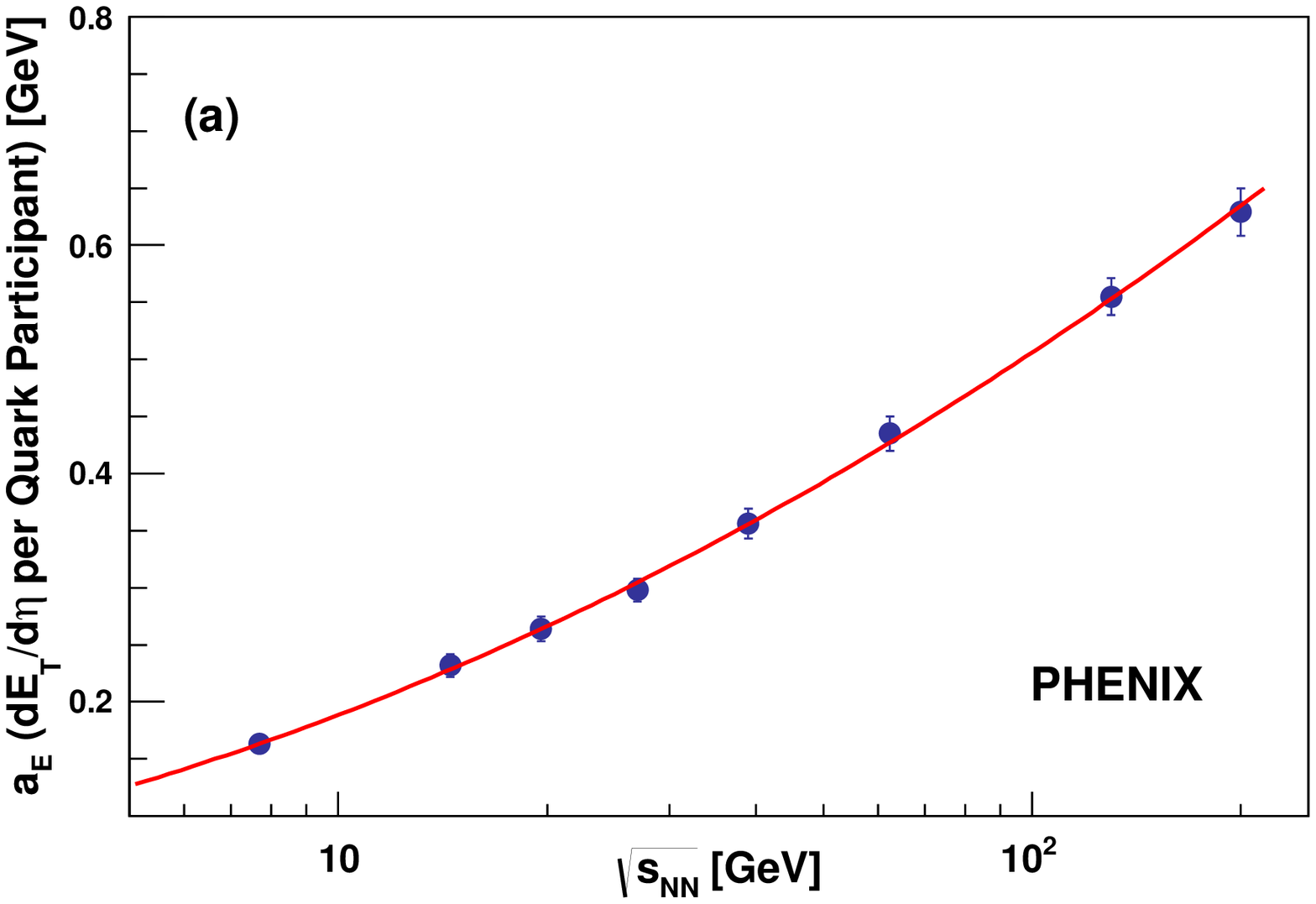}
  \includegraphics[width=0.48\linewidth]{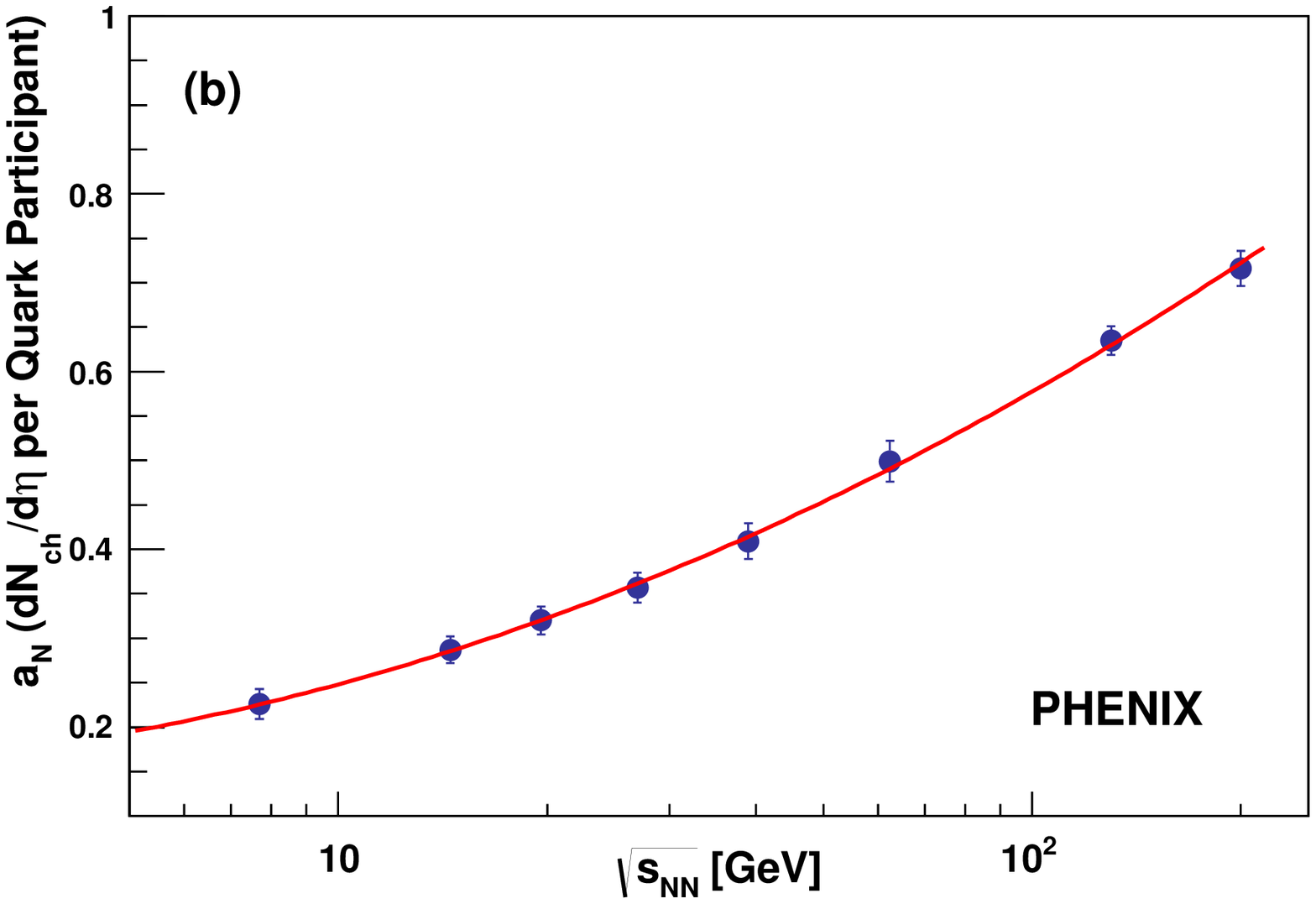}
  \caption{(Color online) 
The slopes of the fit to \dEt, $a_{E}$ (a), and \dNch, $a_{N}$ (b), as a 
function of \Nqp plotted as a function of \sqsn for Au$+$Au collisions. The 
error bars are the uncertainties from the fit. The red line is a 
second-order polynomial fit to the data.}
  \label{fig:detdnPerNqp}
\end{figure*}
\begin{figure*}[!htb]
  \includegraphics[width=0.4\linewidth]{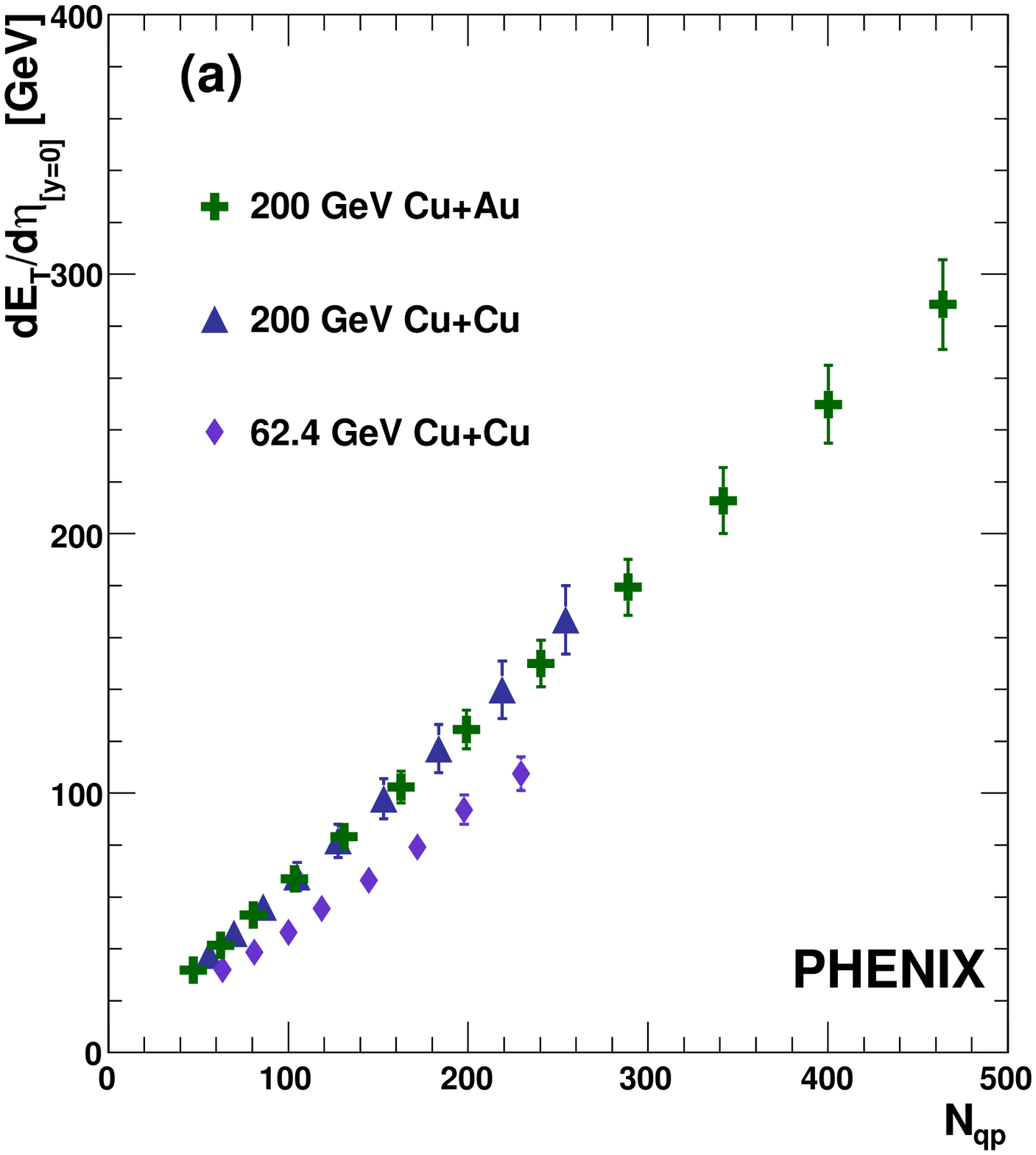}
  \includegraphics[width=0.4\linewidth]{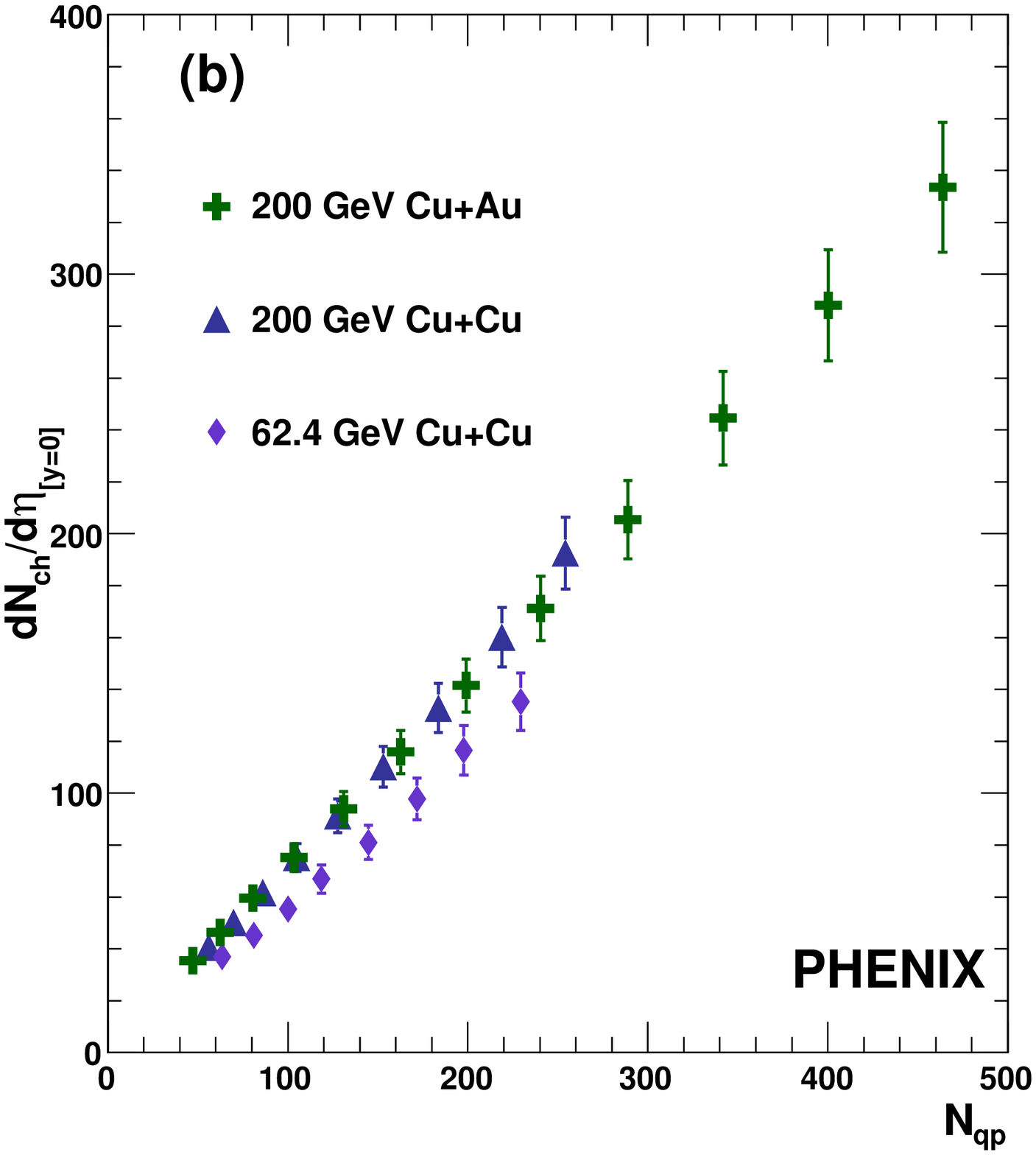}
  \caption{(Color online)
\dEt (a) and \dNch (b) at midrapidity as a function of \Nqp for \cucu and
\cuau collisions. The error bars represent the total statistical and
systematic uncertainties.}
  \label{fig:detdnNquarkCu}
\end{figure*}

\begin{figure*}[!htb]
\begin{minipage}{0.9\linewidth}
  \includegraphics[width=0.4\linewidth]{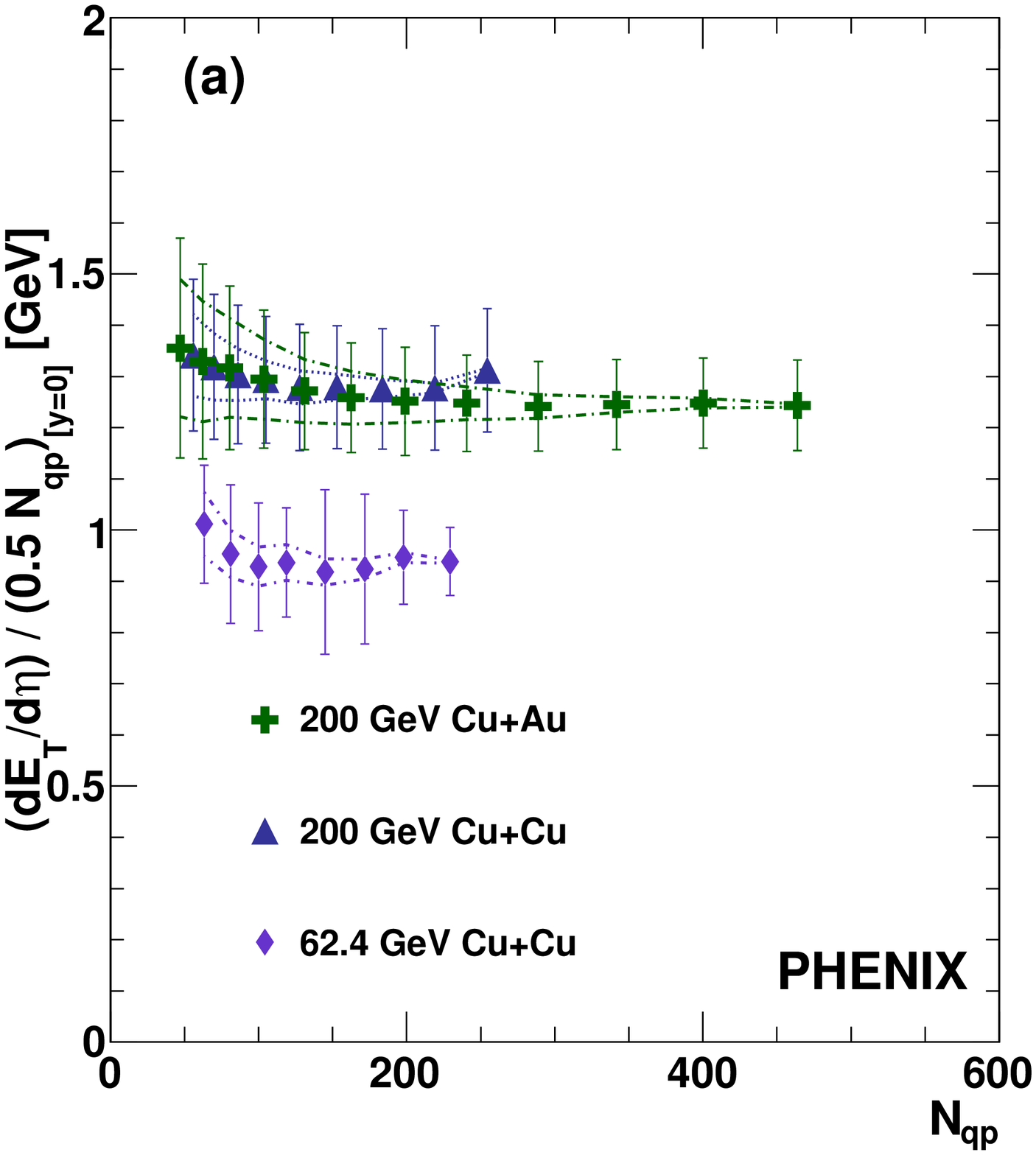}
  \includegraphics[width=0.4\linewidth]{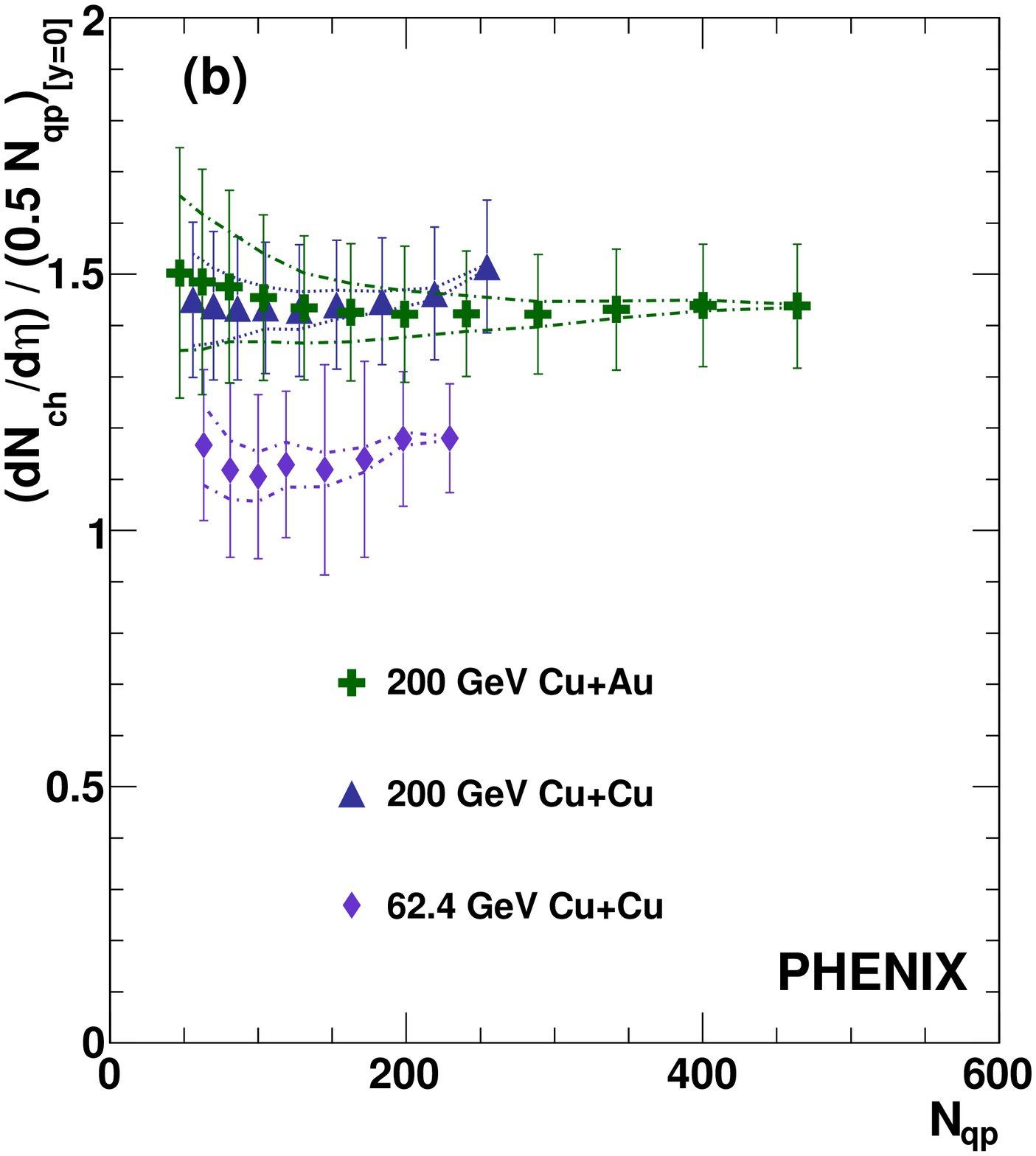}
  \caption{(Color online)
\dEtNormQ (a) and \dNchNormQ (b) at midrapidity as a function of \Nqp for
\cucu and \cuau collisions. The lines bounding the points represent the
trigger efficiency uncertainty within which the points can be tilted. The
error bars represent the remaining total statistical and systematic
uncertainty.}
  \label{fig:detdnNormNquarkCu}
\end{minipage}
\begin{minipage}{0.9\linewidth}
  \includegraphics[width=0.4\linewidth]{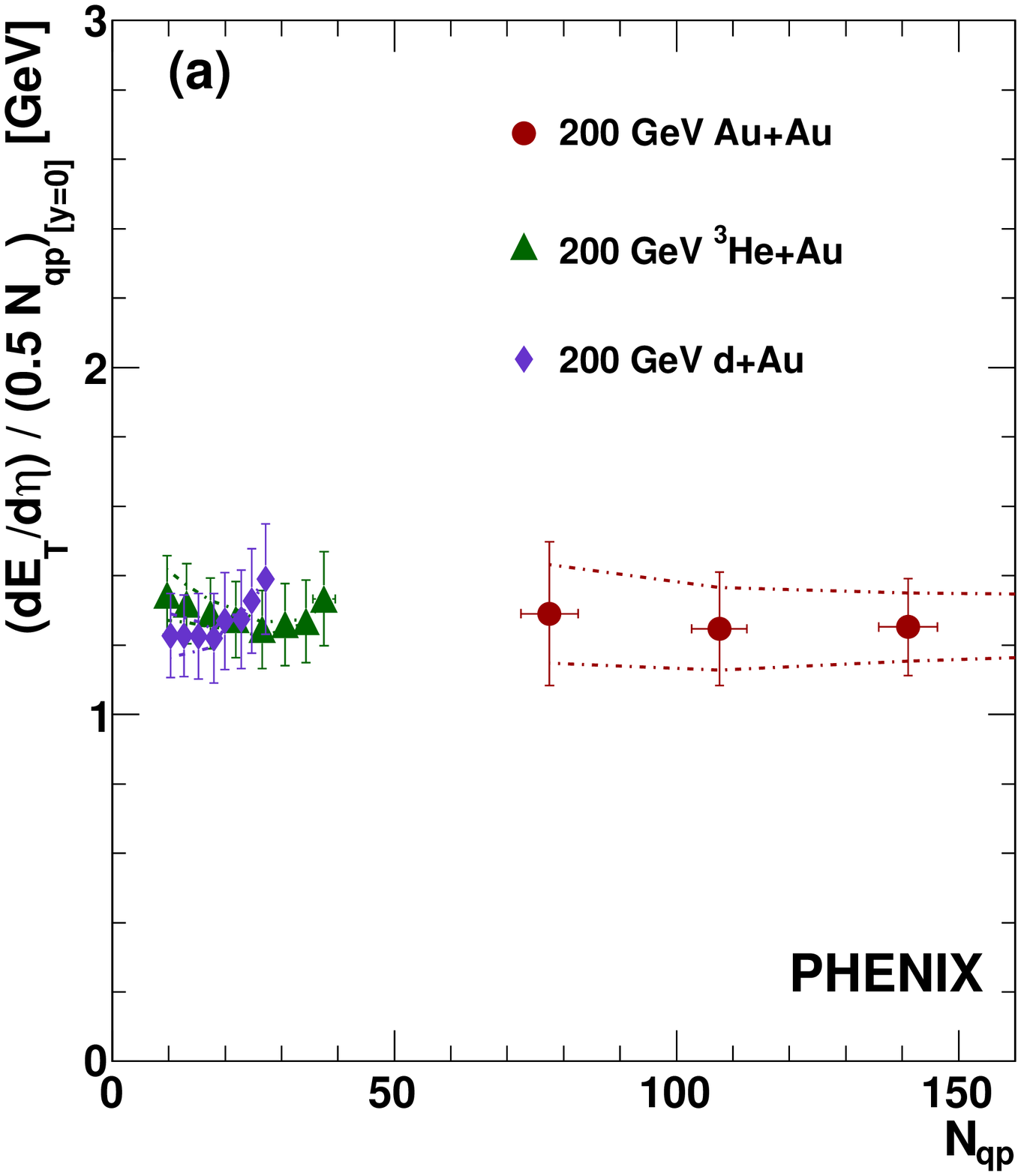}
  \includegraphics[width=0.4\linewidth]{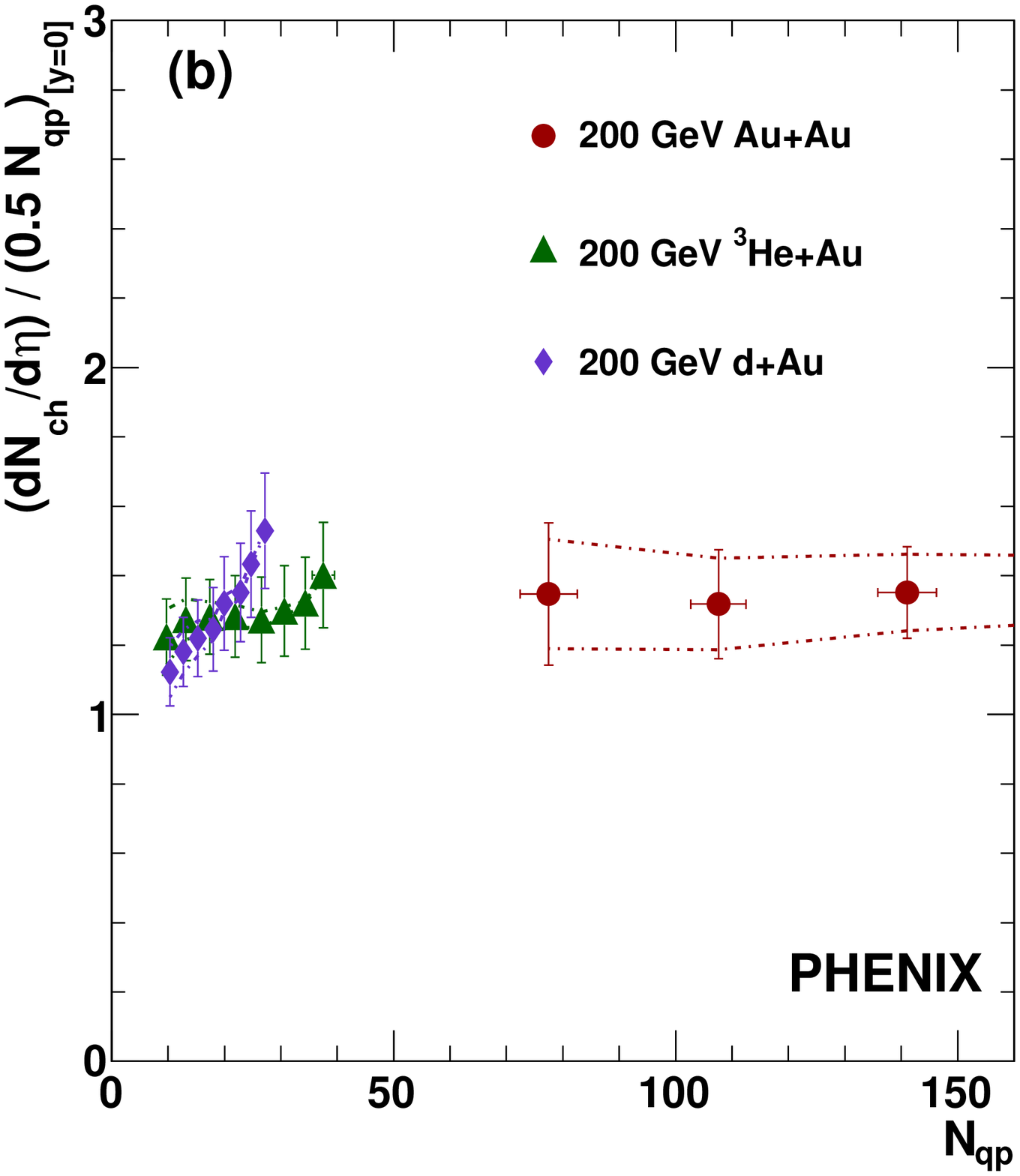}
  \caption{(Color online)
\dEtNormQ (a) and \dNchNormQ (b) at midrapidity as a function of \Nqp for
\dau and \heau collisions. Shown comparison are data from the most
peripheral collisions \auau collisions at \sqsn = 200~GeV. The lines
bounding the points represent the trigger efficiency uncertainty within
which the points can be tilted. The error bars represent the remaining
total statistical and systematic uncertainty.}
  \label{fig:detdnNormNquarkHe}
\end{minipage}
\end{figure*}

Figure~\ref{fig:detdnNormU} shows \dEtNorm and \dNchNorm at midrapidity as 
a function of \Npart for \uu collisions at \sqsn = 193~GeV.  Also shown 
for comparison are the data for \auau collisions at \sqsn = 200~GeV. Both 
the \uu and the \auau data are consistent with each other for all values 
of \Npart.  This behavior is also observed when comparing \auau, \cuau, 
and \cucu data as discussed in the previous section.

		\section{Results for Deuteron+Au and $^3$He+Au Collisions}
		\label{sec:smallResults}

Measurements of \dNch have been published by PHOBOS for \dau collisions at 
\sqsn = 200~GeV\cite{Alver:2010ck}. Here those measurements are extended 
to include measurements of \dEt and the addition of measurements from 
\heau collisions at \sqsn = 200~GeV.

A detailed description of the method used to define the centrality of 
200~GeV \dau collisions using the PHENIX detector can be found 
elsewhere~\cite{Adare:2013nff}. The same method was applied to define the 
centrality in \heau collisions.  Figure~\ref{fig:detdnNormHe} shows 
\dEtNorm and \dNchNorm as a function of \Npart for \dau and \heau 
collisions.  Also shown are the most peripheral \auau points 
at \sqsn = 200~GeV for comparison.  Within the uncertainties, the results 
for 200~GeV \dau and \heau collisions are consistent with each other for 
all values of \Npart. As with the heavier systems, the \Et/\Nch ratio is 
consistent with being independent of \Npart within the uncertainties of 
the measurement as shown in Fig.~\ref{fig:etOverNchHe}.

For minimum-bias \pp collisions at \sqsn=200~GeV, \dEtNorm is 2.27 $\pm$ 
0.19~GeV and \dNchNorm is 2.38 $\pm$ 0.17, where the uncertainties 
represent the total statistical and systematic uncertainties.  These 
measurements are consistent with the most peripheral results from both 
\heau and \dau collisions. The \dNchNorm measurement is also consistent 
with the PHOBOS measurement~\cite{Alver:2010ck}.

		\section{Quark Participant Scaling at Midrapidity}
		\label{sec:quarks}

Thus far, \dEt and \dNch have been discussed in terms of the dependence on 
the number of nucleon participants in the collision.  Here, the behavior 
as a function of the number of quark participants, \Nqp, will be examined. 
PHOBOS \dNch data for \auau collisions at \sqsn = 200 and 130~GeV have 
been analyzed as a function of \Nqp~\cite{Eremin:2003qn}. This analysis 
shows that the data at midrapidity are better described by scaling with 
\Nqp than with \Npart at the top RHIC energies. A separate analysis of the 
PHOBOS \dNch data for \auau collisions extended down to \sqsn = 62.4 and 
19.6~GeV in terms of \Nqp~\cite{Nouicer:2006pr} concludes that \Nqp 
scaling better describes the data than \Npart scaling at those lower 
energies. 
PHENIX compared various models of particle production and verified that \Nqp scaling best describes the midrapidity \dEt measurements in \auau collisions at \sqsn = 200 and 62.4 GeV~\cite{Adler:2013aqf}. Here these analyses are extended to include \dEt and \dNch measurements down to \sqsn = 7.7 GeV.

The number of quark participants is estimated using a Monte Carlo Glauber 
model calculation method~\cite{Miller:2007ri} that has been modified to 
replace nucleons with constituent quarks~\cite{Adler:2013aqf}. The nuclei 
are initially assembled by distributing the centers of the nucleons 
according to a Woods-Saxon distribution.  After a nucleus is fully 
assembled, the nucleons are replaced by three quarks distributed around 
the center of each nucleon. The quarks are distributed radially by 
sampling an empirically determined function:
\begin{equation}
\begin{split}
    f(r) = r^{2}e^{-4.27r} (1.21466-1.888r+2.03r^{2}) \\
    (1+1.0/r - 0.03/r^{2}) (1+0.15r), 
\end{split}
\end{equation}
where $r$ is the radial position of the quark in fm~\cite{dennis}.  The 
azimuthal position of each quark is assigned randomly to achieve a 
spherically symmetric distribution.  Once all of the quark coordinates are 
determined, the center-of-mass of the three quark system is shifted to 
match the center position of the nucleon. The empirical function above is 
chosen such that after the center-of-mass is shifted, the radial 
distribution of the quark positions with respect to the nucleon center 
position reproduces the Fourier transform of the proton form factor as 
measured in electron-proton elastic scattering~\cite{Hofstadter:1956qs}:
\begin{equation}
  \rho^{proton}(r) = \rho_{0}^{proton} \times e^{-ar},
\end{equation}
where a = $\sqrt{12}/r_{m}$ = 4.27 $fm^{-1}$ and $r_{m}$ = 0.81 fm is the 
rms charge radius of the proton ~\footnote{This approach is necessary 
because if $\rho^{proton}(r)$ itself is simply sampled for the quark 
radial coordinates, the re-centering of the three quark system would 
result in a distortion of the radial distribution, which would then be 
calculated with respect to the center of mass of the generated system.}.  
Once all quarks in both nuclei are positioned, the coordinates of the two 
nuclei are shifted relative to each other at random uniformly in the 
impact parameter plane transverse to the beam axis. Interactions between a 
pair of quarks, one from each nucleus, occur if the distance $d$ in this 
plane satisfies the condition:
\begin{equation}
  d < \sqrt{\frac{\sigma^{\rm inel}_{qq}}{\pi}},
\end{equation}
where $\sigma^{\rm inel}_{qq}$ is the inelastic quark-quark cross section. The 
value of $\sigma^{\rm inel}_{qq}$ is set to reproduce the known inelastic 
nucleon-nucleon cross section when running the model for nucleon-nucleon 
collisions at a given collision energy. The inelastic cross sections as a 
function of \sqsn are taken from parametrizations of cross section 
measurements~\cite{Covolan:1996uy}. A summary of $\sigma^{\rm inel}_{qq}$ as a 
function of \sqsn is given in Table~\ref{tab:crossSections}.

The values of midrapidity \dEt and \dNch as a function of \Nqp are shown 
in Fig.~\ref{fig:detdnNquarkBES} for \auau collisions. For all collision 
energies, the dependence on \Nqp is linear.  When \dEtNormQ and \dNchNormQ 
is plotted as a function of \Nqp as shown in 
Fig.~\ref{fig:detdnNormNquarkBES}, the distributions are constant within 
the uncertainties of the measurement, which is not the case when 
centrality is expressed in terms of \Npart, shown in 
Fig.~\ref{fig:detdnNormBES}. For \auau collisions from \sqsn = 200 to 
7.7~GeV, scaling with \Nqp better describes the data than scaling with 
\Npart.

Because there is a linear dependence of \dEt and \dNch with \Nqp, the data 
for each collision energy in Fig.~\ref{fig:detdnNquarkBES} can be fit to 
a straight line \dEt$=a_{E}$\Nqp+$b_{E}$ and \dNch$=a_{N}$\Nqp+$b_{N}$. 
The extracted slopes, $a_{E}$ and $a_{N}$, represent the \dEt and \dNch 
per quark participant, respectively. For all collision energies, the 
intercept of the fit at \sqsn = 0, which is kept as a free parameter in 
the fit, is consistent with zero within at most 1.3 standard deviations 
for all data sets.  Figure~\ref{fig:detdnPerNqp} shows the excitation 
function of the slopes for \auau collisions. The \dEt data can be 
described by a second-order polynomial: $a_{E} = 0.0408 + 0.0273 \times 
log(\sqsn) + 0.0160 \times (log(\sqsn))^{2}$. The \dNch data can be 
described by a second-order polynomial: $a_{N} = 0.153 - 0.0096 \times 
log(\sqsn) + 0.0221 \times (log(\sqsn))^{2}$. The results of the linear fits 
for each collision energy are tabulated in 
Table~\ref{tab:linearFits}~\footnote{Note that the method of generating 
constituent quarks in the present work is slightly different than that of 
Ref.~\cite{Adler:2013aqf}, which did not preserve the center of mass of 
the three quarks. There is a small effect of the different methods 
indicated by the small difference of $\mean{\dEt}/\Nqp = 0.617 \pm 
0.23$~GeV in Ref.~\cite{Adler:2013aqf} compared to the present 
$\mean{\dEt}/\Nqp = 0.629 \pm 0.021$~GeV.}.

The preference of the scaling with \Nqp is also apparent in \cucu and 
\cuau collisions.  This is demonstrated in Fig.~\ref{fig:detdnNquarkCu} 
which shows that \dEt and \dNch increases linearly with increasing \Nqp. 
As previously shown in Fig.~\ref{fig:detdnNormCu}, \dEtNorm and 
\dNchNorm both exhibit a distinct increase as \Npart increases for all 
three systems. This is not the case when comparing to 
Fig.~\ref{fig:detdnNormNquarkCu}, which shows that \dEtNormQ and 
\dNchNormQ exhibits no significant dependence on \Nqp for all three 
systems.  Scaling with \Nqp for \dau and \heau collisions at \sqsn = 
200~GeV is shown in Fig.~\ref{fig:detdnNormNquarkHe} along with a 
comparison to the most peripheral \auau collisions at \sqsn = 200~GeV. As 
seen when scaled with \Npart in Fig.~\ref{fig:detdnNormHe}, \dEtNormQ and 
\dNchNormQ are also consistent with \Nqp scaling, with the exception of 
\dNchNormQ for \dau collisions. There is no significant evidence that 
either \Npart or \Nqp scaling are preferred in \dau and \heau collisions.


		\section{Summary}
		\label{sec:summary}

Midrapidity distributions of transverse energy, \dEt, and charged particle 
multiplicity, \dNch, have been measured for a variety of collision systems 
and energies, including \auau collisions from \sqsn = 7.7 to 200~GeV.  
The centrality dependent distributions are presented in terms of the 
number of nucleon participants, \Npart, and the number of constituent 
quark participants, \Nqp.  The data are better described by scaling with 
\Nqp than scaling with \Npart.  This holds for \auau collisions from \sqsn 
= 200~GeV down to \sqsn = 7.7~GeV, for \cuau collisions at \sqsn = 
200~GeV, and for \cucu collisions at \sqsn = 62.4 and 200~GeV. Although 
comparisons of the data to models such as HIJING, parton saturation models 
like EKRT and KLN, and multiphase transport models such as AMPT are met 
with some success, a simple description using \Nqp scaling describes the 
data very well.

Some of the outstanding features of the data include the following. It is 
observed that measurements of \dEtNorm and \dNchNorm from a variety of 
systems including \auau, \cuau, and \cucu at \sqsn = 200~GeV are all 
consistent with each other as a function of \Npart. The production of \Et 
and \Nch in collisions of symmetric nuclei depends only on the collision 
energy and is independent of the size of the colliding system.  The 
centrality dependent distributions of the Bjorken energy density \ebj show 
an increasing trend with both \Npart and \sqsn.  At \sqsn = 200~GeV, \ebj 
for \cuau and \cucu collisions are consistent with each other for all 
\Npart, again demonstrating that \Et production is independent of the 
system size. The ratio of \dEt to \dNch is found to be constant as a 
function of centrality for all collision systems and energies. There is 
also only a weak dependence of this ratio as function of \sqsn from \sqsn 
= 7.7 to 200~GeV. Taking the ratio of \dEtNorm and \dNchNorm for \sqsn = 
200~GeV to 7.7~GeV shows that the shape of the distributions as a function 
of \Npart do not change significantly over this collision energy range. 
For central \auau collisions from \sqsn = 200 to 7.7~GeV, the value of 
\dEtNorm and \dNchNorm exhibits a power law behavior as a function of 
\sqsn.  Extending this observation, the Bjorken energy density also 
exhibits a power law behavior in central \auau collisions from \sqsn = 200 
to 7.7~GeV. Also calculations of \dEt and \dNch per quark participant are 
observed to increase as \sqsn increases.



\section*{ACKNOWLEDGMENTS}   

We thank the staff of the Collider-Accelerator and Physics
Departments at Brookhaven National Laboratory and the staff of
the other PHENIX participating institutions for their vital
contributions.  
We also thank Adam Bzdak for useful discussions.
We acknowledge support from the 
Office of Nuclear Physics in the
Office of Science of the Department of Energy,
the National Science Foundation, 
Abilene Christian University Research Council, 
Research Foundation of SUNY, and
Dean of the College of Arts and Sciences, Vanderbilt University 
(U.S.A),
Ministry of Education, Culture, Sports, Science, and Technology
and the Japan Society for the Promotion of Science (Japan),
Conselho Nacional de Desenvolvimento Cient\'{\i}fico e
Tecnol{\'o}gico and Funda\c c{\~a}o de Amparo {\`a} Pesquisa do
Estado de S{\~a}o Paulo (Brazil),
Natural Science Foundation of China (P.~R.~China),
Croatian Science Foundation and
Ministry of Science, Education, and Sports (Croatia),
Ministry of Education, Youth and Sports (Czech Republic),
Centre National de la Recherche Scientifique, Commissariat
{\`a} l'{\'E}nergie Atomique, and Institut National de Physique
Nucl{\'e}aire et de Physique des Particules (France),
Bundesministerium f\"ur Bildung und Forschung, Deutscher
Akademischer Austausch Dienst, and Alexander von Humboldt Stiftung (Germany),
National Science Fund, OTKA, K\'aroly R\'obert University College, 
and the Ch. Simonyi Fund (Hungary),
Department of Atomic Energy and Department of Science and Technology (India), 
Israel Science Foundation (Israel), 
Basic Science Research Program through NRF of the Ministry of Education (Korea),
Physics Department, Lahore University of Management Sciences (Pakistan),
Ministry of Education and Science, Russian Academy of Sciences,
Federal Agency of Atomic Energy (Russia),
VR and Wallenberg Foundation (Sweden), 
the U.S. Civilian Research and Development Foundation for the
Independent States of the Former Soviet Union, 
the Hungarian American Enterprise Scholarship Fund,
and the US-Israel Binational Science Foundation.

		\section*{APPENDIX}
		\label{sec:appendix}
 
This Appendix contains data tables for the \dEt and \dNch measurements for 
each of the collision systems.


\begin{table*}[htb]
\caption{Transverse energy results for 200~GeV Au$+$Au collisions. The 
uncertainties include the total statistical and systematic uncertainties.}
\label{tab:etAuAu200}
\begin{ruledtabular}
\begin{tabular}{cccccc}
Centrality & $\langle$\Npart$\rangle$ & $\langle$\Nqp$\rangle$ & \dEt [GeV] & \dEtNorm [GeV] & \dEtNormQ [GeV] \\\hline
0\%--5\%   & 350.9 $\pm$ 4.7 & 924.1 $\pm$ 16.2 & 599.0 $\pm$ 34.7 & 3.41 $\pm$ 0.20 & 1.30 $\pm$ 0.08\\
5\%--10\%  & 297.0 $\pm$ 6.6 & 782.6 $\pm$ 15.3 & 498.7 $\pm$ 28.9 & 3.30 $\pm$ 0.21 & 1.25 $\pm$ 0.08\\
10\%--15\% & 251.0 $\pm$ 7.3 & 644.6 $\pm$ 14.5 & 403.0 $\pm$ 25.0 & 3.21 $\pm$ 0.22 & 1.25 $\pm$ 0.08\\
15\%--20\% & 211.0 $\pm$ 7.3 & 532.9 $\pm$ 12.3 & 332.5 $\pm$ 21.2 & 3.15 $\pm$ 0.23 & 1.25 $\pm$ 0.08\\
20\%--25\% & 176.3 $\pm$ 7.0 & 437.5 $\pm$ 10.4 & 273.6 $\pm$ 18.6 & 3.10 $\pm$ 0.24 & 1.25 $\pm$ 0.09\\
25\%--30\% & 146.8 $\pm$ 7.1 & 356.8 $\pm$ 12.2 & 223.4 $\pm$ 16.4 & 3.04 $\pm$ 0.27 & 1.25 $\pm$ 0.10\\
30\%--35\% & 120.9 $\pm$ 7.0 & 288.3 $\pm$ 11.0 & 180.8 $\pm$ 14.3 & 2.99 $\pm$ 0.29 & 1.25 $\pm$ 0.11\\
35\%--40\% & 98.3  $\pm$ 6.8 & 229.7 $\pm$ 9.2  & 144.5 $\pm$ 12.6 & 2.94 $\pm$ 0.33 & 1.26 $\pm$ 0.12\\
40\%--45\% & 78.7  $\pm$ 6.1 & 181.0 $\pm$ 6.8  & 113.9 $\pm$ 10.9 & 2.89 $\pm$ 0.36 & 1.26 $\pm$ 0.13\\
45\%--50\% & 61.9  $\pm$ 5.2 & 141.1 $\pm$ 5.3  & 88.3  $\pm$ 9.3  & 2.85 $\pm$ 0.38 & 1.25 $\pm$ 0.14\\
50\%--55\% & 47.6  $\pm$ 4.9 & 107.6 $\pm$ 5.5  & 67.1  $\pm$ 8.1  & 2.82 $\pm$ 0.45 & 1.25 $\pm$ 0.16\\
55\%--60\% & 35.6  $\pm$ 5.1 & 77.5  $\pm$ 6.8  & 50.0  $\pm$ 6.7  & 2.81 $\pm$ 0.55 & 1.29 $\pm$ 0.21\\
\end{tabular}
\end{ruledtabular}

\caption{Charged particle multiplicity results for 200~GeV Au$+$Au 
collisions. The uncertainties include the total statistical and 
systematic uncertainties.}
\label{tab:nchAuAu200}
\begin{ruledtabular}
\begin{tabular}{cccccc}
Centrality & $\langle$\Npart$\rangle$ & $\langle$\Nqp$\rangle$ & \dNch  & \dNchNorm  & \dNchNormQ  \\\hline
0\%--5\%   & 350.9 $\pm$ 4.7 & 924.1 $\pm$ 16.2 & 687.4 $\pm$ 36.6 & 3.92 $\pm$ 0.22 & 1.49 $\pm$ 0.08\\
5\%--10\%  & 297.9 $\pm$ 6.6 & 782.6 $\pm$ 15.3 & 560.4 $\pm$ 27.9 & 3.77 $\pm$ 0.21 & 1.43 $\pm$ 0.08\\
10\%--15\% & 251.0 $\pm$ 7.3 & 644.6 $\pm$ 14.5 & 456.8 $\pm$ 22.3 & 3.64 $\pm$ 0.21 & 1.42 $\pm$ 0.08\\
15\%--20\% & 211.0 $\pm$ 7.3 & 532.9 $\pm$ 12.3 & 371.5 $\pm$ 18.2 & 3.52 $\pm$ 0.21 & 1.39 $\pm$ 0.08\\
20\%--25\% & 176.3 $\pm$ 7.0 & 437.5 $\pm$ 10.4 & 302.5 $\pm$ 15.8 & 3.43 $\pm$ 0.22 & 1.38 $\pm$ 0.08\\
25\%--30\% & 146.8 $\pm$ 7.1 & 356.8 $\pm$ 12.2 & 245.6 $\pm$ 13.8 & 3.35 $\pm$ 0.25 & 1.38 $\pm$ 0.09\\
30\%--35\% & 120.9 $\pm$ 7.0 & 288.3 $\pm$ 11.0 & 197.2 $\pm$ 12.2 & 3.26 $\pm$ 0.28 & 1.37 $\pm$ 0.10\\
35\%--40\% & 98.3  $\pm$ 6.8 & 229.7 $\pm$ 9.2  & 156.4 $\pm$ 10.9 & 3.18 $\pm$ 0.31 & 1.36 $\pm$ 0.11\\
40\%--45\% & 78.7  $\pm$ 6.1 & 181.0 $\pm$ 6.8  & 123.5 $\pm$ 9.6  & 3.14 $\pm$ 0.34 & 1.36 $\pm$ 0.12\\
45\%--50\% & 61.9  $\pm$ 5.2 & 141.1 $\pm$ 5.3  & 95.3  $\pm$ 8.6  & 3.08 $\pm$ 0.38 & 1.35 $\pm$ 0.13\\
50\%--55\% & 47.6  $\pm$ 4.9 & 107.6 $\pm$ 5.5  & 70.9  $\pm$ 7.6  & 2.98 $\pm$ 0.44 & 1.32 $\pm$ 0.16\\
55\%--60\% & 35.6  $\pm$ 5.1 & 77.5  $\pm$ 6.8  & 52.2  $\pm$ 6.5  & 2.93 $\pm$ 0.56 & 1.35 $\pm$ 0.20\\
\end{tabular}
\end{ruledtabular}

\caption{Transverse energy results for 130~GeV Au$+$Au collisions. The 
uncertainties include the total statistical and systematic uncertainties.}
\label{tab:etAuAu130}
\begin{ruledtabular}
\begin{tabular}{cccccc}
Centrality & $\langle$\Npart$\rangle$ & $\langle$\Nqp$\rangle$ & \dEt [GeV] & \dEtNorm [GeV] & \dEtNormQ [GeV] \\\hline
0\%--5\%   & 347.7 $\pm$ 10.0 & 914.1 $\pm$ 22.6 & 522.8 $\pm$ 27.3 & 3.01 $\pm$ 0.18 & 1.14 $\pm$ 0.07\\
5\%--10\%  & 294.0 $\pm$ 8.9  & 773.3 $\pm$ 20.3 & 425.2 $\pm$ 22.5 & 2.89 $\pm$ 0.18 & 1.10 $\pm$ 0.07\\
10\%--15\% & 249.5 $\pm$ 8.0  & 633.4 $\pm$ 19.4 & 349.0 $\pm$ 19.0 & 2.80 $\pm$ 0.18 & 1.10 $\pm$ 0.07\\
15\%--20\% & 211.0 $\pm$ 7.2  & 522.6 $\pm$ 18.3 & 287.2 $\pm$ 16.5 & 2.72 $\pm$ 0.18 & 1.10 $\pm$ 0.07\\
20\%--25\% & 178.6 $\pm$ 6.6  & 431.5 $\pm$ 19.0 & 237.1 $\pm$ 14.5 & 2.66 $\pm$ 0.19 & 1.10 $\pm$ 0.08\\
25\%--30\% & 149.7 $\pm$ 6.0  & 353.3 $\pm$ 15.9 & 191.3 $\pm$ 12.5 & 2.56 $\pm$ 0.20 & 1.08 $\pm$ 0.09\\
30\%--35\% & 124.8 $\pm$ 5.5  & 283.0 $\pm$ 13.2 & 153.9 $\pm$ 11.0 & 2.47 $\pm$ 0.21 & 1.09 $\pm$ 0.09\\
35\%--40\% & 102.9 $\pm$ 5.1  & 225.3 $\pm$ 11.0 & 121.8 $\pm$ 9.4  & 2.37 $\pm$ 0.22 & 1.08 $\pm$ 0.10\\
40\%--45\% & 83.2  $\pm$ 4.7  & 179.1 $\pm$ 8.8  & 96.0  $\pm$ 8.8  & 2.31 $\pm$ 0.25 & 1.07 $\pm$ 0.11\\
45\%--50\% & 66.3  $\pm$ 4.3  & 137.1 $\pm$ 7.1  & 73.3  $\pm$ 7.3  & 2.21 $\pm$ 0.26 & 1.07 $\pm$ 0.12\\
50\%--55\% & 52.1  $\pm$ 4.0  & 101.6 $\pm$ 6.5  & 55.5  $\pm$ 6.5  & 2.13 $\pm$ 0.30 & 1.09 $\pm$ 0.15\\
55\%--60\% & 40.1  $\pm$ 3.8  & 74.6  $\pm$ 7.3  & 41.0  $\pm$ 5.5  & 2.04 $\pm$ 0.34 & 1.10 $\pm$ 0.18\\
\end{tabular}
\end{ruledtabular}
\end{table*}

\begin{table*}[htb]
\caption{Charged particle multiplicity results for 130~GeV Au$+$Au 
collisions. The uncertainties include the total statistical and 
systematic uncertainties.}
\label{tab:nchAuAu130}
\begin{ruledtabular}
\begin{tabular}{cccccc}
Centrality & $\langle$\Npart$\rangle$ & $\langle$\Nqp$\rangle$ & \dNch  & \dNchNorm  & \dNchNormQ  \\\hline
0\%--5\%   & 347.7 $\pm$ 10.0 & 914.1 $\pm$ 22.6 & 601.8 $\pm$ 28.4 & 3.46 $\pm$ 0.19 & 1.32 $\pm$ 0.07\\
5\%--10\%  & 294.0 $\pm$ 8.9  & 773.3 $\pm$ 20.3 & 488.5 $\pm$ 21.6 & 3.32 $\pm$ 0.18 & 1.26 $\pm$ 0.07\\
10\%--15\% & 249.5 $\pm$ 8.0  & 633.4 $\pm$ 19.4 & 402.7 $\pm$ 17.4 & 3.23 $\pm$ 0.17 & 1.27 $\pm$ 0.07\\
15\%--20\% & 211.0 $\pm$ 7.2  & 522.6 $\pm$ 18.3 & 328.8 $\pm$ 15.2 & 3.12 $\pm$ 0.18 & 1.26 $\pm$ 0.07\\
20\%--25\% & 178.6 $\pm$ 6.6  & 431.5 $\pm$ 19.0 & 270.5 $\pm$ 12.8 & 3.03 $\pm$ 0.18 & 1.25 $\pm$ 0.08\\
25\%--30\% & 149.7 $\pm$ 6.0  & 353.3 $\pm$ 15.9 & 219.3 $\pm$ 11.4 & 2.93 $\pm$ 0.19 & 1.24 $\pm$ 0.09\\
30\%--35\% & 124.8 $\pm$ 5.5  & 283.0 $\pm$ 13.2 & 175.7 $\pm$ 10.3 & 2.82 $\pm$ 0.21 & 1.24 $\pm$ 0.09\\
35\%--40\% & 102.9 $\pm$ 5.1  & 225.3 $\pm$ 11.0 & 139.0 $\pm$ 9.1  & 2.70 $\pm$ 0.22 & 1.23 $\pm$ 0.10\\
40\%--45\% & 83.2  $\pm$ 4.7  & 179.1 $\pm$ 8.8  & 109.4 $\pm$ 8.4  & 2.63 $\pm$ 0.25 & 1.22 $\pm$ 0.11\\
45\%--50\% & 66.3  $\pm$ 4.3  & 137.1 $\pm$ 7.1  & 84.1  $\pm$ 7.0  & 2.54 $\pm$ 0.27 & 1.23 $\pm$ 0.12\\
50\%--55\% & 52.1  $\pm$ 4    & 101.6 $\pm$ 6.5  & 64.3  $\pm$ 6.3  & 2.47 $\pm$ 0.31 & 1.27 $\pm$ 0.15\\
55\%--60\% & 40.1  $\pm$ 3.8  & 74.6  $\pm$ 7.3  & 48.4  $\pm$ 5.4  & 2.41 $\pm$ 0.35 & 1.30 $\pm$ 0.19\\
\end{tabular}
\end{ruledtabular}

\caption{Transverse energy results for 62.4~GeV Au$+$Au collisions. The 
uncertainties include the total statistical and systematic uncertainties.}
\label{tab:etAuAu062}
\begin{ruledtabular}
\begin{tabular}{cccccc}
Centrality & $\langle$\Npart$\rangle$ & $\langle$\Nqp$\rangle$ & \dEt [GeV] & \dEtNorm [GeV] & \dEtNormQ [GeV] \\\hline
0\%--5\%   & 342.6 $\pm$ 4.9 & 891.7 $\pm$ 26.6 & 389.7 $\pm$ 23.5 & 2.27 $\pm$ 0.14 & 0.87 $\pm$ 0.06\\
5\%--10\%  & 291.3 $\pm$ 7.3 & 730.7 $\pm$ 24.1 & 320.5 $\pm$ 19.3 & 2.20 $\pm$ 0.14 & 0.88 $\pm$ 0.06\\
10\%--15\% & 244.5 $\pm$ 8.9 & 600.6 $\pm$ 21.5 & 260.6 $\pm$ 15.7 & 2.13 $\pm$ 0.15 & 0.87 $\pm$ 0.06\\
15\%--20\% & 205.0 $\pm$ 9.6 & 493.4 $\pm$ 19.6 & 212.1 $\pm$ 12.8 & 2.07 $\pm$ 0.16 & 0.86 $\pm$ 0.06\\
20\%--25\% & 171.3 $\pm$ 8.9 & 403.8 $\pm$ 18.5 & 171.9 $\pm$ 10.4 & 2.01 $\pm$ 0.16 & 0.85 $\pm$ 0.06\\
25\%--30\% & 142.2 $\pm$ 8.5 & 327.0 $\pm$ 16.7 & 138.6 $\pm$ 8.36 & 1.95 $\pm$ 0.17 & 0.85 $\pm$ 0.07\\
30\%--35\% & 116.7 $\pm$ 8.9 & 261.7 $\pm$ 15.7 & 110.4 $\pm$ 6.67 & 1.90 $\pm$ 0.18 & 0.85 $\pm$ 0.07\\
35\%--40\% & 95.2  $\pm$ 7.7 & 206.9 $\pm$ 14.3 & 86.9  $\pm$ 5.25 & 1.83 $\pm$ 0.19 & 0.84 $\pm$ 0.08\\
40\%--45\% & 76.1  $\pm$ 7.7 & 161.4 $\pm$ 13.3 & 67.3  $\pm$ 4.08 & 1.78 $\pm$ 0.21 & 0.84 $\pm$ 0.09\\
45\%--50\% & 59.9  $\pm$ 6.9 & 123.5 $\pm$ 13.2 & 51.2  $\pm$ 3.12 & 1.73 $\pm$ 0.22 & 0.84 $\pm$ 0.10\\
50\%--55\% & 46.8  $\pm$ 5.2 & 92.4  $\pm$ 11.2 & 38.4  $\pm$ 2.33 & 1.65 $\pm$ 0.21 & 0.84 $\pm$ 0.11\\
55\%--60\% & 35.8  $\pm$ 4.6 & 67.8  $\pm$ 9.0  & 28.5  $\pm$ 1.72 & 1.59 $\pm$ 0.23 & 0.84 $\pm$ 0.12\\
\end{tabular}
\end{ruledtabular}

\caption{Charged particle multiplicity results for 62.4~GeV Au$+$Au 
collisions. The uncertainties include the total statistical and 
systematic uncertainties.}
\label{tab:nchAuAu062}
\begin{ruledtabular}
\begin{tabular}{cccccc}
Centrality & $\langle$\Npart$\rangle$ & $\langle$\Nqp$\rangle$ & \dNch  & \dNchNorm  & \dNchNormQ  \\\hline
0\%--5\%   & 342.6 $\pm$ 4.9 & 891.7 $\pm$ 26.6 & 447.5 $\pm$ 38.9 & 2.61 $\pm$ 0.23 & 1.00 $\pm$ 0.09\\
5\%--10\%  & 291.3 $\pm$ 7.3 & 730.7 $\pm$ 24.1 & 367.4 $\pm$ 31.6 & 2.52 $\pm$ 0.23 & 1.01 $\pm$ 0.09\\
10\%--15\% & 244.5 $\pm$ 8.9 & 600.6 $\pm$ 21.5 & 301.8 $\pm$ 25.8 & 2.47 $\pm$ 0.23 & 1.01 $\pm$ 0.09\\
15\%--20\% & 205.0 $\pm$ 9.6 & 493.4 $\pm$ 19.6 & 248.0 $\pm$ 21.0 & 2.42 $\pm$ 0.23 & 1.01 $\pm$ 0.09\\
20\%--25\% & 171.3 $\pm$ 8.9 & 403.8 $\pm$ 18.5 & 203.0 $\pm$ 17.1 & 2.37 $\pm$ 0.24 & 1.01 $\pm$ 0.10\\
25\%--30\% & 142.2 $\pm$ 8.5 & 327.0 $\pm$ 16.7 & 165.1 $\pm$ 13.8 & 2.32 $\pm$ 0.24 & 1.01 $\pm$ 0.10\\
30\%--35\% & 116.7 $\pm$ 8.9 & 261.7 $\pm$ 15.7 & 133.0 $\pm$ 11.1 & 2.28 $\pm$ 0.26 & 1.02 $\pm$ 0.10\\
35\%--40\% & 95.2  $\pm$ 7.7 & 206.9 $\pm$ 14.3 & 105.9 $\pm$ 8.76 & 2.22 $\pm$ 0.26 & 1.02 $\pm$ 0.11\\
40\%--45\% & 76.1  $\pm$ 7.7 & 161.4 $\pm$ 13.3 & 83.0  $\pm$ 6.83 & 2.18 $\pm$ 0.28 & 1.03 $\pm$ 0.12\\
45\%--50\% & 59.9  $\pm$ 6.9 & 123.5 $\pm$ 13.2 & 63.9  $\pm$ 5.24 & 2.13 $\pm$ 0.30 & 1.03 $\pm$ 0.14\\
50\%--55\% & 46.8  $\pm$ 5.2 & 92.4  $\pm$ 11.2 & 48.4  $\pm$ 3.95 & 2.07 $\pm$ 0.29 & 1.05 $\pm$ 0.15\\
55\%--60\% & 35.8  $\pm$ 4.6 & 67.8  $\pm$ 9.0  & 35.8  $\pm$ 2.92 & 2.00 $\pm$ 0.30 & 1.06 $\pm$ 0.16\\
\end{tabular}
\end{ruledtabular}
\end{table*}

\begin{table*}[htb]
\caption{Transverse energy results for 39~GeV Au$+$Au collisions. The 
uncertainties include the total statistical and systematic uncertainties.}
\label{tab:etAuAu039}
\begin{ruledtabular}
\begin{tabular}{cccccc}
Centrality & $\langle$\Npart$\rangle$ & $\langle$\Nqp$\rangle$ & \dEt [GeV] & \dEtNorm [GeV] & \dEtNormQ [GeV] \\\hline
0\%--5\%   & 340.0 $\pm$ 7.4 & 874.6 $\pm$ 42.0 & 303.8 $\pm$ 18.2 & 1.79 $\pm$ 0.11 & 0.69 $\pm$ 0.05\\
5\%--10\%  & 289.6 $\pm$ 8.1 & 726.7 $\pm$ 36.7 & 262.1 $\pm$ 15.7 & 1.81 $\pm$ 0.12 & 0.72 $\pm$ 0.06\\
10\%--15\% & 244.1 $\pm$ 6.4 & 599.1 $\pm$ 26.8 & 216.6 $\pm$ 13.0 & 1.77 $\pm$ 0.12 & 0.72 $\pm$ 0.05\\
15\%--20\% & 206.5 $\pm$ 6.3 & 496.9 $\pm$ 23.7 & 178.5 $\pm$ 10.7 & 1.73 $\pm$ 0.12 & 0.72 $\pm$ 0.06\\
20\%--25\% & 174.1 $\pm$ 6.3 & 410.4 $\pm$ 20.9 & 146.9 $\pm$ 8.8  & 1.69 $\pm$ 0.12 & 0.72 $\pm$ 0.06\\
25\%--30\% & 145.8 $\pm$ 6.2 & 336.8 $\pm$ 22.2 & 120.4 $\pm$ 7.2  & 1.65 $\pm$ 0.12 & 0.72 $\pm$ 0.06\\
30\%--35\% & 120.8 $\pm$ 7.5 & 273.0 $\pm$ 18.1 & 97.7  $\pm$ 5.8  & 1.62 $\pm$ 0.14 & 0.72 $\pm$ 0.06\\
35\%--40\% & 98.6  $\pm$ 6.4 & 217.6 $\pm$ 15.1 & 78.5  $\pm$ 4.7  & 1.59 $\pm$ 0.14 & 0.72 $\pm$ 0.07\\
40\%--45\% & 79.8  $\pm$ 6.0 & 172.0 $\pm$ 14.1 & 62.3  $\pm$ 3.7  & 1.56 $\pm$ 0.15 & 0.72 $\pm$ 0.07\\
45\%--50\% & 63.9  $\pm$ 5.8 & 134.3 $\pm$ 13.1 & 48.6  $\pm$ 2.9  & 1.52 $\pm$ 0.17 & 0.72 $\pm$ 0.08\\
50\%--55\% & 50.3  $\pm$ 5.5 & 103.1 $\pm$ 13.5 & 37.3  $\pm$ 2.2  & 1.48 $\pm$ 0.18 & 0.72 $\pm$ 0.10\\
\end{tabular}
\end{ruledtabular}

\caption{Charged particle multiplicity results for 39~GeV Au$+$Au 
collisions. The uncertainties include the total statistical and systematic 
uncertainties.}
\label{tab:nchAuAu039}
\begin{ruledtabular}
\begin{tabular}{cccccc}
Centrality & $\langle$\Npart$\rangle$ & $\langle$\Nqp$\rangle$ & \dNch  & \dNchNorm  & \dNchNormQ  \\\hline
0\%--5\%   & 340.0 $\pm$ 7.4 & 874.6 $\pm$ 42.0 & 363.2 $\pm$ 31.6 & 2.14 $\pm$ 0.19 & 0.83 $\pm$ 0.08\\
5\%--10\%  & 289.6 $\pm$ 8.1 & 726.7 $\pm$ 36.7 & 297.8 $\pm$ 25.8 & 2.06 $\pm$ 0.19 & 0.82 $\pm$ 0.08\\
10\%--15\% & 244.1 $\pm$ 6.4 & 599.1 $\pm$ 26.8 & 246.6 $\pm$ 21.3 & 2.02 $\pm$ 0.18 & 0.82 $\pm$ 0.08\\
15\%--20\% & 206.5 $\pm$ 6.3 & 496.9 $\pm$ 23.7 & 204.4 $\pm$ 17.5 & 1.98 $\pm$ 0.18 & 0.82 $\pm$ 0.08\\
20\%--25\% & 174.1 $\pm$ 6.3 & 410.4 $\pm$ 20.9 & 168.9 $\pm$ 14.4 & 1.94 $\pm$ 0.18 & 0.82 $\pm$ 0.08\\
25\%--30\% & 145.8 $\pm$ 6.2 & 336.8 $\pm$ 22.2 & 138.3 $\pm$ 11.8 & 1.90 $\pm$ 0.18 & 0.82 $\pm$ 0.09\\
30\%--35\% & 120.8 $\pm$ 7.5 & 273.0 $\pm$ 18.1 & 112.6 $\pm$ 9.6  & 1.86 $\pm$ 0.20 & 0.83 $\pm$ 0.09\\
35\%--40\% & 98.6  $\pm$ 6.4 & 217.6 $\pm$ 15.1 & 90.6  $\pm$ 7.7  & 1.84 $\pm$ 0.20 & 0.83 $\pm$ 0.09\\
40\%--45\% & 79.8  $\pm$ 6.0 & 172.0 $\pm$ 14.1 & 72.1  $\pm$ 6.1  & 1.81 $\pm$ 0.20 & 0.84 $\pm$ 0.10\\
45\%--50\% & 63.9  $\pm$ 5.8 & 134.3 $\pm$ 13.1 & 56.8  $\pm$ 4.8  & 1.78 $\pm$ 0.22 & 0.85 $\pm$ 0.11\\
50\%--55\% & 50.3  $\pm$ 5.5 & 103.1 $\pm$ 13.5 & 43.7  $\pm$ 3.7  & 1.73 $\pm$ 0.24 & 0.85 $\pm$ 0.13\\
\end{tabular}
\end{ruledtabular}

\caption{Transverse energy results for 27~GeV Au$+$Au collisions. The 
uncertainties include the total statistical and systematic uncertainties.}
\label{tab:etAuAu027}
\begin{ruledtabular}
\begin{tabular}{cccccc}
Centrality & $\langle$\Npart$\rangle$ & $\langle$\Nqp$\rangle$ & \dEt [GeV] & \dEtNorm [GeV] & \dEtNormQ [GeV] \\\hline
0\%--5\%   & 338.9 $\pm$ 3.1 & 863.7 $\pm$ 23.5 & 265.6 $\pm$ 16.4 & 1.57 $\pm$ 0.10 & 0.62 $\pm$ 0.04\\
5\%--10\%  & 288.8 $\pm$ 4.7 & 718.8 $\pm$ 22.7 & 217.3 $\pm$ 13.4 & 1.50 $\pm$ 0.10 & 0.61 $\pm$ 0.04\\
10\%--15\% & 244.3 $\pm$ 6.5 & 595.0 $\pm$ 23.7 & 179.7 $\pm$ 11.1 & 1.47 $\pm$ 0.10 & 0.60 $\pm$ 0.04\\
15\%--20\% & 205.7 $\pm$ 5.8 & 490.7 $\pm$ 19.4 & 148.9 $\pm$ 9.2  & 1.45 $\pm$ 0.10 & 0.61 $\pm$ 0.04\\
20\%--25\% & 173.0 $\pm$ 5.5 & 404.6 $\pm$ 16.7 & 122.8 $\pm$ 7.6  & 1.42 $\pm$ 0.10 & 0.61 $\pm$ 0.05\\
25\%--30\% & 144.6 $\pm$ 6.2 & 330.8 $\pm$ 17.7 & 100.7 $\pm$ 6.2  & 1.39 $\pm$ 0.10 & 0.61 $\pm$ 0.05\\
30\%--35\% & 119.4 $\pm$ 6.1 & 267.4 $\pm$ 16.2 & 81.9  $\pm$ 5.1  & 1.37 $\pm$ 0.11 & 0.61 $\pm$ 0.05\\
35\%--40\% & 97.6  $\pm$ 5.8 & 213.6 $\pm$ 14.6 & 65.8  $\pm$ 4.1  & 1.35 $\pm$ 0.12 & 0.62 $\pm$ 0.06\\
40\%--45\% & 77.9  $\pm$ 5.7 & 166.0 $\pm$ 13.7 & 52.1  $\pm$ 3.2  & 1.34 $\pm$ 0.13 & 0.63 $\pm$ 0.06\\
45\%--50\% & 60.8  $\pm$ 6.0 & 125.9 $\pm$ 13.8 & 40.8  $\pm$ 2.5  & 1.34 $\pm$ 0.16 & 0.65 $\pm$ 0.08\\
\end{tabular}
\end{ruledtabular}

\caption{Charged particle multiplicity results for 27~GeV Au$+$Au 
collisions. The uncertainties include the total statistical and 
systematic uncertainties.}
\label{tab:nchAuAu027}
\begin{ruledtabular}
\begin{tabular}{cccccc}
Centrality & $\langle$\Npart$\rangle$ & $\langle$\Nqp$\rangle$ & \dNch  & \dNchNorm  & \dNchNormQ  \\\hline
0\%--5\%   & 338.9 $\pm$ 3.1 & 863.7 $\pm$ 23.5 & 321.2 $\pm$ 28.1 & 1.90 $\pm$ 0.17 & 0.74 $\pm$ 0.07\\
5\%--10\%  & 288.8 $\pm$ 4.7 & 718.8 $\pm$ 22.7 & 258.7 $\pm$ 22.5 & 1.79 $\pm$ 0.16 & 0.72 $\pm$ 0.07\\
10\%--15\% & 244.3 $\pm$ 6.5 & 595.0 $\pm$ 23.7 & 212.6 $\pm$ 18.5 & 1.74 $\pm$ 0.16 & 0.72 $\pm$ 0.07\\
15\%--20\% & 205.7 $\pm$ 5.8 & 490.7 $\pm$ 19.4 & 175.0 $\pm$ 15.1 & 1.70 $\pm$ 0.16 & 0.71 $\pm$ 0.07\\
20\%--25\% & 173.0 $\pm$ 5.5 & 404.6 $\pm$ 16.7 & 143.5 $\pm$ 12.4 & 1.66 $\pm$ 0.15 & 0.71 $\pm$ 0.07\\
25\%--30\% & 144.6 $\pm$ 6.2 & 330.8 $\pm$ 17.7 & 116.7 $\pm$ 10.0 & 1.61 $\pm$ 0.16 & 0.71 $\pm$ 0.07\\
30\%--35\% & 119.4 $\pm$ 6.1 & 267.4 $\pm$ 16.2 & 94.2  $\pm$ 8.1  & 1.58 $\pm$ 0.16 & 0.70 $\pm$ 0.07\\
35\%--40\% & 97.6  $\pm$ 5.8 & 213.6 $\pm$ 14.6 & 75.0  $\pm$ 6.4  & 1.54 $\pm$ 0.16 & 0.70 $\pm$ 0.08\\
40\%--45\% & 77.9  $\pm$ 5.7 & 166.0 $\pm$ 13.7 & 59.0  $\pm$ 5.0  & 1.51 $\pm$ 0.17 & 0.71 $\pm$ 0.08\\
45\%--50\% & 60.8  $\pm$ 6.0 & 125.9 $\pm$ 13.8 & 45.7  $\pm$ 3.9  & 1.50 $\pm$ 0.20 & 0.73 $\pm$ 0.10\\
\end{tabular}
\end{ruledtabular}
\end{table*}

\begin{table*}[htb]
\caption{Transverse energy results for 19.6~GeV Au$+$Au collisions. The 
uncertainties include the total statistical and systematic uncertainties.}
\label{tab:etAuAu019}
\begin{ruledtabular}
\begin{tabular}{cccccc}
Centrality & $\langle$\Npart$\rangle$ & $\langle$\Nqp$\rangle$ & \dEt [GeV] & \dEtNorm [GeV] & \dEtNormQ [GeV] \\\hline
0\%--5\%   & 338.5 $\pm$ 4.4 & 858.8 $\pm$ 27.7 & 233.1 $\pm$ 15.3 & 1.38 $\pm$ 0.09 & 0.54 $\pm$ 0.04\\
5\%--10\%  & 288.3 $\pm$ 6.0 & 714.1 $\pm$ 25.0 & 190.7 $\pm$ 12.5 & 1.32 $\pm$ 0.09 & 0.53 $\pm$ 0.04\\
10\%--15\% & 242.4 $\pm$ 6.1 & 587.3 $\pm$ 23.2 & 157.8 $\pm$ 10.3 & 1.30 $\pm$ 0.09 & 0.54 $\pm$ 0.04\\
15\%--20\% & 204.3 $\pm$ 5.7 & 485.0 $\pm$ 20.0 & 130.8 $\pm$ 8.6  & 1.28 $\pm$ 0.09 & 0.54 $\pm$ 0.04\\
20\%--25\% & 172.4 $\pm$ 7.3 & 401.4 $\pm$ 22.0 & 108.2 $\pm$ 7.1  & 1.25 $\pm$ 0.10 & 0.54 $\pm$ 0.05\\
25\%--30\% & 143.5 $\pm$ 6.6 & 326.9 $\pm$ 18.4 & 88.9  $\pm$ 5.8  & 1.24 $\pm$ 0.10 & 0.54 $\pm$ 0.05\\
30\%--35\% & 117.9 $\pm$ 6.7 & 262.5 $\pm$ 17.7 & 72.5  $\pm$ 4.8  & 1.23 $\pm$ 0.11 & 0.55 $\pm$ 0.05\\
35\%--40\% & 95.7  $\pm$ 6.9 & 208.0 $\pm$ 17.0 & 58.5  $\pm$ 3.8  & 1.22 $\pm$ 0.12 & 0.56 $\pm$ 0.06\\
40\%--45\% & 77.4  $\pm$ 5.7 & 164.2 $\pm$ 13.5 & 46.6  $\pm$ 3.1  & 1.21 $\pm$ 0.12 & 0.57 $\pm$ 0.06\\
45\%--50\% & 61.7  $\pm$ 4.8 & 127.7 $\pm$ 11.3 & 36.6  $\pm$ 2.4  & 1.19 $\pm$ 0.12 & 0.57 $\pm$ 0.06\\
\end{tabular}
\end{ruledtabular}

\caption{Charged particle multiplicity results for 19.6~GeV Au$+$Au collisions. The uncertainties include the total statistical and systematic uncertainties.}
\label{tab:nchAuAu019}
\begin{ruledtabular}
\begin{tabular}{cccccc}
Centrality & $\langle$\Npart$\rangle$ & $\langle$\Nqp$\rangle$ & \dNch  & \dNchNorm  & \dNchNormQ  \\\hline
0\%--5\%   & 338.5 $\pm$ 4.4 & 858.8 $\pm$ 27.7 & 285.3 $\pm$ 25.1 & 1.69 $\pm$ 0.15 & 0.66 $\pm$ 0.06\\
5\%--10\%  & 288.3 $\pm$ 6.0 & 714.1 $\pm$ 25.0 & 229.3 $\pm$ 20.1 & 1.59 $\pm$ 0.14 & 0.64 $\pm$ 0.06\\
10\%--15\% & 242.4 $\pm$ 6.1 & 587.3 $\pm$ 23.2 & 188.8 $\pm$ 16.5 & 1.56 $\pm$ 0.14 & 0.64 $\pm$ 0.06\\
15\%--20\% & 204.3 $\pm$ 5.7 & 485.0 $\pm$ 20.0 & 155.7 $\pm$ 13.5 & 1.52 $\pm$ 0.14 & 0.64 $\pm$ 0.06\\
20\%--25\% & 172.4 $\pm$ 7.3 & 401.4 $\pm$ 22.0 & 128.2 $\pm$ 11.1 & 1.49 $\pm$ 0.14 & 0.64 $\pm$ 0.07\\
25\%--30\% & 143.5 $\pm$ 6.6 & 326.9 $\pm$ 18.4 & 104.8 $\pm$ 9.1  & 1.46 $\pm$ 0.14 & 0.64 $\pm$ 0.07\\
30\%--35\% & 117.9 $\pm$ 6.7 & 262.5 $\pm$ 17.7 & 85.1  $\pm$ 7.4  & 1.44 $\pm$ 0.15 & 0.65 $\pm$ 0.07\\
35\%--40\% & 95.7  $\pm$ 6.9 & 208.0 $\pm$ 17.0 & 68.4  $\pm$ 5.9  & 1.43 $\pm$ 0.16 & 0.66 $\pm$ 0.08\\
40\%--45\% & 77.4  $\pm$ 5.7 & 164.2 $\pm$ 13.5 & 54.3  $\pm$ 4.7  & 1.40 $\pm$ 0.16 & 0.66 $\pm$ 0.08\\
45\%--50\% & 61.7  $\pm$ 4.8 & 127.7 $\pm$ 11.3 & 42.4  $\pm$ 3.7  & 1.37 $\pm$ 0.16 & 0.66 $\pm$ 0.08\\
\end{tabular}
\end{ruledtabular}

\caption{Transverse energy results for 14.5~GeV Au$+$Au collisions. The 
uncertainties include the total statistical and systematic uncertainties.}
\label{tab:etAuAu015}
\begin{ruledtabular}
\begin{tabular}{cccccc}
Centrality & $\langle$\Npart$\rangle$ & $\langle$\Nqp$\rangle$ & \dEt [GeV] & \dEtNorm [GeV] & \dEtNormQ [GeV] \\\hline
0\%--5\%   & 337.3 $\pm$ 4.2 & 852.0 $\pm$ 27.5 & 200.4 $\pm$ 14.0 & 1.19 $\pm$ 0.08 & 0.47 $\pm$ 0.04\\
5\%--10\%  & 287.7 $\pm$ 4.9 & 710.1 $\pm$ 23.4 & 164.0 $\pm$ 11.5 & 1.14 $\pm$ 0.08 & 0.46 $\pm$ 0.04\\
10\%--15\% & 242.5 $\pm$ 5.5 & 585.6 $\pm$ 22.0 & 134.9 $\pm$ 9.4  & 1.11 $\pm$ 0.08 & 0.46 $\pm$ 0.04\\
15\%--20\% & 205.1 $\pm$ 5.9 & 485.5 $\pm$ 19.7 & 111.0 $\pm$ 7.8  & 1.08 $\pm$ 0.08 & 0.46 $\pm$ 0.04\\
20\%--25\% & 172.6 $\pm$ 6.4 & 400.4 $\pm$ 19.6 & 91.1  $\pm$ 6.4  & 1.06 $\pm$ 0.08 & 0.46 $\pm$ 0.04\\
25\%--30\% & 143.6 $\pm$ 7.8 & 325.9 $\pm$ 21.7 & 74.4  $\pm$ 5.2  & 1.04 $\pm$ 0.09 & 0.46 $\pm$ 0.04\\
30\%--35\% & 119.2 $\pm$ 7.2 & 264.9 $\pm$ 19.2 & 60.2  $\pm$ 4.2  & 1.01 $\pm$ 0.09 & 0.45 $\pm$ 0.05\\
35\%--40\% & 98.3  $\pm$ 5.8 & 213.7 $\pm$ 14.8 & 48.2  $\pm$ 3.4  & 0.98 $\pm$ 0.09 & 0.45 $\pm$ 0.04\\
40\%--45\% & 80.2  $\pm$ 5.6 & 170.2 $\pm$ 13.6 & 38.2  $\pm$ 2.7  & 0.95 $\pm$ 0.09 & 0.45 $\pm$ 0.05\\
45\%--50\% & 63.9  $\pm$ 4.7 & 132.2 $\pm$ 11.0 & 29.7  $\pm$ 2.1  & 0.93 $\pm$ 0.09 & 0.45 $\pm$ 0.05\\
\end{tabular}
\end{ruledtabular}


\caption{Charged particle multiplicity results for 14.5~GeV Au$+$Au 
collisions. The uncertainties include the total statistical and systematic 
uncertainties.}
\label{tab:nchAuAu015}
\begin{ruledtabular}
\begin{tabular}{cccccc}
Centrality & $\langle$\Npart$\rangle$ & $\langle$\Nqp$\rangle$ & \dNch  & \dNchNorm  & \dNchNormQ  \\\hline
0\%--5\%   & 337.3 $\pm$ 4.2 & 852.0 $\pm$ 27.5 & 250.9 $\pm$ 22.2 & 1.49 $\pm$ 0.13 & 0.59 $\pm$ 0.06\\
5\%--10\%  & 287.7 $\pm$ 4.9 & 710.1 $\pm$ 23.4 & 201.2 $\pm$ 17.7 & 1.40 $\pm$ 0.13 & 0.57 $\pm$ 0.05\\
10\%--15\% & 242.5 $\pm$ 5.5 & 585.6 $\pm$ 22.0 & 164.5 $\pm$ 14.5 & 1.36 $\pm$ 0.12 & 0.56 $\pm$ 0.05\\
15\%--20\% & 205.1 $\pm$ 5.9 & 485.5 $\pm$ 19.7 & 134.7 $\pm$ 11.8 & 1.31 $\pm$ 0.12 & 0.56 $\pm$ 0.05\\
20\%--25\% & 172.6 $\pm$ 6.4 & 400.4 $\pm$ 19.6 & 110.0 $\pm$ 9.6  & 1.28 $\pm$ 0.12 & 0.55 $\pm$ 0.06\\
25\%--30\% & 143.6 $\pm$ 7.8 & 325.9 $\pm$ 21.7 & 89.4  $\pm$ 7.8  & 1.25 $\pm$ 0.13 & 0.55 $\pm$ 0.06\\
30\%--35\% & 119.2 $\pm$ 7.2 & 264.9 $\pm$ 19.2 & 72.0  $\pm$ 6.3  & 1.21 $\pm$ 0.13 & 0.54 $\pm$ 0.06\\
35\%--40\% & 98.3  $\pm$ 5.8 & 213.7 $\pm$ 14.8 & 57.4  $\pm$ 5.0  & 1.17 $\pm$ 0.12 & 0.55 $\pm$ 0.06\\
40\%--45\% & 80.2  $\pm$ 5.6 & 170.2 $\pm$ 13.6 & 45.2  $\pm$ 3.9  & 1.13 $\pm$ 0.13 & 0.53 $\pm$ 0.06\\
45\%--50\% & 63.9  $\pm$ 4.7 & 132.2 $\pm$ 11.0 & 34.9  $\pm$ 3.0  & 1.09 $\pm$ 0.12 & 0.53 $\pm$ 0.06\\
\end{tabular}
\end{ruledtabular}
\end{table*}

\begin{table*}[htb]
\caption{Transverse energy results for 7.7~GeV Au$+$Au collisions. The 
uncertainties include the total statistical and systematic uncertainties.}
\label{tab:etAuAu007}
\begin{ruledtabular}
\begin{tabular}{cccccc}
Centrality & $\langle$\Npart$\rangle$ & $\langle$\Nqp$\rangle$ & \dEt [GeV] & \dEtNorm [GeV] & \dEtNormQ [GeV] \\\hline
0\%--5\%   & 332.1 $\pm$ 5.4 & 830.4 $\pm$ 33.9 & 137.7 $\pm$ 9.1 & 0.83 $\pm$ 0.06 & 0.33 $\pm$ 0.03\\
5\%--10\%  & 283.2 $\pm$ 5.9 & 692.3 $\pm$ 27.0 & 114.3 $\pm$ 7.5 & 0.81 $\pm$ 0.06 & 0.33 $\pm$ 0.03\\
10\%--15\% & 240.1 $\pm$ 5.7 & 574.4 $\pm$ 24.0 & 93.3  $\pm$ 6.2 & 0.78 $\pm$ 0.05 & 0.33 $\pm$ 0.03\\
15\%--20\% & 204.1 $\pm$ 5.7 & 479.0 $\pm$ 20.6 & 76.2  $\pm$ 5.0 & 0.75 $\pm$ 0.05 & 0.32 $\pm$ 0.03\\
20\%--25\% & 172.9 $\pm$ 6.7 & 398.0 $\pm$ 19.8 & 62.0  $\pm$ 4.1 & 0.72 $\pm$ 0.05 & 0.31 $\pm$ 0.03\\
25\%--30\% & 145.5 $\pm$ 7.2 & 328.1 $\pm$ 19.8 & 50.0  $\pm$ 3.3 & 0.69 $\pm$ 0.06 & 0.30 $\pm$ 0.03\\
30\%--35\% & 121.0 $\pm$ 7.3 & 267.1 $\pm$ 19.0 & 40.1  $\pm$ 2.6 & 0.66 $\pm$ 0.06 & 0.30 $\pm$ 0.03\\
35\%--40\% & 98.2  $\pm$ 7.0 & 211.6 $\pm$ 17.8 & 31.8  $\pm$ 2.1 & 0.66 $\pm$ 0.06 & 0.30 $\pm$ 0.03\\
40\%--45\% & 78.8  $\pm$ 6.7 & 165.6 $\pm$ 16.3 & 24.8  $\pm$ 1.6 & 0.63 $\pm$ 0.07 & 0.30 $\pm$ 0.04\\
45\%--50\% & 61.8  $\pm$ 6.5 & 126.4 $\pm$ 14.7 & 19.2  $\pm$ 1.3 & 0.62 $\pm$ 0.08 & 0.30 $\pm$ 0.04\\
\end{tabular}
\end{ruledtabular}

\caption{Charged particle multiplicity results for 7.7~GeV Au$+$Au 
collisions. The uncertainties include the total statistical and systematic 
uncertainties.}
\label{tab:nchAuAu007}
\begin{ruledtabular}
\begin{tabular}{cccccc}
Centrality & $\langle$\Npart$\rangle$ & $\langle$\Nqp$\rangle$ & \dNch  & \dNchNorm  & \dNchNormQ  \\\hline
0\%--5\%   & 332.1 $\pm$ 5.4 & 830.4 $\pm$ 33.9 & 192.4 $\pm$ 16.9 & 1.16 $\pm$ 0.10 & 0.46 $\pm$ 0.04\\
5\%--10\%  & 283.2 $\pm$ 5.9 & 692.3 $\pm$ 27.0 & 159.2 $\pm$ 14.0 & 1.12 $\pm$ 0.10 & 0.46 $\pm$ 0.04\\
10\%--15\% & 240.1 $\pm$ 5.7 & 574.4 $\pm$ 24.0 & 129.3 $\pm$ 11.3 & 1.08 $\pm$ 0.10 & 0.45 $\pm$ 0.04\\
15\%--20\% & 204.1 $\pm$ 5.7 & 479.0 $\pm$ 20.6 & 105.4 $\pm$ 9.2  & 1.03 $\pm$ 0.09 & 0.44 $\pm$ 0.04\\
20\%--25\% & 172.9 $\pm$ 6.7 & 398.0 $\pm$ 19.8 & 85.6  $\pm$ 7.5  & 0.99 $\pm$ 0.09 & 0.43 $\pm$ 0.04\\
25\%--30\% & 145.5 $\pm$ 7.2 & 328.1 $\pm$ 19.8 & 68.8  $\pm$ 6.0  & 0.95 $\pm$ 0.09 & 0.42 $\pm$ 0.04\\
30\%--35\% & 121.0 $\pm$ 7.3 & 267.1 $\pm$ 19.0 & 55.0  $\pm$ 4.8  & 0.91 $\pm$ 0.10 & 0.41 $\pm$ 0.05\\
35\%--40\% & 98.2  $\pm$ 7.0 & 211.6 $\pm$ 17.8 & 43.5  $\pm$ 3.8  & 0.89 $\pm$ 0.10 & 0.41 $\pm$ 0.05\\
40\%--45\% & 78.8  $\pm$ 6.7 & 165.6 $\pm$ 16.3 & 33.9  $\pm$ 3.0  & 0.86 $\pm$ 0.11 & 0.41 $\pm$ 0.05\\
45\%--50\% & 61.8  $\pm$ 6.5 & 126.4 $\pm$ 14.7 & 26.1  $\pm$ 2.3  & 0.85 $\pm$ 0.11 & 0.41 $\pm$ 0.06\\
\end{tabular}
\end{ruledtabular}


\caption{Transverse energy results for 200~GeV Cu$+$Cu collisions. The 
uncertainties include the total statistical and systematic uncertainties.}
\label{tab:etCuCu200}
\begin{ruledtabular}
\begin{tabular}{cccccc}
Centrality & $\langle$\Npart$\rangle$ & $\langle$\Nqp$\rangle$ & \dEt [GeV] & \dEtNorm [GeV] & \dEtNormQ [GeV] \\\hline
0\%--5\%   & 105.6 $\pm$ 2.5 & 254.3 $\pm$ 11.8 & 166.8 $\pm$ 13.2 & 3.16 $\pm$ 0.26 & 1.31 $\pm$ 0.12\\
5\%--10\%  & 93.1  $\pm$ 3.0 & 219.0 $\pm$ 11.4 & 139.9 $\pm$ 11.1 & 3.01 $\pm$ 0.26 & 1.28 $\pm$ 0.12\\
10\%--15\% & 80.1  $\pm$ 2.4 & 183.6 $\pm$ 8.6  & 117.1 $\pm$ 9.3  & 2.92 $\pm$ 0.25 & 1.28 $\pm$ 0.12\\
15\%--20\% & 68.4  $\pm$ 2.5 & 153.0 $\pm$ 7.7  & 97.9  $\pm$ 7.8  & 2.86 $\pm$ 0.25 & 1.28 $\pm$ 0.12\\
20\%--25\% & 58.4  $\pm$ 2.3 & 127.7 $\pm$ 7.0  & 81.6  $\pm$ 6.5  & 2.80 $\pm$ 0.25 & 1.28 $\pm$ 0.12\\
25\%--30\% & 49.2  $\pm$ 2.1 & 104.9 $\pm$ 5.7  & 67.8  $\pm$ 5.4  & 2.76 $\pm$ 0.25 & 1.29 $\pm$ 0.12\\
30\%--35\% & 41.3  $\pm$ 2.2 & 86.0  $\pm$ 5.8  & 56.1  $\pm$ 4.4  & 2.72 $\pm$ 0.26 & 1.30 $\pm$ 0.13\\
35\%--40\% & 34.3  $\pm$ 2.0 & 69.8  $\pm$ 5.0  & 46.0  $\pm$ 3.6  & 2.68 $\pm$ 0.26 & 1.32 $\pm$ 0.14\\
40\%--45\% & 28.1  $\pm$ 1.8 & 55.9  $\pm$ 4.3  & 37.5  $\pm$ 3.0  & 2.67 $\pm$ 0.27 & 1.34 $\pm$ 0.15\\
\end{tabular}
\end{ruledtabular}

\caption{Charged particle multiplicity results for 200~GeV Cu$+$Cu 
collisions. The uncertainties include the total statistical and systematic 
uncertainties.}
\label{tab:nchCuCu200}
\begin{ruledtabular}
\begin{tabular}{cccccc}
Centrality & $\langle$\Npart$\rangle$ & $\langle$\Nqp$\rangle$ & \dNch  & \dNchNorm  & \dNchNormQ  \\\hline
0\%--5\%   & 105.6 $\pm$ 2.5 & 254.3 $\pm$ 11.8 & 192.6 $\pm$ 13.9 & 3.65 $\pm$ 0.28 & 1.51 $\pm$ 0.13\\
5\%--10\%  & 93.1  $\pm$ 3.0 & 219.0 $\pm$ 11.4 & 160.1 $\pm$ 11.5 & 3.44 $\pm$ 0.27 & 1.46 $\pm$ 0.13\\
10\%--15\% & 80.1  $\pm$ 2.4 & 183.6 $\pm$ 8.6  & 132.8 $\pm$ 9.5  & 3.32 $\pm$ 0.26 & 1.45 $\pm$ 0.12\\
15\%--20\% & 68.4  $\pm$ 2.5 & 153.0 $\pm$ 7.7  & 110.2 $\pm$ 7.9  & 3.22 $\pm$ 0.26 & 1.44 $\pm$ 0.12\\
20\%--25\% & 58.4  $\pm$ 2.3 & 127.7 $\pm$ 7.0  & 91.3  $\pm$ 6.5  & 3.13 $\pm$ 0.25 & 1.43 $\pm$ 0.13\\
25\%--30\% & 49.2  $\pm$ 2.1 & 104.9 $\pm$ 5.7  & 75.2  $\pm$ 5.3  & 3.06 $\pm$ 0.25 & 1.43 $\pm$ 0.13\\
30\%--35\% & 41.3  $\pm$ 2.2 & 86.0  $\pm$ 5.8  & 61.7  $\pm$ 4.4  & 2.99 $\pm$ 0.27 & 1.43 $\pm$ 0.14\\
35\%--40\% & 34.3  $\pm$ 2.0 & 69.8  $\pm$ 5.0  & 50.2  $\pm$ 3.5  & 2.93 $\pm$ 0.27 & 1.44 $\pm$ 0.14\\
40\%--45\% & 28.1  $\pm$ 1.8 & 55.9  $\pm$ 4.3  & 40.6  $\pm$ 2.9  & 2.89 $\pm$ 0.28 & 1.45 $\pm$ 0.15\\
\end{tabular}
\end{ruledtabular}
\end{table*}

\begin{table*}[htb]
\caption{Transverse energy results for 62.4~GeV Cu$+$Cu collisions. The 
uncertainties include the total statistical and systematic uncertainties.}
\label{tab:etCuCu062}
\begin{ruledtabular}
\begin{tabular}{cccccc}
Centrality & $\langle$\Npart$\rangle$ & $\langle$\Nqp$\rangle$ & \dEt [GeV] & \dEtNorm [GeV] & \dEtNormQ [GeV] \\\hline
0\%--5\%   & 100.5 $\pm$ 4.5 & 229.3 $\pm$ 8.5  & 107.6 $\pm$ 6.5 & 2.14 $\pm$ 0.16 & 0.94 $\pm$ 0.07\\
5\%--10\%  & 88.3  $\pm$ 4.8 & 197.8 $\pm$ 15.0 & 93.6  $\pm$ 5.6 & 2.12 $\pm$ 0.17 & 0.95 $\pm$ 0.09\\
10\%--15\% & 78.2  $\pm$ 4.3 & 171.7 $\pm$ 25.2 & 79.3  $\pm$ 4.8 & 2.03 $\pm$ 0.17 & 0.92 $\pm$ 0.15\\
15\%--20\% & 67.4  $\pm$ 4.3 & 144.8 $\pm$ 23.8 & 66.5  $\pm$ 4.0 & 1.97 $\pm$ 0.17 & 0.92 $\pm$ 0.16\\
20\%--25\% & 56.6  $\pm$ 4.4 & 118.7 $\pm$ 11.5 & 55.6  $\pm$ 3.3 & 1.96 $\pm$ 0.19 & 0.94 $\pm$ 0.11\\
25\%--30\% & 48.7  $\pm$ 4.9 & 100.0 $\pm$ 12.0 & 46.4  $\pm$ 2.8 & 1.91 $\pm$ 0.22 & 0.93 $\pm$ 0.12\\
30\%--35\% & 40.4  $\pm$ 4.5 & 81.1  $\pm$ 10.4 & 38.6  $\pm$ 2.3 & 1.91 $\pm$ 0.24 & 0.95 $\pm$ 0.13\\
35\%--40\% & 32.3  $\pm$ 4.1 & 63.3  $\pm$ 6.1  & 32.0  $\pm$ 1.9 & 1.98 $\pm$ 0.28 & 1.01 $\pm$ 0.11\\
\end{tabular}
\end{ruledtabular}
\end{table*}

\begin{table*}[htb]
\caption{Charged particle multiplicity results for 62.4~GeV Cu$+$Cu 
collisions. The uncertainties include the total statistical and systematic 
uncertainties.}
\label{tab:nchCuCu062}
\begin{ruledtabular}
\begin{tabular}{cccccc}
Centrality & $\langle$\Npart$\rangle$ & $\langle$\Nqp$\rangle$ & \dNch  & \dNchNorm  & \dNchNormQ  \\\hline
0\%--5\%   & 100.5 $\pm$ 4.5 & 229.3 $\pm$ 8.5  & 135.3 $\pm$ 11.1 & 2.69 $\pm$ 0.25 & 1.18 $\pm$ 0.11\\
5\%--10\%  & 88.3  $\pm$ 4.8 & 197.8 $\pm$ 15.0 & 116.6 $\pm$ 9.5  & 2.64 $\pm$ 0.26 & 1.18 $\pm$ 0.13\\
10\%--15\% & 78.2  $\pm$ 4.3 & 171.7 $\pm$ 25.2 & 97.8  $\pm$ 8.0  & 2.50 $\pm$ 0.25 & 1.14 $\pm$ 0.19\\
15\%--20\% & 67.4  $\pm$ 4.3 & 144.8 $\pm$ 23.8 & 81.0  $\pm$ 6.6  & 2.40 $\pm$ 0.25 & 1.12 $\pm$ 0.20\\
20\%--25\% & 56.6  $\pm$ 4.4 & 118.7 $\pm$ 11.5 & 67.0  $\pm$ 5.5  & 2.37 $\pm$ 0.27 & 1.13 $\pm$ 0.14\\
25\%--30\% & 48.7  $\pm$ 4.9 & 100.0 $\pm$ 12.0 & 55.3  $\pm$ 4.5  & 2.27 $\pm$ 0.30 & 1.11 $\pm$ 0.16\\
30\%--35\% & 40.4  $\pm$ 4.5 & 81.1  $\pm$ 10.4 & 45.3  $\pm$ 3.7  & 2.24 $\pm$ 0.31 & 1.12 $\pm$ 0.17\\
35\%--40\% & 32.3  $\pm$ 4.1 & 63.3  $\pm$ 6.11 & 36.9  $\pm$ 3.0  & 2.28 $\pm$ 0.35 & 1.17 $\pm$ 0.15\\
\end{tabular}
\end{ruledtabular}
\end{table*}

\begin{table*}[htb]
\caption{Transverse energy results for 200~GeV Cu$+$Au collisions. The 
uncertainties include the total statistical and systematic uncertainties.}
\label{tab:etCuAu200}
\begin{ruledtabular}
\begin{tabular}{cccccc}
Centrality & $\langle$\Npart$\rangle$ & $\langle$\Nqp$\rangle$ & \dEt [GeV] & \dEtNorm [GeV] & \dEtNormQ [GeV] \\\hline
0\%--5\%   & 189.0 $\pm$ 5.2 & 463.8 $\pm$ 17.6 & 288.3 $\pm$ 17.3 & 3.05 $\pm$ 0.20 & 1.24 $\pm$ 0.09\\
5\%--10\%  & 164.2 $\pm$ 4.3 & 400.3 $\pm$ 14.8 & 249.8 $\pm$ 15.0 & 3.04 $\pm$ 0.20 & 1.25 $\pm$ 0.09\\
10\%--15\% & 142.4 $\pm$ 3.7 & 341.7 $\pm$ 12.7 & 212.8 $\pm$ 12.8 & 2.99 $\pm$ 0.20 & 1.25 $\pm$ 0.09\\
15\%--20\% & 122.6 $\pm$ 3.3 & 288.9 $\pm$ 10.7 & 179.4 $\pm$ 10.8 & 2.93 $\pm$ 0.19 & 1.24 $\pm$ 0.09\\
20\%--25\% & 104.5 $\pm$ 3.5 & 240.5 $\pm$ 11.0 & 150.0 $\pm$ 9.0  & 2.87 $\pm$ 0.20 & 1.25 $\pm$ 0.09\\
25\%--30\% & 88.5  $\pm$ 4.0 & 199.0 $\pm$ 11.8 & 124.5 $\pm$ 7.5  & 2.81 $\pm$ 0.21 & 1.25 $\pm$ 0.11\\
30\%--35\% & 73.8  $\pm$ 3.6 & 162.6 $\pm$ 9.8  & 102.3 $\pm$ 6.1  & 2.77 $\pm$ 0.21 & 1.26 $\pm$ 0.11\\
35\%--40\% & 60.9  $\pm$ 3.6 & 131.0 $\pm$ 8.8  & 83.3  $\pm$ 5.0  & 2.74 $\pm$ 0.23 & 1.27 $\pm$ 0.11\\
40\%--45\% & 49.7  $\pm$ 3.2 & 103.4 $\pm$ 8.8  & 67.0  $\pm$ 4.0  & 2.69 $\pm$ 0.24 & 1.29 $\pm$ 0.13\\
45\%--50\% & 39.9  $\pm$ 3.1 & 80.6  $\pm$ 8.5  & 53.1  $\pm$ 3.2  & 2.66 $\pm$ 0.26 & 1.32 $\pm$ 0.16\\
50\%--55\% & 31.4  $\pm$ 3.2 & 62.3  $\pm$ 8.1  & 41.4  $\pm$ 2.5  & 2.64 $\pm$ 0.31 & 1.33 $\pm$ 0.19\\
55\%--60\% & 24.3  $\pm$ 2.8 & 47.1  $\pm$ 6.9  & 31.9  $\pm$ 1.9  & 2.63 $\pm$ 0.34 & 1.36 $\pm$ 0.21\\
\end{tabular}
\end{ruledtabular}
\end{table*}

\begin{table*}[htb]
\caption{Charged particle multiplicity results for 200~GeV Cu$+$Au 
collisions. The uncertainties include the total statistical and systematic 
uncertainties.}
\label{tab:nchCuAu200}
\begin{ruledtabular}
\begin{tabular}{cccccc}
Centrality & $\langle$\Npart$\rangle$ & $\langle$\Nqp$\rangle$ & \dNch  & \dNchNorm  & \dNchNormQ  \\\hline
0\%--5\%   & 189.0 $\pm$ 5.2 & 463.8 $\pm$ 17.6 & 333.5 $\pm$ 25.0 & 3.53 $\pm$ 0.28 & 1.44 $\pm$ 0.12\\
5\%--10\%  & 164.2 $\pm$ 4.3 & 400.3 $\pm$ 14.8 & 288.0 $\pm$ 21.4 & 3.51 $\pm$ 0.28 & 1.44 $\pm$ 0.12\\
10\%--15\% & 142.4 $\pm$ 3.7 & 341.7 $\pm$ 12.7 & 244.5 $\pm$ 18.1 & 3.43 $\pm$ 0.27 & 1.43 $\pm$ 0.11\\
15\%--20\% & 122.6 $\pm$ 3.3 & 288.9 $\pm$ 10.7 & 205.4 $\pm$ 15.1 & 3.35 $\pm$ 0.26 & 1.42 $\pm$ 0.11\\
20\%--25\% & 104.5 $\pm$ 3.5 & 240.5 $\pm$ 11.0 & 171.2 $\pm$ 12.5 & 3.28 $\pm$ 0.26 & 1.42 $\pm$ 0.12\\
25\%--30\% & 88.5  $\pm$ 4.0 & 199.0 $\pm$ 11.8 & 141.5 $\pm$ 10.2 & 3.20 $\pm$ 0.27 & 1.42 $\pm$ 0.13\\
30\%--35\% & 73.8  $\pm$ 3.6 & 162.6 $\pm$ 9.8  & 115.9 $\pm$ 8.3  & 3.14 $\pm$ 0.27 & 1.43 $\pm$ 0.13\\
35\%--40\% & 60.9  $\pm$ 3.6 & 131.0 $\pm$ 8.8  & 94.0  $\pm$ 6.7  & 3.09 $\pm$ 0.29 & 1.43 $\pm$ 0.14\\
40\%--45\% & 49.7  $\pm$ 3.2 & 103.4 $\pm$ 8.8  & 75.2  $\pm$ 5.4  & 3.03 $\pm$ 0.29 & 1.45 $\pm$ 0.16\\
45\%--50\% & 39.9  $\pm$ 3.1 & 80.6  $\pm$ 8.5  & 59.5  $\pm$ 4.2  & 2.98 $\pm$ 0.31 & 1.48 $\pm$ 0.19\\
50\%--55\% & 31.4  $\pm$ 3.2 & 62.3  $\pm$ 8.1  & 46.3  $\pm$ 3.3  & 2.95 $\pm$ 0.37 & 1.48 $\pm$ 0.22\\
55\%--60\% & 24.3  $\pm$ 2.8 & 47.1  $\pm$ 6.9  & 35.4  $\pm$ 2.5  & 2.91 $\pm$ 0.39 & 1.50 $\pm$ 0.24\\
\end{tabular}
\end{ruledtabular}
\end{table*}

\begin{table*}[htb]
\caption{Transverse energy results for 193~GeV U$+$U collisions. The 
uncertainties include the total statistical and systematic errors.}
\label{tab:etUU193}
\begin{ruledtabular}
\begin{tabular}{cccc}
Centrality & $\langle$\Npart$\rangle$ & \dEt [GeV] & \dEtNorm [GeV] \\\hline
0\%--5\%   & 418.8 $\pm$ 5.0 & 783.0 $\pm$ 46.1 & 3.74 $\pm$ 0.22 \\
5\%--10\%  & 353.2 $\pm$ 6.0 & 625.6 $\pm$ 36.9 & 3.54 $\pm$ 0.22 \\
10\%--15\% & 296.7 $\pm$ 6.1 & 504.0 $\pm$ 29.7 & 3.40 $\pm$ 0.21 \\
15\%--20\% & 248.9 $\pm$ 6.8 & 406.2 $\pm$ 23.9 & 3.26 $\pm$ 0.21 \\
20\%--25\% & 207.6 $\pm$ 6.7 & 325.9 $\pm$ 19.2 & 3.14 $\pm$ 0.21 \\
25\%--30\% & 172.5 $\pm$ 6.5 & 259.2 $\pm$ 15.3 & 3.00 $\pm$ 0.21 \\
30\%--35\% & 141.6 $\pm$ 6.8 & 203.7 $\pm$ 12.0 & 2.88 $\pm$ 0.22 \\
35\%--40\% & 114.9 $\pm$ 6.9 & 157.8 $\pm$ 9.3  & 2.75 $\pm$ 0.23 \\
40\%--45\% & 91.8  $\pm$ 6.4 & 119.9 $\pm$ 7.1  & 2.61 $\pm$ 0.24 \\
45\%--50\% & 72.0  $\pm$ 6.2 & 89.16 $\pm$ 5.3  & 2.48 $\pm$ 0.26 \\
\end{tabular}
\end{ruledtabular}
\end{table*}

\begin{table*}[htb]
\caption{Charged particle multiplicity results for 193~GeV U$+$U 
collisions. 
The uncertainties include the total statistical and systematic errors.}
\label{tab:nchUU193}
\begin{ruledtabular}
\begin{tabular}{cccc}
Centrality & $\langle$\Npart$\rangle$ & \dNch & \dNchNorm  \\\hline
0\%--5\%   & 418.8 $\pm$ 5.0 & 830.4 $\pm$ 67.8 & 3.97 $\pm$ 0.33 \\
5\%--10\%  & 353.2 $\pm$ 6.0 & 689.2 $\pm$ 55.5 & 3.90 $\pm$ 0.32 \\
10\%--15\% & 296.7 $\pm$ 6.1 & 565.5 $\pm$ 44.9 & 3.81 $\pm$ 0.31 \\
15\%--20\% & 248.9 $\pm$ 6.8 & 459.6 $\pm$ 36.1 & 3.69 $\pm$ 0.31 \\
20\%--25\% & 207.6 $\pm$ 6.7 & 369.7 $\pm$ 28.7 & 3.56 $\pm$ 0.30 \\
25\%--30\% & 172.5 $\pm$ 6.5 & 293.9 $\pm$ 22.6 & 3.41 $\pm$ 0.29 \\
30\%--35\% & 141.6 $\pm$ 6.8 & 230.6 $\pm$ 17.5 & 3.26 $\pm$ 0.29 \\
35\%--40\% & 114.9 $\pm$ 6.9 & 178.1 $\pm$ 13.4 & 3.10 $\pm$ 0.30 \\
40\%--45\% & 91.8  $\pm$ 6.4 & 135.0 $\pm$ 10.1 & 2.94 $\pm$ 0.30 \\
45\%--50\% & 72.0  $\pm$ 6.2 & 100.0 $\pm$ 7.4  & 2.78 $\pm$ 0.32 \\
\end{tabular}
\end{ruledtabular}
\end{table*}


\begin{table*}[htb]
\caption{Transverse energy results for 200~GeV $d$$+$Au collisions. The 
uncertainties include the total statistical and systematic uncertainties.}
\label{tab:etdAu200}
\begin{ruledtabular}
\begin{tabular}{cccccc}
Centrality & $\langle$\Npart$\rangle$ & $\langle$\Nqp$\rangle$ & \dEt [GeV] & \dEtNorm [GeV] & \dEtNormQ [GeV] \\\hline
0\%--5\%   & 17.8 $\pm$ 1.2 & 27.2 $\pm$ 2.3 & 20.3 $\pm$ 1.7 & 2.29 $\pm$ 0.24 & 1.39 $\pm$ 0.16\\
5\%--10\%  & 15.6 $\pm$ 1.0 & 24.7 $\pm$ 2.0 & 17.4 $\pm$ 1.5 & 2.24 $\pm$ 0.23 & 1.33 $\pm$ 0.15\\
10\%--20\% & 14.1 $\pm$ 0.9 & 22.9 $\pm$ 1.8 & 15.4 $\pm$ 1.3 & 2.18 $\pm$ 0.23 & 1.27 $\pm$ 0.14\\
20\%--30\% & 11.9 $\pm$ 0.7 & 20.0 $\pm$ 1.5 & 13.2 $\pm$ 1.1 & 2.21 $\pm$ 0.22 & 1.27 $\pm$ 0.14\\
30\%--40\% & 10.5 $\pm$ 0.6 & 18.0 $\pm$ 1.2 & 11.3 $\pm$ 0.9 & 2.16 $\pm$ 0.22 & 1.22 $\pm$ 0.13\\
40\%--50\% & 8.7  $\pm$ 0.5 & 15.3 $\pm$ 0.9 & 9.5  $\pm$ 0.8 & 2.20 $\pm$ 0.22 & 1.23 $\pm$ 0.12\\
50\%--60\% & 7.1  $\pm$ 0.4 & 12.7 $\pm$ 0.6 & 7.8  $\pm$ 0.7 & 2.19 $\pm$ 0.23 & 1.23 $\pm$ 0.12\\
60\%--70\% & 5.7  $\pm$ 0.4 & 10.4 $\pm$ 0.5 & 6.3  $\pm$ 0.5 & 2.21 $\pm$ 0.23 & 1.23 $\pm$ 0.12\\
\end{tabular}
\end{ruledtabular}
\end{table*}

\begin{table*}[htb]
\caption{Charged particle multiplicity results for 200~GeV $d$$+$Au 
collisions. The uncertainties include the total statistical and systematic 
uncertainties.}
\label{tab:nchdAu200}
\begin{ruledtabular}
\begin{tabular}{cccccc}
Centrality & $\langle$\Npart$\rangle$ & $\langle$\Nqp$\rangle$ & \dNch  & \dNchNorm  & \dNchNormQ  \\\hline
0\%--5\%   & 17.8 $\pm$ 1.2 & 27.2 $\pm$ 2.3 & 20.8 $\pm$ 1.5 & 2.43 $\pm$ 0.23 & 1.53 $\pm$ 0.17\\
5\%--10\%  & 15.6 $\pm$ 1.0 & 24.7 $\pm$ 2.0 & 17.7 $\pm$ 1.2 & 2.36 $\pm$ 0.22 & 1.43 $\pm$ 0.15\\
10\%--20\% & 14.1 $\pm$ 0.9 & 22.9 $\pm$ 1.8 & 15.5 $\pm$ 1.1 & 2.28 $\pm$ 0.21 & 1.35 $\pm$ 0.14\\
20\%--30\% & 11.9 $\pm$ 0.7 & 20.0 $\pm$ 1.5 & 13.2 $\pm$ 0.9 & 2.30 $\pm$ 0.21 & 1.32 $\pm$ 0.13\\
30\%--40\% & 10.5 $\pm$ 0.6 & 18.0 $\pm$ 1.2 & 11.2 $\pm$ 0.8 & 2.22 $\pm$ 0.20 & 1.25 $\pm$ 0.12\\
40\%--50\% & 8.7  $\pm$ 0.5 & 15.3 $\pm$ 0.9 & 9.3  $\pm$ 0.7 & 2.23 $\pm$ 0.20 & 1.22 $\pm$ 0.11\\
50\%--60\% & 7.1  $\pm$ 0.4 & 12.7 $\pm$ 0.6 & 7.5  $\pm$ 0.5 & 2.18 $\pm$ 0.20 & 1.18 $\pm$ 0.10\\
60\%--70\% & 5.7  $\pm$ 0.4 & 10.4 $\pm$ 0.5 & 5.8  $\pm$ 0.4 & 2.12 $\pm$ 0.20 & 1.12 $\pm$ 0.10\\
\end{tabular}
\end{ruledtabular}
\end{table*}

\begin{table*}[htb]
\caption{Transverse energy results for 200~GeV $^3$He$+$Au 
collisions. The uncertainties include the total statistical 
and systematic uncertainties.}
\label{tab:etHeAu200}
\begin{ruledtabular}
\begin{tabular}{cccccc}
Centrality & $\langle$\Npart$\rangle$ & $\langle$\Nqp$\rangle$ & \dEt [GeV] & \dEtNorm [GeV] & \dEtNormQ [GeV] \\\hline
0\%--5\%   & 25.0 $\pm$ 2.0 & 37.5 $\pm$ 3.1 & 26.7 $\pm$ 1.8 & 2.13 $\pm$ 0.19 & 1.33 $\pm$ 0.14\\
5\%--10\%  & 22.6 $\pm$ 1.3 & 34.3 $\pm$ 2.4 & 23.2 $\pm$ 1.5 & 2.06 $\pm$ 0.18 & 1.27 $\pm$ 0.12\\
10\%--20\% & 19.9 $\pm$ 1.1 & 30.6 $\pm$ 2.2 & 20.6 $\pm$ 1.4 & 2.07 $\pm$ 0.18 & 1.26 $\pm$ 0.12\\
20\%--30\% & 17.0 $\pm$ 1.0 & 26.6 $\pm$ 1.8 & 17.7 $\pm$ 1.2 & 2.08 $\pm$ 0.18 & 1.25 $\pm$ 0.11\\
30\%--40\% & 13.8 $\pm$ 0.7 & 21.9 $\pm$ 1.3 & 14.9 $\pm$ 1.0 & 2.16 $\pm$ 0.18 & 1.27 $\pm$ 0.11\\
40\%--50\% & 10.9 $\pm$ 0.7 & 17.4 $\pm$ 0.8 & 12.0 $\pm$ 0.8 & 2.22 $\pm$ 0.20 & 1.29 $\pm$ 0.10\\
50\%--60\% & 8.16 $\pm$ 0.5 & 13.2 $\pm$ 0.8 & 9.3  $\pm$ 0.6 & 2.29 $\pm$ 0.21 & 1.32 $\pm$ 0.12\\
60\%--70\% & 6.01 $\pm$ 0.4 & 9.7  $\pm$ 0.5 & 7.0  $\pm$ 0.5 & 2.33 $\pm$ 0.21 & 1.34 $\pm$ 0.11\\
\end{tabular}
\end{ruledtabular}
\end{table*}

\begin{table*}[htb]
\caption{Charged particle multiplicity results for 200~GeV $^3$He$+$Au collisions. The uncertainties include the total statistical and systematic uncertainties.}
\label{tab:nchHeAu200}
\begin{ruledtabular}
\begin{tabular}{cccccc}
Centrality & $\langle$\Npart$\rangle$ & $\langle$\Nqp$\rangle$ & \dNch  & \dNchNorm  & \dNchNormQ  \\\hline
0\%--5\%   & 25.0 $\pm$ 2.0 & 37.5 $\pm$ 3.1 & 26.3 $\pm$ 1.8 & 2.10 $\pm$ 0.22 & 1.40 $\pm$ 0.15\\
5\%--10\%  & 22.6 $\pm$ 1.3 & 34.3 $\pm$ 2.4 & 22.7 $\pm$ 1.6 & 2.01 $\pm$ 0.18 & 1.32 $\pm$ 0.13\\
10\%--20\% & 19.9 $\pm$ 1.1 & 30.6 $\pm$ 2.2 & 19.9 $\pm$ 1.4 & 2.00 $\pm$ 0.18 & 1.30 $\pm$ 0.13\\
20\%--30\% & 17.0 $\pm$ 1.0 & 26.6 $\pm$ 1.8 & 16.9 $\pm$ 1.2 & 1.99 $\pm$ 0.18 & 1.27 $\pm$ 0.12\\
30\%--40\% & 13.8 $\pm$ 0.8 & 21.9 $\pm$ 1.3 & 14.0 $\pm$ 1.0 & 2.04 $\pm$ 0.18 & 1.28 $\pm$ 0.12\\
40\%--50\% & 10.9 $\pm$ 0.7 & 17.4 $\pm$ 0.8 & 11.2 $\pm$ 0.8 & 2.06 $\pm$ 0.19 & 1.28 $\pm$ 0.11\\
50\%--60\% & 8.16 $\pm$ 0.5 & 13.2 $\pm$ 0.8 & 8.4  $\pm$ 0.6 & 2.06 $\pm$ 0.20 & 1.27 $\pm$ 0.12\\
60\%--70\% & 6.01 $\pm$ 0.4 & 9.72 $\pm$ 0.5 & 5.9  $\pm$ 0.4 & 1.98 $\pm$ 0.19 & 1.22 $\pm$ 0.11\\
\end{tabular}
\end{ruledtabular}
\end{table*}


\clearpage



\begin{thebibliography}{51}%
\makeatletter
\providecommand \@ifxundefined [1]{%
 \@ifx{#1\undefined}
}%
\providecommand \@ifnum [1]{%
 \ifnum #1\expandafter \@firstoftwo
 \else \expandafter \@secondoftwo
 \fi
}%
\providecommand \@ifx [1]{%
 \ifx #1\expandafter \@firstoftwo
 \else \expandafter \@secondoftwo
 \fi
}%
\providecommand \natexlab [1]{#1}%
\providecommand \enquote  [1]{``#1''}%
\providecommand \bibnamefont  [1]{#1}%
\providecommand \bibfnamefont [1]{#1}%
\providecommand \citenamefont [1]{#1}%
\providecommand \href@noop [0]{\@secondoftwo}%
\providecommand \href [0]{\begingroup \@sanitize@url \@href}%
\providecommand \@href[1]{\@@startlink{#1}\@@href}%
\providecommand \@@href[1]{\endgroup#1\@@endlink}%
\providecommand \@sanitize@url [0]{\catcode `\\12\catcode `\$12\catcode
  `\&12\catcode `\#12\catcode `\^12\catcode `\_12\catcode `\%12\relax}%
\providecommand \@@startlink[1]{}%
\providecommand \@@endlink[0]{}%
\providecommand \url  [0]{\begingroup\@sanitize@url \@url }%
\providecommand \@url [1]{\endgroup\@href {#1}{\urlprefix }}%
\providecommand \urlprefix  [0]{URL }%
\providecommand \Eprint [0]{\href }%
\providecommand \doibase [0]{http://dx.doi.org/}%
\providecommand \selectlanguage [0]{\@gobble}%
\providecommand \bibinfo  [0]{\@secondoftwo}%
\providecommand \bibfield  [0]{\@secondoftwo}%
\providecommand \translation [1]{[#1]}%
\providecommand \BibitemOpen [0]{}%
\providecommand \bibitemStop [0]{}%
\providecommand \bibitemNoStop [0]{.\EOS\space}%
\providecommand \EOS [0]{\spacefactor3000\relax}%
\providecommand \BibitemShut  [1]{\csname bibitem#1\endcsname}%
\let\auto@bib@innerbib\@empty
\bibitem [{\citenamefont {Wang}\ and\ \citenamefont
  {Gyulassy}(2001)}]{Wang:2000bf}%
  \BibitemOpen
  \bibfield  {author} {\bibinfo {author} {\bibfnamefont {X.-N.}\ \bibnamefont
  {Wang}}\ and\ \bibinfo {author} {\bibfnamefont {M.}~\bibnamefont
  {Gyulassy}},\ }\bibfield  {title} {\enquote {\bibinfo {title} {{Energy and
  centrality dependence of rapidity densities at RHIC}},}\ }\href {\doibase
  10.1103/PhysRevLett.86.3496} {\bibfield  {journal} {\bibinfo  {journal}
  {Phys. Rev. Lett.}\ }\textbf {\bibinfo {volume} {86}},\ \bibinfo {pages}
  {3496} (\bibinfo {year} {2001})}\BibitemShut {NoStop}%
\bibitem [{\citenamefont {Kharzeev}\ and\ \citenamefont
  {Nardi}(2001)}]{Kharzeev:2000ph}%
  \BibitemOpen
  \bibfield  {author} {\bibinfo {author} {\bibfnamefont {D.}~\bibnamefont
  {Kharzeev}}\ and\ \bibinfo {author} {\bibfnamefont {M.}~\bibnamefont
  {Nardi}},\ }\bibfield  {title} {\enquote {\bibinfo {title} {{Hadron
  production in nuclear collisions at RHIC and high density QCD}},}\ }\href
  {\doibase 10.1016/S0370-2693(01)00457-9} {\bibfield  {journal} {\bibinfo
  {journal} {Phys. Lett. B}\ }\textbf {\bibinfo {volume} {507}},\ \bibinfo
  {pages} {121} (\bibinfo {year} {2001})}\BibitemShut {NoStop}%
\bibitem [{\citenamefont {Adler}\ \emph {et~al.}(2005)\citenamefont {Adler}
  \emph {et~al.}}]{Adler:2004zn}%
  \BibitemOpen
  \bibfield  {author} {\bibinfo {author} {\bibfnamefont {S.~S.}\ \bibnamefont
  {Adler}} \emph {et~al.} (\bibinfo {collaboration} {PHENIX Collaboration}),\
  }\bibfield  {title} {\enquote {\bibinfo {title} {{Systematic studies of the
  centrality and $\sqrt{s_{NN}}$ dependence of the d E(T) / d eta and d (N(ch)
  / d eta in heavy ion collisions at mid-rapidity}},}\ }\href {\doibase
  10.1103/PhysRevC.71.049901, 10.1103/PhysRevC.71.034908} {\bibfield  {journal}
  {\bibinfo  {journal} {Phys. Rev. C}\ }\textbf {\bibinfo {volume} {71}},\
  \bibinfo {pages} {034908} (\bibinfo {year} {2005})},\ \bibinfo {note}
  {[Erratum: Phys. Rev. C71,049901(2005)]}\BibitemShut {NoStop}%
\bibitem [{\citenamefont {Wang}\ and\ \citenamefont
  {Gyulassy}(1991)}]{Wang:1991hta}%
  \BibitemOpen
  \bibfield  {author} {\bibinfo {author} {\bibfnamefont {X.-N.}\ \bibnamefont
  {Wang}}\ and\ \bibinfo {author} {\bibfnamefont {M.}~\bibnamefont
  {Gyulassy}},\ }\bibfield  {title} {\enquote {\bibinfo {title} {{HIJING: A
  Monte Carlo model for multiple jet production in $pp$, $pA$ and $AA$
  collisions}},}\ }\href {\doibase 10.1103/PhysRevD.44.3501} {\bibfield
  {journal} {\bibinfo  {journal} {Phys. Rev. D}\ }\textbf {\bibinfo {volume}
  {44}},\ \bibinfo {pages} {3501} (\bibinfo {year} {1991})}\BibitemShut
  {NoStop}%
\bibitem [{\citenamefont {Eskola}\ \emph {et~al.}(2000)\citenamefont {Eskola},
  \citenamefont {Kajantie}, \citenamefont {Ruuskanen},\ and\ \citenamefont
  {Tuominen}}]{Eskola:1999fc}%
  \BibitemOpen
  \bibfield  {author} {\bibinfo {author} {\bibfnamefont {K.~J.}\ \bibnamefont
  {Eskola}}, \bibinfo {author} {\bibfnamefont {K.}~\bibnamefont {Kajantie}},
  \bibinfo {author} {\bibfnamefont {P.~V.}\ \bibnamefont {Ruuskanen}}, \ and\
  \bibinfo {author} {\bibfnamefont {Kimmo}\ \bibnamefont {Tuominen}},\
  }\bibfield  {title} {\enquote {\bibinfo {title} {{Scaling of transverse
  energies and multiplicities with atomic number and energy in
  ultrarelativistic nuclear collisions}},}\ }\href {\doibase
  10.1016/S0550-3213(99)00720-8} {\bibfield  {journal} {\bibinfo  {journal}
  {Nucl. Phys. B}\ }\textbf {\bibinfo {volume} {570}},\ \bibinfo {pages} {379}
  (\bibinfo {year} {2000})}\BibitemShut {NoStop}%
\bibitem [{\citenamefont {Lin}\ \emph {et~al.}(2001)\citenamefont {Lin},
  \citenamefont {Pal}, \citenamefont {Ko}, \citenamefont {Li},\ and\
  \citenamefont {Zhang}}]{Lin:2000cx}%
  \BibitemOpen
  \bibfield  {author} {\bibinfo {author} {\bibfnamefont {Z.-W.}\ \bibnamefont
  {Lin}}, \bibinfo {author} {\bibfnamefont {S.}~\bibnamefont {Pal}}, \bibinfo
  {author} {\bibfnamefont {C.~M.}\ \bibnamefont {Ko}}, \bibinfo {author}
  {\bibfnamefont {B.-A.}\ \bibnamefont {Li}}, \ and\ \bibinfo {author}
  {\bibfnamefont {Bin}\ \bibnamefont {Zhang}},\ }\bibfield  {title} {\enquote
  {\bibinfo {title} {{Charged particle rapidity distributions at relativistic
  energies}},}\ }\href {\doibase 10.1103/PhysRevC.64.011902} {\bibfield
  {journal} {\bibinfo  {journal} {Phys. Rev. C}\ }\textbf {\bibinfo {volume}
  {64}},\ \bibinfo {pages} {011902} (\bibinfo {year} {2001})}\BibitemShut
  {NoStop}%
\bibitem [{\citenamefont {Eremin}\ and\ \citenamefont
  {Voloshin}(2003)}]{Eremin:2003qn}%
  \BibitemOpen
  \bibfield  {author} {\bibinfo {author} {\bibfnamefont {S.}~\bibnamefont
  {Eremin}}\ and\ \bibinfo {author} {\bibfnamefont {S.}~\bibnamefont
  {Voloshin}},\ }\bibfield  {title} {\enquote {\bibinfo {title} {{Nucleon
  participants or quark participants?}}}\ }\href {\doibase
  10.1103/PhysRevC.67.064905} {\bibfield  {journal} {\bibinfo  {journal} {Phys.
  Rev. C}\ }\textbf {\bibinfo {volume} {67}},\ \bibinfo {pages} {064905}
  (\bibinfo {year} {2003})}\BibitemShut {NoStop}%
\bibitem [{\citenamefont {Adler}\ \emph {et~al.}(2014)\citenamefont {Adler}
  \emph {et~al.}}]{Adler:2013aqf}%
  \BibitemOpen
  \bibfield  {author} {\bibinfo {author} {\bibfnamefont {S.~S.}\ \bibnamefont
  {Adler}} \emph {et~al.} (\bibinfo {collaboration} {PHENIX Collaboration}),\
  }\bibfield  {title} {\enquote {\bibinfo {title} {{Transverse-energy
  distributions at midrapidity in p+p , d+Au , and Au+Au collisions at
  $\sqrt{s_{NN}}=62.4–200$ GeV and implications for particle-production
  models}},}\ }\href {\doibase 10.1103/PhysRevC.89.044905} {\bibfield
  {journal} {\bibinfo  {journal} {Phys. Rev. C}\ }\textbf {\bibinfo {volume}
  {89}},\ \bibinfo {pages} {044905} (\bibinfo {year} {2014})}\BibitemShut
  {NoStop}%
\bibitem [{\citenamefont {Bialas}\ \emph {et~al.}(1976)\citenamefont {Bialas},
  \citenamefont {Bleszynski},\ and\ \citenamefont {Czyz}}]{Bialas:1976ed}%
  \BibitemOpen
  \bibfield  {author} {\bibinfo {author} {\bibfnamefont {A.}~\bibnamefont
  {Bialas}}, \bibinfo {author} {\bibfnamefont {M.}~\bibnamefont {Bleszynski}},
  \ and\ \bibinfo {author} {\bibfnamefont {W.}~\bibnamefont {Czyz}},\
  }\bibfield  {title} {\enquote {\bibinfo {title} {{Multiplicity Distributions
  in Nucleus-Nucleus Collisions at High-Energies}},}\ }\href {\doibase
  10.1016/0550-3213(76)90329-1} {\bibfield  {journal} {\bibinfo  {journal}
  {Nucl. Phys. B}\ }\textbf {\bibinfo {volume} {111}},\ \bibinfo {pages} {461}
  (\bibinfo {year} {1976})}\BibitemShut {NoStop}%
\bibitem [{\citenamefont {Aoki}\ \emph {et~al.}(2006)\citenamefont {Aoki},
  \citenamefont {Endrodi}, \citenamefont {Fodor}, \citenamefont {Katz},\ and\
  \citenamefont {Szabo}}]{Aoki:2006we}%
  \BibitemOpen
  \bibfield  {author} {\bibinfo {author} {\bibfnamefont {Y.}~\bibnamefont
  {Aoki}}, \bibinfo {author} {\bibfnamefont {G.}~\bibnamefont {Endrodi}},
  \bibinfo {author} {\bibfnamefont {Z.}~\bibnamefont {Fodor}}, \bibinfo
  {author} {\bibfnamefont {S.~D.}\ \bibnamefont {Katz}}, \ and\ \bibinfo
  {author} {\bibfnamefont {K.~K.}\ \bibnamefont {Szabo}},\ }\bibfield  {title}
  {\enquote {\bibinfo {title} {{The Order of the quantum chromodynamics
  transition predicted by the standard model of particle physics}},}\ }\href
  {\doibase 10.1038/nature05120} {\bibfield  {journal} {\bibinfo  {journal}
  {Nature}\ }\textbf {\bibinfo {volume} {443}},\ \bibinfo {pages} {675}
  (\bibinfo {year} {2006})}\BibitemShut {NoStop}%
\bibitem [{\citenamefont {Ejiri}(2008)}]{Ejiri:2008xt}%
  \BibitemOpen
  \bibfield  {author} {\bibinfo {author} {\bibfnamefont {S.}~\bibnamefont
  {Ejiri}},\ }\bibfield  {title} {\enquote {\bibinfo {title} {{Canonical
  partition function and finite density phase transition in lattice QCD}},}\
  }\href {\doibase 10.1103/PhysRevD.78.074507} {\bibfield  {journal} {\bibinfo
  {journal} {Phys. Rev. D}\ }\textbf {\bibinfo {volume} {78}},\ \bibinfo
  {pages} {074507} (\bibinfo {year} {2008})}\BibitemShut {NoStop}%
\bibitem [{\citenamefont {Stephanov}\ \emph {et~al.}(1998)\citenamefont
  {Stephanov}, \citenamefont {Rajagopal},\ and\ \citenamefont
  {Shuryak}}]{Stephanov:1998dy}%
  \BibitemOpen
  \bibfield  {author} {\bibinfo {author} {\bibfnamefont {M.~A.}\ \bibnamefont
  {Stephanov}}, \bibinfo {author} {\bibfnamefont {K.}~\bibnamefont
  {Rajagopal}}, \ and\ \bibinfo {author} {\bibfnamefont {E.~V.}\ \bibnamefont
  {Shuryak}},\ }\bibfield  {title} {\enquote {\bibinfo {title} {{Signatures of
  the tricritical point in QCD}},}\ }\href {\doibase
  10.1103/PhysRevLett.81.4816} {\bibfield  {journal} {\bibinfo  {journal}
  {Phys. Rev. Lett.}\ }\textbf {\bibinfo {volume} {81}},\ \bibinfo {pages}
  {4816} (\bibinfo {year} {1998})}\BibitemShut {NoStop}%
\bibitem [{\citenamefont {Adcox}\ \emph
  {et~al.}(2001{\natexlab{a}})\citenamefont {Adcox} \emph
  {et~al.}}]{Adcox:2000sp}%
  \BibitemOpen
  \bibfield  {author} {\bibinfo {author} {\bibfnamefont {K.}~\bibnamefont
  {Adcox}} \emph {et~al.} (\bibinfo {collaboration} {PHENIX Collaboration}),\
  }\bibfield  {title} {\enquote {\bibinfo {title} {{Centrality dependence of
  charged particle multiplicity in Au$-$Au collisions at
  $\sqrt{s_{NN}}=130$~GeV}},}\ }\href {\doibase 10.1103/PhysRevLett.86.3500}
  {\bibfield  {journal} {\bibinfo  {journal} {Phys. Rev. Lett.}\ }\textbf
  {\bibinfo {volume} {86}},\ \bibinfo {pages} {3500} (\bibinfo {year}
  {2001}{\natexlab{a}})}\BibitemShut {NoStop}%
\bibitem [{\citenamefont {Adcox}\ \emph
  {et~al.}(2001{\natexlab{b}})\citenamefont {Adcox} \emph
  {et~al.}}]{Adcox:2001ry}%
  \BibitemOpen
  \bibfield  {author} {\bibinfo {author} {\bibfnamefont {K.}~\bibnamefont
  {Adcox}} \emph {et~al.} (\bibinfo {collaboration} {PHENIX Collaboration}),\
  }\bibfield  {title} {\enquote {\bibinfo {title} {{Measurement of the
  midrapidity transverse energy distribution from $\sqrt{s_{NN}}=130$~GeV
  Au$+$Au collisions at RHIC}},}\ }\href {\doibase
  10.1103/PhysRevLett.87.052301} {\bibfield  {journal} {\bibinfo  {journal}
  {Phys. Rev. Lett.}\ }\textbf {\bibinfo {volume} {87}},\ \bibinfo {pages}
  {052301} (\bibinfo {year} {2001}{\natexlab{b}})}\BibitemShut {NoStop}%
\bibitem [{\citenamefont {Arsene}\ \emph {et~al.}(2005)\citenamefont {Arsene}
  \emph {et~al.}}]{Arsene:2004fa}%
  \BibitemOpen
  \bibfield  {author} {\bibinfo {author} {\bibfnamefont {I.}~\bibnamefont
  {Arsene}} \emph {et~al.} (\bibinfo {collaboration} {BRAHMS Collaboration}),\
  }\bibfield  {title} {\enquote {\bibinfo {title} {{Quark gluon plasma and
  color glass condensate at RHIC? The Perspective from the BRAHMS
  experiment}},}\ }\href {\doibase 10.1016/j.nuclphysa.2005.02.130} {\bibfield
  {journal} {\bibinfo  {journal} {Nucl. Phys. A}\ }\textbf {\bibinfo {volume}
  {757}},\ \bibinfo {pages} {1} (\bibinfo {year} {2005})}\BibitemShut {NoStop}%
\bibitem [{\citenamefont {Adler}\ \emph {et~al.}(2001)\citenamefont {Adler}
  \emph {et~al.}}]{Adler:2001yq}%
  \BibitemOpen
  \bibfield  {author} {\bibinfo {author} {\bibfnamefont {C.}~\bibnamefont
  {Adler}} \emph {et~al.} (\bibinfo {collaboration} {STAR Collaboration}),\
  }\bibfield  {title} {\enquote {\bibinfo {title} {{Multiplicity distribution
  and spectra of negatively charged hadrons in Au+Au collisions at
  $\sqrt{s_{NN}}=130$~GeV}},}\ }\href {\doibase 10.1103/PhysRevLett.87.112303}
  {\bibfield  {journal} {\bibinfo  {journal} {Phys. Rev. Lett.}\ }\textbf
  {\bibinfo {volume} {87}},\ \bibinfo {pages} {112303} (\bibinfo {year}
  {2001})}\BibitemShut {NoStop}%
\bibitem [{\citenamefont {Alver}\ \emph {et~al.}(2011)\citenamefont {Alver}
  \emph {et~al.}}]{Alver:2010ck}%
  \BibitemOpen
  \bibfield  {author} {\bibinfo {author} {\bibfnamefont {B.}~\bibnamefont
  {Alver}} \emph {et~al.} (\bibinfo {collaboration} {PHOBOS Collaboration}),\
  }\bibfield  {title} {\enquote {\bibinfo {title} {{Phobos results on charged
  particle multiplicity and pseudorapidity distributions in Au+Au, Cu+Cu, d+Au,
  and p+p collisions at ultra-relativistic energies}},}\ }\href {\doibase
  10.1103/PhysRevC.83.024913} {\bibfield  {journal} {\bibinfo  {journal} {Phys.
  Rev. C}\ }\textbf {\bibinfo {volume} {83}},\ \bibinfo {pages} {024913}
  (\bibinfo {year} {2011})}\BibitemShut {NoStop}%
\bibitem [{\citenamefont {Adams}\ \emph {et~al.}(2004)\citenamefont {Adams}
  \emph {et~al.}}]{Adams:2004cb}%
  \BibitemOpen
  \bibfield  {author} {\bibinfo {author} {\bibfnamefont {J.}~\bibnamefont
  {Adams}} \emph {et~al.} (\bibinfo {collaboration} {STAR Collaboration}),\
  }\bibfield  {title} {\enquote {\bibinfo {title} {{Measurements of transverse
  energy distributions in Au + Au collisions at $\sqrt{s_{NN}}$ = 200~GeV}},}\
  }\href {\doibase 10.1103/PhysRevC.70.054907} {\bibfield  {journal} {\bibinfo
  {journal} {Phys. Rev. C}\ }\textbf {\bibinfo {volume} {70}},\ \bibinfo
  {pages} {054907} (\bibinfo {year} {2004})}\BibitemShut {NoStop}%
\bibitem [{\citenamefont {Adcox}\ \emph
  {et~al.}(2003{\natexlab{a}})\citenamefont {Adcox} \emph
  {et~al.}}]{Adcox:2003zm}%
  \BibitemOpen
  \bibfield  {author} {\bibinfo {author} {\bibfnamefont {K.}~\bibnamefont
  {Adcox}} \emph {et~al.} (\bibinfo {collaboration} {PHENIX Collaboration}),\
  }\bibfield  {title} {\enquote {\bibinfo {title} {{PHENIX detector
  overview}},}\ }\href {\doibase 10.1016/S0168-9002(02)01950-2} {\bibfield
  {journal} {\bibinfo  {journal} {Nucl. Instrum. Methods Phys. Res., Sec. A}\
  }\textbf {\bibinfo {volume} {499}},\ \bibinfo {pages} {469} (\bibinfo {year}
  {2003}{\natexlab{a}})}\BibitemShut {NoStop}%
\bibitem [{\citenamefont {Adcox}\ \emph
  {et~al.}(2003{\natexlab{b}})\citenamefont {Adcox} \emph
  {et~al.}}]{Adcox:2003zp}%
  \BibitemOpen
  \bibfield  {author} {\bibinfo {author} {\bibfnamefont {K.}~\bibnamefont
  {Adcox}} \emph {et~al.} (\bibinfo {collaboration} {PHENIX Collaboration}),\
  }\bibfield  {title} {\enquote {\bibinfo {title} {{PHENIX central arm tracking
  detectors}},}\ }\href {\doibase 10.1016/S0168-9002(02)01952-6} {\bibfield
  {journal} {\bibinfo  {journal} {Nucl. Instrum. Methods Phys. Res., Sec. A}\
  }\textbf {\bibinfo {volume} {499}},\ \bibinfo {pages} {489} (\bibinfo {year}
  {2003}{\natexlab{b}})}\BibitemShut {NoStop}%
\bibitem [{\citenamefont {Aphecetche}\ \emph {et~al.}(2003)\citenamefont
  {Aphecetche} \emph {et~al.}}]{Aphecetche:2003zr}%
  \BibitemOpen
  \bibfield  {author} {\bibinfo {author} {\bibfnamefont {L.}~\bibnamefont
  {Aphecetche}} \emph {et~al.} (\bibinfo {collaboration} {PHENIX
  Collaboration}),\ }\bibfield  {title} {\enquote {\bibinfo {title} {{PHENIX
  calorimeter}},}\ }\href {\doibase 10.1016/S0168-9002(02)01954-X} {\bibfield
  {journal} {\bibinfo  {journal} {Nucl. Instrum. Methods Phys. Res., Sec. A}\
  }\textbf {\bibinfo {volume} {499}},\ \bibinfo {pages} {521} (\bibinfo {year}
  {2003})}\BibitemShut {NoStop}%
\bibitem [{\citenamefont {Allen}\ \emph {et~al.}(2003)\citenamefont {Allen}
  \emph {et~al.}}]{Allen:2003zt}%
  \BibitemOpen
  \bibfield  {author} {\bibinfo {author} {\bibfnamefont {M.}~\bibnamefont
  {Allen}} \emph {et~al.} (\bibinfo {collaboration} {PHENIX Collaboration}),\
  }\bibfield  {title} {\enquote {\bibinfo {title} {{PHENIX inner detectors}},}\
  }\href {\doibase 10.1016/S0168-9002(02)01956-3} {\bibfield  {journal}
  {\bibinfo  {journal} {Nucl. Instrum. Methods Phys. Res., Sec. A}\ }\textbf
  {\bibinfo {volume} {499}},\ \bibinfo {pages} {549} (\bibinfo {year}
  {2003})}\BibitemShut {NoStop}%
\bibitem [{\citenamefont {Back}\ \emph {et~al.}(2004)\citenamefont {Back} \emph
  {et~al.}}]{Back:2004dy}%
  \BibitemOpen
  \bibfield  {author} {\bibinfo {author} {\bibfnamefont {B.~B.}\ \bibnamefont
  {Back}} \emph {et~al.} (\bibinfo {collaboration} {PHOBOS Collaboration}),\
  }\bibfield  {title} {\enquote {\bibinfo {title} {{Collision geometry scaling
  of Au+Au pseudorapidity density from $\sqrt{s_{NN}}$ = 19.6~GeV to
  200-GeV}},}\ }\href {\doibase 10.1103/PhysRevC.70.021902} {\bibfield
  {journal} {\bibinfo  {journal} {Phys. Rev. C}\ }\textbf {\bibinfo {volume}
  {70}},\ \bibinfo {pages} {021902} (\bibinfo {year} {2004})}\BibitemShut
  {NoStop}%
\bibitem [{\citenamefont {Richardson}\ \emph {et~al.}(2011)\citenamefont
  {Richardson} \emph {et~al.}}]{Richardson:2010hm}%
  \BibitemOpen
  \bibfield  {author} {\bibinfo {author} {\bibfnamefont {E.}~\bibnamefont
  {Richardson}} \emph {et~al.} (\bibinfo {collaboration} {PHENIX
  Collaboration}),\ }\bibfield  {title} {\enquote {\bibinfo {title} {{A
  Reaction Plane Detector for PHENIX at RHIC}},}\ }\href {\doibase
  10.1016/j.nima.2011.01.034} {\bibfield  {journal} {\bibinfo  {journal} {Nucl.
  Instrum. Methods Phys. Res., Sec. A}\ }\textbf {\bibinfo {volume} {636}},\
  \bibinfo {pages} {99} (\bibinfo {year} {2011})}\BibitemShut {NoStop}%
\bibitem [{\citenamefont {Bleicher}\ \emph {et~al.}(1999)\citenamefont
  {Bleicher} \emph {et~al.}}]{Bleicher:1999xi}%
  \BibitemOpen
  \bibfield  {author} {\bibinfo {author} {\bibfnamefont {M.}~\bibnamefont
  {Bleicher}} \emph {et~al.},\ }\bibfield  {title} {\enquote {\bibinfo {title}
  {{Relativistic hadron hadron collisions in the ultrarelativistic quantum
  molecular dynamics model}},}\ }\href {\doibase 10.1088/0954-3899/25/9/308}
  {\bibfield  {journal} {\bibinfo  {journal} {J. Phys. G}\ }\textbf {\bibinfo
  {volume} {25}},\ \bibinfo {pages} {1859} (\bibinfo {year}
  {1999})}\BibitemShut {NoStop}%
\bibitem [{\citenamefont {Miller}\ \emph {et~al.}(2007)\citenamefont {Miller},
  \citenamefont {Reygers}, \citenamefont {Sanders},\ and\ \citenamefont
  {Steinberg}}]{Miller:2007ri}%
  \BibitemOpen
  \bibfield  {author} {\bibinfo {author} {\bibfnamefont {M.~L.}\ \bibnamefont
  {Miller}}, \bibinfo {author} {\bibfnamefont {K.}~\bibnamefont {Reygers}},
  \bibinfo {author} {\bibfnamefont {S.~J.}\ \bibnamefont {Sanders}}, \ and\
  \bibinfo {author} {\bibfnamefont {P.}~\bibnamefont {Steinberg}},\ }\bibfield
  {title} {\enquote {\bibinfo {title} {{Glauber modeling in high energy nuclear
  collisions}},}\ }\href {\doibase 10.1146/annurev.nucl.57.090506.123020}
  {\bibfield  {journal} {\bibinfo  {journal} {Ann. Rev. Nucl. Part. Sci.}\
  }\textbf {\bibinfo {volume} {57}},\ \bibinfo {pages} {205} (\bibinfo {year}
  {2007})}\BibitemShut {NoStop}%
\bibitem [{\citenamefont {Reisdorf}\ \emph {et~al.}(1997)\citenamefont
  {Reisdorf} \emph {et~al.}}]{Reisdorf:1996qj}%
  \BibitemOpen
  \bibfield  {author} {\bibinfo {author} {\bibfnamefont {W.}~\bibnamefont
  {Reisdorf}} \emph {et~al.} (\bibinfo {collaboration} {FOPI Collaboration}),\
  }\bibfield  {title} {\enquote {\bibinfo {title} {{Central collisions of Au on
  Au at 150, 250 and 400 MeV/nucleon}},}\ }\href {\doibase
  10.1016/S0375-9474(96)00388-0} {\bibfield  {journal} {\bibinfo  {journal}
  {Nucl. Phys. A}\ }\textbf {\bibinfo {volume} {612}},\ \bibinfo {pages} {493}
  (\bibinfo {year} {1997})}\BibitemShut {NoStop}%
\bibitem [{\citenamefont {Ahle}\ \emph {et~al.}(1999)\citenamefont {Ahle} \emph
  {et~al.}}]{Ahle:1999jm}%
  \BibitemOpen
  \bibfield  {author} {\bibinfo {author} {\bibfnamefont {L.}~\bibnamefont
  {Ahle}} \emph {et~al.} (\bibinfo {collaboration} {E802 Collaboration}),\
  }\bibfield  {title} {\enquote {\bibinfo {title} {{Simultaneous multiplicity
  and forward energy characterization of particle spectra in Au + Au collisions
  at 11.6-A~GeV/c}},}\ }\href {\doibase 10.1103/PhysRevC.59.2173} {\bibfield
  {journal} {\bibinfo  {journal} {Phys. Rev. C}\ }\textbf {\bibinfo {volume}
  {59}},\ \bibinfo {pages} {2173} (\bibinfo {year} {1999})}\BibitemShut
  {NoStop}%
\bibitem [{\citenamefont {Afanasiev}\ \emph {et~al.}(2003)\citenamefont
  {Afanasiev} \emph {et~al.}}]{Afanasiev:2002fk}%
  \BibitemOpen
  \bibfield  {author} {\bibinfo {author} {\bibfnamefont {S.~V.}\ \bibnamefont
  {Afanasiev}} \emph {et~al.},\ }\bibfield  {title} {\enquote {\bibinfo {title}
  {{Recent results on spectra and yields from NA49 Collaboration}},}\
  }\bibfield  {booktitle} {\emph {\bibinfo {booktitle} {{Proceedings, 16th
  International Conference on Ultra-Relativistic nucleus nucleus collisions
  (Quark Matter 2012)}}},\ }\href {\doibase 10.1016/S0375-9474(02)01424-0}
  {\bibfield  {journal} {\bibinfo  {journal} {Nucl. Phys. A}\ }\textbf
  {\bibinfo {volume} {715}},\ \bibinfo {pages} {161} (\bibinfo {year}
  {2003})}\BibitemShut {NoStop}%
\bibitem [{\citenamefont {Chatrchyan}\ \emph {et~al.}(2012)\citenamefont
  {Chatrchyan} \emph {et~al.}}]{Chatrchyan:2012mb}%
  \BibitemOpen
  \bibfield  {author} {\bibinfo {author} {\bibfnamefont {S.}~\bibnamefont
  {Chatrchyan}} \emph {et~al.} (\bibinfo {collaboration} {CMS Collaboration}),\
  }\bibfield  {title} {\enquote {\bibinfo {title} {{Measurement of the
  pseudorapidity and centrality dependence of the transverse energy density in
  PbPb collisions at $\sqrt{s_{NN}}$=2.76 TeV}},}\ }\href {\doibase
  10.1103/PhysRevLett.109.152303} {\bibfield  {journal} {\bibinfo  {journal}
  {Phys. Rev. Lett.}\ }\textbf {\bibinfo {volume} {109}},\ \bibinfo {pages}
  {152303} (\bibinfo {year} {2012})}\BibitemShut {NoStop}%
\bibitem [{\citenamefont {Ahle}\ \emph {et~al.}(1998)\citenamefont {Ahle} \emph
  {et~al.}}]{Ahle:1998gv}%
  \BibitemOpen
  \bibfield  {author} {\bibinfo {author} {\bibfnamefont {L.}~\bibnamefont
  {Ahle}} \emph {et~al.} (\bibinfo {collaboration} {E802 Collaboration}),\
  }\bibfield  {title} {\enquote {\bibinfo {title} {{Kaon production in Au + Au
  collisions at 11.6-A~GeV/c}},}\ }\href {\doibase 10.1103/PhysRevC.58.3523}
  {\bibfield  {journal} {\bibinfo  {journal} {Phys. Rev. C}\ }\textbf {\bibinfo
  {volume} {58}},\ \bibinfo {pages} {3523} (\bibinfo {year}
  {1998})}\BibitemShut {NoStop}%
\bibitem [{\citenamefont {Ahle}\ \emph {et~al.}(2000)\citenamefont {Ahle} \emph
  {et~al.}}]{Ahle:2000wq}%
  \BibitemOpen
  \bibfield  {author} {\bibinfo {author} {\bibfnamefont {L.}~\bibnamefont
  {Ahle}} \emph {et~al.} (\bibinfo {collaboration} {E917 and E866
  Collaborations}),\ }\bibfield  {title} {\enquote {\bibinfo {title} {{An
  Excitation function of $K^-$ and $K^+$ production in Au$+$Au reactions at the
  AGS}},}\ }\href {\doibase 10.1016/S0370-2693(00)00916-3} {\bibfield
  {journal} {\bibinfo  {journal} {Phys. Lett. B}\ }\textbf {\bibinfo {volume}
  {490}},\ \bibinfo {pages} {53} (\bibinfo {year} {2000})}\BibitemShut
  {NoStop}%
\bibitem [{\citenamefont {Abelev}\ \emph {et~al.}(2010)\citenamefont {Abelev}
  \emph {et~al.}}]{Abelev:2009bw}%
  \BibitemOpen
  \bibfield  {author} {\bibinfo {author} {\bibfnamefont {B.~I.}\ \bibnamefont
  {Abelev}} \emph {et~al.} (\bibinfo {collaboration} {STAR Collaboration}),\
  }\bibfield  {title} {\enquote {\bibinfo {title} {{Identified particle
  production, azimuthal anisotropy, and interferometry measurements in Au+Au
  collisions at $\sqrt{s_{NN}}$ = 9.2~GeV}},}\ }\href {\doibase
  10.1103/PhysRevC.81.024911} {\bibfield  {journal} {\bibinfo  {journal} {Phys.
  Rev. C}\ }\textbf {\bibinfo {volume} {81}},\ \bibinfo {pages} {024911}
  (\bibinfo {year} {2010})}\BibitemShut {NoStop}%
\bibitem [{\citenamefont {Aamodt}\ \emph {et~al.}(2011)\citenamefont {Aamodt}
  \emph {et~al.}}]{Aamodt:2010cz}%
  \BibitemOpen
  \bibfield  {author} {\bibinfo {author} {\bibfnamefont {K.}~\bibnamefont
  {Aamodt}} \emph {et~al.} (\bibinfo {collaboration} {ALICE Collaboration}),\
  }\bibfield  {title} {\enquote {\bibinfo {title} {{Centrality dependence of
  the charged-particle multiplicity density at mid-rapidity in Pb-Pb collisions
  at $\sqrt{s_{NN}}$=2.76 TeV}},}\ }\href {\doibase
  10.1103/PhysRevLett.106.032301} {\bibfield  {journal} {\bibinfo  {journal}
  {Phys. Rev. Lett.}\ }\textbf {\bibinfo {volume} {106}},\ \bibinfo {pages}
  {032301} (\bibinfo {year} {2011})}\BibitemShut {NoStop}%
\bibitem [{\citenamefont {Aad}\ \emph {et~al.}(2012)\citenamefont {Aad} \emph
  {et~al.}}]{ATLAS:2011ag}%
  \BibitemOpen
  \bibfield  {author} {\bibinfo {author} {\bibfnamefont {G.}~\bibnamefont
  {Aad}} \emph {et~al.} (\bibinfo {collaboration} {ATLAS Collaboration}),\
  }\bibfield  {title} {\enquote {\bibinfo {title} {{Measurement of the
  centrality dependence of the charged particle pseudorapidity distribution in
  lead-lead collisions at $\sqrt{s_{NN}}$=2.76 TeV with the ATLAS detector}},}\
  }\href {\doibase 10.1016/j.physletb.2012.02.045} {\bibfield  {journal}
  {\bibinfo  {journal} {Phys. Lett. B}\ }\textbf {\bibinfo {volume} {710}},\
  \bibinfo {pages} {363} (\bibinfo {year} {2012})}\BibitemShut {NoStop}%
\bibitem [{\citenamefont {Busza}\ \emph {et~al.}(1975)\citenamefont {Busza},
  \citenamefont {Elias}, \citenamefont {Jacobs}, \citenamefont {Swartz},
  \citenamefont {Young},\ and\ \citenamefont {Sogard}}]{Busza:1975df}%
  \BibitemOpen
  \bibfield  {author} {\bibinfo {author} {\bibfnamefont {W.}~\bibnamefont
  {Busza}}, \bibinfo {author} {\bibfnamefont {J.~E.}\ \bibnamefont {Elias}},
  \bibinfo {author} {\bibfnamefont {D.~F.}\ \bibnamefont {Jacobs}}, \bibinfo
  {author} {\bibfnamefont {P.~A.}\ \bibnamefont {Swartz}}, \bibinfo {author}
  {\bibfnamefont {C.~C.}\ \bibnamefont {Young}}, \ and\ \bibinfo {author}
  {\bibfnamefont {M.~R.}\ \bibnamefont {Sogard}},\ }\bibfield  {title}
  {\enquote {\bibinfo {title} {{Charged Particle Multiplicity in pi- Nucleus
  Interactions at 100~GeV/c and 175-GeV/c.}}}\ }\href {\doibase
  10.1103/PhysRevLett.34.836} {\bibfield  {journal} {\bibinfo  {journal} {Phys.
  Rev. Lett.}\ }\textbf {\bibinfo {volume} {34}},\ \bibinfo {pages} {836}
  (\bibinfo {year} {1975})}\BibitemShut {NoStop}%
\bibitem [{\citenamefont {Bjorken}(1983)}]{Bjorken:1982qr}%
  \BibitemOpen
  \bibfield  {author} {\bibinfo {author} {\bibfnamefont {J.~D.}\ \bibnamefont
  {Bjorken}},\ }\bibfield  {title} {\enquote {\bibinfo {title} {{Highly
  Relativistic Nucleus-Nucleus Collisions: The Central Rapidity Region}},}\
  }\href {\doibase 10.1103/PhysRevD.27.140} {\bibfield  {journal} {\bibinfo
  {journal} {Phys. Rev. D}\ }\textbf {\bibinfo {volume} {27}},\ \bibinfo
  {pages} {140} (\bibinfo {year} {1983})}\BibitemShut {NoStop}%
\bibitem [{\citenamefont {Karsch}(2002)}]{Karsch:2001cy}%
  \BibitemOpen
  \bibfield  {author} {\bibinfo {author} {\bibfnamefont {F.}~\bibnamefont
  {Karsch}},\ }\bibfield  {title} {\enquote {\bibinfo {title} {{Lattice QCD at
  high temperature and density}},}\ }\bibfield  {booktitle} {\emph {\bibinfo
  {booktitle} {{Lectures on quark matter. Proceedings, 40. International
  Universitätswochen for theoretical physics, 40th Winter School, IUKT 40}}},\
  }\href {\doibase 10.1007/3-540-45792-5_6} {\bibfield  {journal} {\bibinfo
  {journal} {Lect. Notes Phys.}\ }\textbf {\bibinfo {volume} {583}},\ \bibinfo
  {pages} {209} (\bibinfo {year} {2002})},\ \Eprint
  {http://arxiv.org/abs/hep-lat/0106019} {hep-lat/0106019} \BibitemShut
  {NoStop}%
\bibitem [{\citenamefont {Martinez}()}]{Martinez:2013xka}%
  \BibitemOpen
  \bibfield  {author} {\bibinfo {author} {\bibfnamefont {G.}~\bibnamefont
  {Martinez}},\ }\href@noop {} {\enquote {\bibinfo {title} {{Advances in Quark
  Gluon Plasma}},}\ }\bibinfo {note} {{arXiv:1304.1452}}\BibitemShut {NoStop}%
\bibitem [{\citenamefont {Nouicer}(2007)}]{Nouicer:2006pr}%
  \BibitemOpen
  \bibfield  {author} {\bibinfo {author} {\bibfnamefont {R.}~\bibnamefont
  {Nouicer}},\ }\bibfield  {title} {\enquote {\bibinfo {title} {{Charged
  particle multiplicities in A+A and $p^+ p$ collisions in the constituent
  quarks framework}},}\ }\bibfield  {booktitle} {\emph {\bibinfo {booktitle}
  {{Proceedings, Workshop for Young Scientists on the Phys. of
  Ultrarelativistic Nucleus-Nucleus Collisions (Hot Quarks 2006)}}},\ }\href
  {\doibase 10.1140/epjc/s10052-006-0128-z} {\bibfield  {journal} {\bibinfo
  {journal} {Eur. Phys. J.}\ }\textbf {\bibinfo {volume} {49}},\ \bibinfo
  {pages} {281} (\bibinfo {year} {2007})}\BibitemShut {NoStop}%
\bibitem [{\citenamefont {Heinz}\ and\ \citenamefont
  {Kuhlman}(2005)}]{Heinz:2004ir}%
  \BibitemOpen
  \bibfield  {author} {\bibinfo {author} {\bibfnamefont {U.~W.}\ \bibnamefont
  {Heinz}}\ and\ \bibinfo {author} {\bibfnamefont {A.}~\bibnamefont
  {Kuhlman}},\ }\bibfield  {title} {\enquote {\bibinfo {title} {{Anisotropic
  flow and jet quenching in ultrarelativistic U + U collisions}},}\ }\href
  {\doibase 10.1103/PhysRevLett.94.132301} {\bibfield  {journal} {\bibinfo
  {journal} {Phys. Rev. Lett.}\ }\textbf {\bibinfo {volume} {94}},\ \bibinfo
  {pages} {132301} (\bibinfo {year} {2005})}\BibitemShut {NoStop}%
\bibitem [{\citenamefont {Hirano}\ \emph {et~al.}(2011)\citenamefont {Hirano},
  \citenamefont {Huovinen},\ and\ \citenamefont {Nara}}]{Hirano:2010jg}%
  \BibitemOpen
  \bibfield  {author} {\bibinfo {author} {\bibfnamefont {T.}~\bibnamefont
  {Hirano}}, \bibinfo {author} {\bibfnamefont {P.}~\bibnamefont {Huovinen}}, \
  and\ \bibinfo {author} {\bibfnamefont {Y.}~\bibnamefont {Nara}},\ }\bibfield
  {title} {\enquote {\bibinfo {title} {{Elliptic flow in U+U collisions at
  $\sqrt{s_{NN}}$=200 GeV and in Pb+Pb collisions at $\sqrt{s_{NN}}$=2.76 TeV:
  Prediction from a hybrid approach}},}\ }\href {\doibase
  10.1103/PhysRevC.83.021902} {\bibfield  {journal} {\bibinfo  {journal} {Phys.
  Rev. C}\ }\textbf {\bibinfo {volume} {83}},\ \bibinfo {pages} {021902}
  (\bibinfo {year} {2011})}\BibitemShut {NoStop}%
\bibitem [{\citenamefont {Haque}\ \emph {et~al.}(2012)\citenamefont {Haque},
  \citenamefont {Lin},\ and\ \citenamefont {Mohanty}}]{Haque:2011aa}%
  \BibitemOpen
  \bibfield  {author} {\bibinfo {author} {\bibfnamefont {M.~R.}\ \bibnamefont
  {Haque}}, \bibinfo {author} {\bibfnamefont {Z.-W.}\ \bibnamefont {Lin}}, \
  and\ \bibinfo {author} {\bibfnamefont {B.}~\bibnamefont {Mohanty}},\
  }\bibfield  {title} {\enquote {\bibinfo {title} {{Multiplicity, average
  transverse momentum and azimuthal anisotropy in U+U collisions at
  $\sqrt{s_{NN}}$ = 200 GeV using AMPT model}},}\ }\href {\doibase
  10.1103/PhysRevC.85.034905} {\bibfield  {journal} {\bibinfo  {journal} {Phys.
  Rev. C}\ }\textbf {\bibinfo {volume} {85}},\ \bibinfo {pages} {034905}
  (\bibinfo {year} {2012})}\BibitemShut {NoStop}%
\bibitem [{\citenamefont {Schenke}\ \emph {et~al.}(2014)\citenamefont
  {Schenke}, \citenamefont {Tribedy},\ and\ \citenamefont
  {Venugopalan}}]{Schenke:2014tga}%
  \BibitemOpen
  \bibfield  {author} {\bibinfo {author} {\bibfnamefont {B.}~\bibnamefont
  {Schenke}}, \bibinfo {author} {\bibfnamefont {P.}~\bibnamefont {Tribedy}}, \
  and\ \bibinfo {author} {\bibfnamefont {R.}~\bibnamefont {Venugopalan}},\
  }\bibfield  {title} {\enquote {\bibinfo {title} {{Initial-state geometry and
  fluctuations in Au$+$Au, Cu$+$Au, and U$+$U collisions at energies available
  at the BNL Relativistic Heavy Ion Collider}},}\ }\href {\doibase
  10.1103/PhysRevC.89.064908} {\bibfield  {journal} {\bibinfo  {journal} {Phys.
  Rev. C}\ }\textbf {\bibinfo {volume} {89}},\ \bibinfo {pages} {064908}
  (\bibinfo {year} {2014})}\BibitemShut {NoStop}%
\bibitem [{\citenamefont {Kuhlman}\ and\ \citenamefont
  {Heinz}(2005)}]{Kuhlman:2005ts}%
  \BibitemOpen
  \bibfield  {author} {\bibinfo {author} {\bibfnamefont {A.~J.}\ \bibnamefont
  {Kuhlman}}\ and\ \bibinfo {author} {\bibfnamefont {U.~W.}\ \bibnamefont
  {Heinz}},\ }\bibfield  {title} {\enquote {\bibinfo {title} {{Multiplicity
  distribution and source deformation in full-overlap U+U collisions}},}\
  }\href {\doibase 10.1103/PhysRevC.72.037901} {\bibfield  {journal} {\bibinfo
  {journal} {Phys. Rev. C}\ }\textbf {\bibinfo {volume} {72}},\ \bibinfo
  {pages} {037901} (\bibinfo {year} {2005})}\BibitemShut {NoStop}%
\bibitem [{\citenamefont {Masui}\ \emph {et~al.}(2009)\citenamefont {Masui},
  \citenamefont {Mohanty},\ and\ \citenamefont {Xu}}]{Masui:2009qk}%
  \BibitemOpen
  \bibfield  {author} {\bibinfo {author} {\bibfnamefont {H.}~\bibnamefont
  {Masui}}, \bibinfo {author} {\bibfnamefont {B.}~\bibnamefont {Mohanty}}, \
  and\ \bibinfo {author} {\bibfnamefont {N.}~\bibnamefont {Xu}},\ }\bibfield
  {title} {\enquote {\bibinfo {title} {{Predictions of elliptic flow and
  nuclear modification factor from 200 GeV U + U collisions at RHIC}},}\ }\href
  {\doibase 10.1016/j.physletb.2009.08.025} {\bibfield  {journal} {\bibinfo
  {journal} {Phys. Lett. B}\ }\textbf {\bibinfo {volume} {679}},\ \bibinfo
  {pages} {440} (\bibinfo {year} {2009})}\BibitemShut {NoStop}%
\bibitem [{\citenamefont {Shou}\ \emph {et~al.}(2015)\citenamefont {Shou},
  \citenamefont {Ma}, \citenamefont {Sorensen}, \citenamefont {Tang},
  \citenamefont {Videbaek},\ and\ \citenamefont {Wang}}]{Shou:2014eya}%
  \BibitemOpen
  \bibfield  {author} {\bibinfo {author} {\bibfnamefont {Q.~Y.}\ \bibnamefont
  {Shou}}, \bibinfo {author} {\bibfnamefont {Y.~G.}\ \bibnamefont {Ma}},
  \bibinfo {author} {\bibfnamefont {P.}~\bibnamefont {Sorensen}}, \bibinfo
  {author} {\bibfnamefont {A.~H.}\ \bibnamefont {Tang}}, \bibinfo {author}
  {\bibfnamefont {F.}~\bibnamefont {Videbaek}}, \ and\ \bibinfo {author}
  {\bibfnamefont {H.}~\bibnamefont {Wang}},\ }\bibfield  {title} {\enquote
  {\bibinfo {title} {{Parameterization of Deformed Nuclei for Glauber Modeling
  in Relativistic Heavy Ion Collisions}},}\ }\href {\doibase
  10.1016/j.physletb.2015.07.078} {\bibfield  {journal} {\bibinfo  {journal}
  {Phys. Lett. B}\ }\textbf {\bibinfo {volume} {749}},\ \bibinfo {pages} {215}
  (\bibinfo {year} {2015})}\BibitemShut {NoStop}%
\bibitem [{\citenamefont {Adare}\ \emph {et~al.}(2014)\citenamefont {Adare}
  \emph {et~al.}}]{Adare:2013nff}%
  \BibitemOpen
  \bibfield  {author} {\bibinfo {author} {\bibfnamefont {A.}~\bibnamefont
  {Adare}} \emph {et~al.} (\bibinfo {collaboration} {PHENIX Collaboration}),\
  }\bibfield  {title} {\enquote {\bibinfo {title} {{Centrality categorization
  for $R_{p(d)A}$ in high-energy collisions}},}\ }\href {\doibase
  10.1103/PhysRevC.90.034902} {\bibfield  {journal} {\bibinfo  {journal} {Phys.
  Rev. C}\ }\textbf {\bibinfo {volume} {90}},\ \bibinfo {pages} {034902}
  (\bibinfo {year} {2014})}\BibitemShut {NoStop}%
\bibitem [{\citenamefont {V.}(2015)}]{dennis}%
  \BibitemOpen
  \bibfield  {author} {\bibinfo {author} {\bibfnamefont {Perepelitsa.~D.}\
  \bibnamefont {V.}},\ }\href@noop {} {}\bibinfo {howpublished} {private
  communication} (\bibinfo {year} {2015})\BibitemShut {NoStop}%
\bibitem [{\citenamefont {Hofstadter}(1956)}]{Hofstadter:1956qs}%
  \BibitemOpen
  \bibfield  {author} {\bibinfo {author} {\bibfnamefont {R.}~\bibnamefont
  {Hofstadter}},\ }\bibfield  {title} {\enquote {\bibinfo {title} {{Electron
  scattering and nuclear structure}},}\ }\href {\doibase
  10.1103/RevModPhys.28.214} {\bibfield  {journal} {\bibinfo  {journal} {Rev.
  Mod. Phys.}\ }\textbf {\bibinfo {volume} {28}},\ \bibinfo {pages} {214}
  (\bibinfo {year} {1956})}\BibitemShut {NoStop}%
\bibitem [{\citenamefont {Covolan}\ \emph {et~al.}(1996)\citenamefont
  {Covolan}, \citenamefont {Montanha},\ and\ \citenamefont
  {Goulianos}}]{Covolan:1996uy}%
  \BibitemOpen
  \bibfield  {author} {\bibinfo {author} {\bibfnamefont {R.~J.~M.}\
  \bibnamefont {Covolan}}, \bibinfo {author} {\bibfnamefont {J.}~\bibnamefont
  {Montanha}}, \ and\ \bibinfo {author} {\bibfnamefont {Konstantin~A.}\
  \bibnamefont {Goulianos}},\ }\bibfield  {title} {\enquote {\bibinfo {title}
  {{A New determination of the soft pomeron intercept}},}\ }\href {\doibase
  10.1016/S0370-2693(96)01362-7} {\bibfield  {journal} {\bibinfo  {journal}
  {Phys. Lett. B}\ }\textbf {\bibinfo {volume} {389}},\ \bibinfo {pages} {176}
  (\bibinfo {year} {1996})}\BibitemShut {NoStop}%
\end{thebibliography}

%
 
\end{document}